\documentclass[twocolumn,aps,showpacs,amssymb,superscriptaddress,nofootinbib,10pt]{revtex4-1}

\usepackage{graphicx,color}
\usepackage{subfigure}
\usepackage{setspace}

\usepackage{amsmath,amssymb,ascmac}
\usepackage{type1cm}
\usepackage{bm}

\begin{document}

\title{Properties of a kaon-condensed phase in hyperon-mixed matter \\ with three-baryon forces}

\author{Takumi Muto}
\affiliation{Department of Physics, Chiba Institute of Technology, 2-1-1 Shibazono, Narashino, Chiba 275-0023, Japan}

\date{\today}

\begin{abstract}
Coexistent phase of kaon condensates and hyperons [($Y$+$K$) phase] in beta equilibrium with electrons and muons is investigated as a possible form of dense hadronic phase with multi-strangeness. The effective chiral Lagrangian for kaon-baryon and kaon-kaon interactions is utilized within chiral symmetry approach in combination with the interaction model between baryons. 
For the baryon-baryon interactions, one adopts the minimal relativistic mean-field theory with exchange of scalar mesons ($\sigma$, $\sigma^\ast$) and vector mesons ($\omega$, $\rho$, $\phi$) between baryons without including the nonlinear self-interacting meson field terms. In addition, the universal three-baryon repulsion and the phenomenological three-nucleon attraction are introduced as density-dependent effective two-body potentials. 

The repulsive effects leading to stiff equation of state at high densities consist of both the two-baryon repulsion via the vector-meson exchange and the universal three-baryon repulsion. 
Interplay of kaon condensates with hyperons through chiral dynamics in dense matter is clarified, and resulting onset mechanisms of kaon condensation in hyperon-mixed matter and the equation of state with the ($Y$+$K$) phase and characteristic features of the system are presented.
 
It is shown that the slope $L$ of the symmetry energy controls the two-baryon repulsion beyond the saturation density
 and resulting stiffness of the equation of state. The stiffness of the equation of state in turn controls admixture of hyperons and the onset and development of kaon condensates as a result of competing effect between kaon condensates and hyperons. 
 The equation of state with the ($Y$+$K$) phase becomes stiff enough to be consistent with recent observations of massive neutron stars. Static properties of neutron stars with the ($Y$+$K$) phase are discussed such as gravitational mass to radius, mass to baryon number density relations, and density distribution in the core with the ($Y$+$K$) phase for typical masses of neutron stars. 
\end{abstract}


\maketitle

\section{Introduction}
It has been a long standing issue to clarify properties of various phases, phase structures, and equation of state (EOS) of strongly interacting matter at high density and/or high temperature based on underlying quantum 
chromodynamics (QCD)~\cite{baym2018}. 
Condensation of Nambu-Goldstone (NG) bosons in strongly interacting system, 
which may be realized in dense hadronic matter such as inner core of neutron stars, has been paid much attention as unique exotic phases.
Historically pion condensation (PC) has been considered extensively from both theoretical and observational points of view, although its existence has not yet been established~\cite{sawyer1972,migdal1978,baym1979,kmttt1993}. 
Subsequently, a possible existence of kaon condensation (KC), the Bose-Einstein condensation of $K^-$ meson, has been suggested~\cite{kn86,t88,mt92,m93,mtt93,tpl94,kvk95,lbm95,lee1996,prakash1997,tstw98,fmmt1996}. 

While a driving force of PC is the $p$-wave $\pi$-nucleon ($N$) interaction, KC is realized 
by making use of the $s$-wave $K$-nucleon ($N$) interaction~\cite{kn86,t88,mt92}: the $s$-wave scalar interaction simulated by the $KN$ sigma term $\Sigma_{KN}$ and the $s$-wave vector interaction corresponding to the Tomozawa-Weinberg term. The former is the explicit chiral symmetry breaking term and is larger than the $\pi N$ sigma term reflecting the relatively large breaking of SU(3) flavor symmetry as compared to SU(2) case. 
The strength of the latter is specified by the V-spin charge, $Q_V^b$ $\equiv \frac{1}{2}\left(I_3^{(b)}+\frac{3}{2}Y^{(b)}\right)$ for baryon $b$ with $I_3^{(b)}$ and $Y^{(b)}$ being the third component of the isospin and hypercharge, respectively, and it works attractively for the nucleon sector (1 for $p$ and 1/2 for $n$). 
Thus meson-baryon dynamics specified by chiral symmetry is responsible to formation and properties of such meson-condensed phases~\cite{baym1979,kmttt1993,kn86,t88,mt92}.
It has also been shown that weak interactions relevant to rapid cooling mechanisms of neutron stars via enhanced neutrino emissions in the presence of meson condensates can be considered in a universal framework based on chiral symmetry~\cite{maxwell1977,t88,bkpp1988,fmtt1994}. 
 
Since the possible existence of kaon condensation in neutron stars was suggested, researches on deeply bound kaonic nuclear states have been promoted in the context of strangeness nuclear physics where $\bar K$ mesons are involved\cite{ay02,yda04,ynoh2005,agnello2005,zs2013,bbg2012,ghm2016}. 
The possibility of multi-antikaonic nuclei, where several kaons are embedded in nuclei, 
and their connection to kaon condensation in neutron stars have been discussed~\cite{mmt2009,mmt2014,gfgm2009}. 
Pertinent kaon dynamics in nuclear medium has also been extensively studied with reference to experiments with meson beams at J-PARC, with heavy-ion collisions at GSI and so forth~\cite{bbg2012,hj2012,ghm2016,tolos2020,song2021}. 
Formation of basic kaonic clusters, ``$K^- p p$'', has been reported in the E27 and E15 experiments at J-PARC, and their production mechanisms and structures have been elucidated~\cite{ichikawa2015,sada2016,a2019,yamaga2020,sekihara2016}. 

It has been shown that the EOS with KC is largely softened at high densities~\cite{tpl94,fmmt1996}. In general, the softer EOS makes neutron stars more compact leading to reduce both the maximum gravitational mass and its radius. 
Following the results concerning the softening of the EOS with KC, static and dynamic properties of kaon-condensed neutron stars have been elucidated; 
effects of KC on mass-radius relations of compact stars, formation of mixed phase of kaon condensates immersed in the normal phase due to the first-order phase transition~\cite{g2001,mtvt2006}, and a low-mass black hole scenario by a delayed collapse of hot neutron stars to kaon-condensed stars~\cite{bb1994} are such examples. The delayed collapse of protoneutron stars has also been studied by taking into account thermal effects and neutrino degeneracy~\cite{ty1998,pons2000}. 

Since massive neutron stars as large as 2~$M_\odot$ ($M_\odot$ being the solar mass) were detected~\cite{demo10,ant13,c2020}, precise measurements of mass and radius of neutron stars have been made possible by the development of observational facilities. From $X$-ray observation by Neutron star Interior Composition ExploreR (NICER)~\cite{riley2019,miller2019}, the mass and radius for the pulsar PSR~J0740+6620 has been detected as $R$ = (12.35$\pm$0.75)~km for $M$ = 2.08~$M_\odot$~\cite{miller2021} and $R$ = (12.39$^{+1.30}_{-0.98}$)~km for $M$ = (2.072~$^{+0.067}_{-0.066}$) $M_\odot$~\cite{riley2021}. 
Multi-messenger observations of gravitational waves from neutron-star mergers (GW170817)~\cite{abbott2018}, 
electromagnetic waves from a gamma-ray burst and kilonova emission have brought about important information on the EOS of dense matter~\cite{margalit2017,horowitz2018,shibata2019}. 

Toward understanding a realistic form of hadronic matter, a possible coexistence of KC and hyperons $Y$ ($Y$=$\Lambda$, $\Sigma^-$, $\Xi^-$, $\cdots$) in the ground state of neutron-star matter [abbreviated to ($Y$+$K$) phase
] has been investigated~\cite{m2008,mmtt2019,mmt2021}. In the ($Y$+$K$) phase, softening of the EOS results from combined effects of decreasing energy by the $s$-wave $K$-baryon ($B$) attraction and avoiding the $N$-$N$ repulsion at high densities by hyperon admixture~\cite{nyt02}. Although the onset density of KC is in general pushed up to a higher density in the hyperon-mixed matter than the case of nucleon matter~\cite{ekp95,kpe95,sm96}, the energy gain becomes large once KC occurs in the hyperon-mixed matter. 
In Ref.~\cite{m2008}, both kaon condensates and hyperon-mixing give large energy gain, leading to a local minimum of the energy with respect to $\rho_{\rm B}$, which suggests self-bound objects with ($Y+K$) phase. 
However, such EOS including the local energy minimum cannot reproduce massive compact stars as large as 2~$M_\odot$, so that the self-bound star with ($Y$+$K$) phase is unlikely to exist. 
Most of the other works on coexistent phase of KC and hyperon-mixed matter were not able to exceed the maximum mass as large as 1.85 $M_\odot$~\cite{m2008} except for the case of density-dependent relativistic mean-field (RMF) theory,  where specific density-dependence of the coupling constants between mediating mesons and baryons are assumed~\cite{cb2014,ts2020,mbb2021}. 

Whereas roles of nuclear three-body force have been considered in case of kaon condensation in neutron-star matter~\cite{zuo2004,Li2006},
a large part of softening effect of the EOS with the ($Y$+$K$) phase comes from mixing of hyperons which necessarily appear at high densities~\cite{nyt02}, and the resulting problem associated with the dramatic softening is known as the ``hyperon puzzle''~\cite{nyt02,burgio2021}.
In order to resolve the hyperon puzzle and to make the EOS reconcile with the recent observations of massive neutron stars, introduction of the universal three-baryon repulsion (UTBR) among not only nucleons ($NNN$) but also hyperons and nucleons ($YYY$, $YYN$, $YNN$) on an equal footing has been shown to be indispensable~\cite{nyt02}.  
Subsequently there appeared several works taking into account three-body and multi-body forces between baryons such as multi-pomeron exchange potential~\cite{yamamoto2014,yamamoto2017} and 
baryon-meson ($M$)$M$, $MMM$ type diagrams in the RMF models~\cite{to12}. Further the $\Lambda N$ and $\Lambda N N$ interactions have been studied by the use of the diffusion Monte Carlo method~\cite{lonardoni2015}. 

Recently, theoretical approaches in the chiral effective field theory ($\chi$EFT) have been applied for the $YNN$ 
interaction~\cite{petschauer2020,kohno2018}. As for $\Lambda$ in matter, a possibility of the strongly repulsive $\Lambda$ potential at high density and resulting suppression of the $\Lambda$-mixing has been suggested with reference to the $\chi$EFT~\cite{kohno2018,gkw2020,jinno2023}. It has also been pointed out that the $\Lambda$-suppression at high densities is consistent with the $\Lambda$ directed flow data  in heavy-ion collisions~\cite{nara2022}. 
Therefore there is still controversy about the $\Lambda$ potential at high densities. It is expected that the problem is to be resolved by forthcoming hypernuclear experiments~\cite{taskforce2022}. In this paper, the $B$-$B$ interaction 
model is based on the RMF theory, where $\Lambda$ hyperons necessarily appear at $\rho_{\rm B}$ = (2.5$-$3.0) $\rho_0$ with the nuclear saturation density $\rho_0$ (= 0.16 fm$^{-3}$). 

In the early stages of a series of our works, we considered the ($Y$+$K$) phase by the use of the interaction model  where $K$-$B$ and $K$-$K$ interactions are described by the effective chiral Lagrangian (abbreviated to ChL) and the $B$-$B$ interactions 
are given by exchange of mesons [$\sigma$, $\sigma^\ast$ ($\sim \bar s s$) for scalar mesons, $\omega$, $\rho$, $\phi$ ($\sim \bar s\gamma^0 s$ on the assumption of an ideal mixing between $\omega$ and $\phi$ mesons) for vector mesons] in the RMF with the nonlinear self-interacting (NLSI) meson fields~\cite{mmtt2019,mmt2022,bb77}. 
As is discussed in Ref.~\cite{sw1997}, many-body effects, in particular, all of the leading terms at each power of baryon density arising from the self-consistency condition are repulsive.
The NLSI terms generate many-baryon forces through the equations of motion for the meson mean fields. However, they are introduced solely to reproduce the ground state properties of symmetric nuclear matter (SNM) such as the incompressibility and slope of the symmetry energy around the saturation density. 
With regard to saturation mechanisms of the SNM, it has been pointed out that large cancellation between the repulsive NLSI terms and two-body attraction maintains the saturation energy at $\rho_0$~\cite{mmt2022}. [See also Appendix~\ref{subsec:a4}.] Therefore, the many-baryon forces induced by the NLSI terms lead to quite different picture for the saturation mechanisms compared to the case of the conventional nuclear matter theory, where two-body attraction has a main contribution to the saturation energy and the three-nucleon attraction (TNA) and three-nucleon repulsion (TNR) have complementary contribution to the energy. But both TNA and TNR  play an indispensable role below and beyond $\rho_0$, respectively, in order to reproduce the empirical saturation point of the SNM~\cite{lp1981,wiringa1988,apr1998}. Furthermore, it has been shown that the NLSI terms are not relevant to repulsive energy corresponding to the UTBR leading to the stiff EOS at high densities and that it cannot be compensated with large attractive energy due to the appearance of the ($Y$+$K$) phase~\cite{mmt2022}.  

In order to keep from the significant softening of the EOS with the ($Y$+$K$) phase, another prescription for introducing many-baryon forces has been proposed, standing on the conventional nuclear matter theory around the saturation density. 
 One adopts the RMF model for two-body $B$-$B$ interaction mediated by meson-exchange, removing the NLSI meson potentials. One calls this model a ``minimal RMF'' (abbreviated to MRMF) throughout this paper. 
 Instead of the NLSI terms, one introduces the density-dependent effective two-body potentials for the UTBR, which has been derived from the string-junction model by Tamagaki~\cite{t2008} (SJM2) and originally applied to $Y$-mixed matter by Tamagaki, Takatsuka and Nishizaki\cite{tnt2008}. 
 Together with the UTBR, phenomenological three-nucleon attraction (TNA) has been taken into account, and one has obtained the baryon interaction model that reproduces saturation properties of SNM together with empirical values of incompressibility and symmetry energy at $\rho_0$. In addition, by specifying the slope $L$ of the symmetry energy within an acceptable range, one sets stiffness of the EOS coming from two-baryon repulsion (Two-BR) through the vector meson exchange. Then 
one has investigated the ($Y$+$K$) phase based upon the ChL coupled with this baryon interaction model (MRMF+UTBR+TNA)~\cite{mmt2021}.  
It has been shown that the EOS and the resulting gravitational mass and radius of compact stars with the ($Y$+$K$) phase are consistent with recent observations of massive neutron stars. 

In this paper, one further elucidates the onset mechanisms of KC in hyperon-mixed matter and the EOS with the 
($Y$+$K$) phase in detail with the use of the (ChL+MRMF+UTBR+TNA) model, following up the results of Ref.~\cite{mmt2021}. 
Primary points of view being addressed in the paper are 
interplay between KC and hyperons through chiral dynamics in dense hadronic matter and 
combined effect of the UTBR and Two-BR on the stiffness of the EOS at high densities. 

The paper is organized as follows. In Sec.~\ref{sec:kaon}, chiral symmetry approach for kaon condensation based on the effective chiral Lagrangian is introduced. The ``$K$-$B$ sigma terms'' are estimated by taking into account of the nonlinear 
effect with respect to the strange quark mass beyond chiral perturbation in the next to leading order. 
In Sec.~\ref{sec:Bforce}, the MRMF is explained in meson-exchange picture for $B$-$B$ interaction. In addition, the UTBR (SJM2 as a specific model) and TNA are introduced phenomenologically. Section~\ref{sec:ground-state} gives description of the ground state for the ($Y$+$K$) phase: energy density expression for the ($Y$+$K$) phase, together with classical field equations for KC and meson mean fields, and ground state conditions. In Sec.~\ref{sec:SNM}, the results on the properties of symmetric nuclear matter and pure neutron matter with the present interaction model are given, prior to the results on KC in hyperon-mixed matter. The dependence of the EOS on the slope $L$ is also discussed. Section~\ref{sec:KP} is devoted to the description of pure hyperon-mixed matter, which is followed by the discussion of kaon properties in hyperon-mixed matter and onset of KC. In Sec.~\ref{sec:KC-EOS}, results of the ground state properties of the ($Y$+$K$) phase and the EOS are given. In Sec.~\ref{sec:MR}, static properties of neutron stars with the ($Y$+$K$) phase are discussed with the results such as gravitational mass to radius, mass to baryon number density relations, and density distribution in the core with the ($Y$+$K$) phase for typical masses of neutron stars.   Summary and concluding remarks are given in Sec.~\ref{sec:summary}. In Appendix~\ref{subsec:appendixA}, the expressions of the baryon masses and the $K$-$B$ sigma terms are given by the use of the Feynman-Hellmann theorem for the effective chiral Lagrangian. In Appendix~\ref{subsec:appendixB}, the expressions for the $s$-wave on-shell $K$-$N$ scattering lengths are given in order to determine the coefficients of the range terms and coupling constants associated with the $\Lambda$~(1405). The expressions for quantities related to saturation properties in SNM, i.~e. the incompressibility, the symmetry energy, and the slope $L$ of the symmetry energy, are given in Appendix~\ref{subsec:appendixC}. In Appendix~\ref{subsec:a4}, the RMF model with the NLSI terms [(MRMF+NLSI) model] is overviewed for comparison with the (MRMF+UTBR+TNA) model.

\section{Chiral symmetry approach for kaon condensation}
\label{sec:kaon}  

The ($Y$+$K$) phase is composed of kaon condensates and hyperon-mixed baryonic matter together with  leptons, being kept in beta equilibrium, charge neutrality, and baryon number conservation. In the following, one simply takes into account protons, neutrons, $\Lambda$, $\Sigma^-$, and $\Xi^-$ hyperons for baryons and electrons and muons for leptons. 

\subsection{Kaon-baryon and multi-kaon interactions}
\label{subsec:kb}

The model for $K$-$B$ and $K$-$K$ interactions is based upon the effective chiral SU(3)$_{\rm L}$$\times$SU(3)$_{\rm R}$ Lagrangian, which is given by
\begin{eqnarray}
{\cal L}_{K,B}&=&\frac{1}{4}f^2 \ {\rm Tr} 
\partial^\mu U^\dagger\partial_\mu U 
+\frac{1}{2}f^2\Lambda_{\chi{\rm SB}}({\rm Tr}M(U-1)+{\rm h.c.}) \cr\cr
&+&{\rm Tr}\overline{\Psi}(i{\gamma^\mu\partial_\mu}-M_B)\Psi 
+{\rm Tr}\overline{\Psi} \gamma^\mu\lbrack V_\mu, \Psi\rbrack \cr\cr
&+& D {\rm Tr}\overline{\Psi}\gamma^\mu \gamma^5\lbrace A_\mu, \Psi\rbrace
+F {\rm Tr}\overline{\Psi}\gamma^\mu \gamma^5\lbrack A_\mu, \Psi\rbrack \cr\cr
&+& a_1{\rm Tr}\overline{\Psi}(\xi M^\dagger\xi+{\rm h.c.})\Psi 
+a_2{\rm Tr}\overline{\Psi}\Psi(\xi M^\dagger\xi+{\rm h.c.}) \cr\cr
&+& a_3({\rm Tr}MU +{\rm h.c.}){\rm Tr}\overline{\Psi}\Psi \cr\cr
&+& 2d_1{\rm Tr}(A^\mu A_\mu) {\rm Tr}(\overline\Psi\Psi)
+4d_2{\rm Tr}(\overline\Psi A^\mu A_\mu\Psi) \ , 
\label{eq:lagkb}
\end{eqnarray}
where the first and second terms on the r.~h.~s. of Eq.~(\ref{eq:lagkb}) are the kinetic and mass terms of the kaons $U$ in the nonlinear representation, respectively, and $\Lambda_{\chi{\rm SB}}$ is the chiral symmetry breaking scale ($\approx$ 1 GeV), $M$ the quark mass matrix,  
$M\equiv {\rm diag}(m_u, m_d, m_s)$ with the quark masses $m_i$\cite{kn86}. 
The nonlinear charged kaon field $U$ is represented by a chiral rotation of the meson vacuum ${\bf 1}_{\rm vac}$ as $U=g_{\rm L}{\bf 1}_{\rm vac}g_{\rm R}^\dagger=\xi^2$ with $g_{\rm L}=g_{\rm R}^\dagger=\xi=\exp[\sqrt{2}i(K^+T_{4+i5}+K^-T_{4-i5})/f]$, where $T_{4\pm i5}$ ($\equiv T_4\pm iT_5$) is the flavor SU(3) generators and $f$ is the meson decay constant. The value of $f$ is set to be the one of the pion decay constant ($f_\pi$ = 93 MeV)  in the lowest-order in chiral perturbation theory.  
The kaon rest mass is identified with  $m_K=\lbrack\Lambda_{\chi{\rm SB}}(m_u+m_s)\rbrack^{1/2}$ and is set to be the empirical value (=493.677 MeV).  

The third term in Eq.~(\ref{eq:lagkb}) denotes the kinetic and mass terms for the octet baryon field $\Psi$, where $M_{\rm B}$ is a  baryon mass generated by spontaneous chiral symmetry breaking in the chiral limit.   
The fourth term in Eq.~(\ref{eq:lagkb}) gives the $s$-wave $K$-$B$ vector interaction corresponding to the Tomozawa-Weinberg term with $V_\mu$ being the vector current of the kaons defined by $V_\mu\equiv 
i/2(\xi^\dagger\partial_\mu\xi+\xi\partial_\mu\xi^\dagger)$. 
The fifth and sixth terms are the $K$-$B$ axial-vector interaction with the mesonic axial-vector current defined by $A_\mu\equiv i/2(\xi^\dagger\partial_\mu\xi-\xi\partial_\mu\xi^\dagger)$. Throughout this paper, one simply omits these axial-vector coupling terms and retain only the $s$-wave $K$-$B$ vector interaction in order to figure out the consequences from $s$-wave kaon condensation.
 
 The remaining terms in Eq.~(\ref{eq:lagkb}) give the terms in the next to leading-order $O(m_K^2)$ in chiral expansion. The terms with coefficients $a_1 \sim a_3$ bring about the 
$s$-wave $K$-$B$ scalar interaction which explicitly breaks chiral symmetry. The last two terms with $d_1$ and $d_2$ are range terms. In addition to (\ref{eq:lagkb}), one further takes into account phenomenologically the pole term of the $\Lambda$~(1405) (abbreviated to $\Lambda^\ast$), which lies by about 30 MeV below the $K^- p$ threshold. 
The $\Lambda^\ast$ pole contribution is introduced by the Lagrangian for $K^- p\Lambda^\ast$ interaction:
\begin{equation}
{\cal L}_{K^- p\Lambda^\ast}=-g_{\Lambda^\ast}\bar\Lambda^\ast \gamma^\mu p \ \partial_\mu K^- +{\rm h.~c.} 
\label{eq:lagLambda-star}
\end{equation}
with the coupling constant $g_{\Lambda^\ast}$. 

The condensed kaon field is assumed to be spatially uniform with kaon spatial momentum ${\bf k}=0$~\cite{t88,mt92,mmt2021} and represented classically as 
 \begin{equation}
K^\pm =\frac{f}{\sqrt{2}}\theta\exp(\pm i\mu_K t) \ , 
\label{eq:kfield}
\end{equation}
where $\theta$ is the chiral angle, and $\mu_K$ is the $K^-$ chemical potential. 
By the use of Eq.~(\ref{eq:kfield}), the  Lagrangian density (\ref{eq:lagkb}) with (\ref{eq:lagLambda-star}) is separated into the  kaon part ${\cal L}_K$ and the baryon part ${\cal L}_B$ in the mean-field approximation: 
${\cal L}_{K,B}={\cal L}_K+{\cal L}_B$. 
For ${\cal L}_K$ one reads~\cite{mmt2021,mmt2022}
\begin{equation}
{\cal L}_K=f^2\Big\lbrack\frac{1}{2}(\mu_K\sin\theta)^2 - m_K^2(1-\cos\theta)
+2 \mu_K X_0 (1-\cos\theta)\Big\rbrack \ ,
\label{eq:lagk}
\end{equation}
where the last term in the bracket stands for the $s$-wave $K$-$B$ vector interaction with $X_0$ being given by 
\begin{eqnarray}
X_0&\equiv&\frac{1}{2f^2}\sum_{b=p,n,\Lambda, \Sigma^-, \Xi^-} Q_V^b\rho_b \cr
&=& \frac{1}{2f^2}\left(\rho_p+\frac{1}{2}\rho_n-\frac{1}{2}\rho_{\Sigma^-}-\rho_{\Xi^-} \right)  \ , 
\label{eq:x0}
\end{eqnarray}
where $\rho_b$ and $Q_V^b$ 
are the number density and V-spin charge, respectively for baryon species $b$. The form of Eq.~(\ref{eq:x0}) for $X_0$ is specified model-independently within chiral symmetry. 
From Eqs.~(\ref{eq:lagk}) and (\ref{eq:x0}), one can see that the $s$-wave $K$-$B$ vector interaction works attractively for protons and neutrons, while repulsively for $\Sigma^-$ and $\Xi^-$ hyperons, as far as $\mu_K > 0$, and there is no $s$-wave  $K$-$\Lambda$ vector interaction.  

For ${\cal L}_B$ one reads 
\begin{equation}
 {\cal L}_B = \sum_{b=p,n,\Lambda, \Sigma^-, \Xi^-}\overline{\psi}_b (i\gamma^\mu \partial_\mu-M_b^\ast ) \psi_b \ , 
 \label{eq:lagb}
\end{equation} 
where $\psi_b$ is each baryon field $b$ and $M_b^\ast$ is the effective baryon mass :
\begin{widetext}
\begin{equation}
M_b^\ast =M_b -\Sigma_{Kb}(1-\cos\theta) 
-\frac{1}{2}(\mu_K\sin\theta)^2\Bigg\lbrace \frac{d_b}{m_K}+\delta_{b,p}\frac{g_{\Lambda^\ast}^2}{2}\frac{\delta M_{\Lambda^\ast p} -\mu_K}{(\delta M_{\Lambda^\ast p} -\mu_K)^2+\gamma_{\Lambda^\ast}^2}\Bigg\rbrace \ ,
\label{eq:effbm}
\end{equation}
\end{widetext}
where $M_b$ is the empirical baryon mass, i.~e.~, $M_p$$\approx$$M_n$ = 939.57 MeV, $M_\Lambda$=1115.68 MeV, $M_{\Sigma^-}$=1197.45 MeV, and $M_{\Xi^-}$=1321.71 MeV. 
The second term on the r.~h.~s. in Eq.~(\ref{eq:effbm}) represents modification of the free baryon mass $M_b$ through the $s$-wave $K$-$B$ scalar interaction simulated by  the ``kaon-baryon sigma terms'' $\Sigma_{Kb}$ ($b=(p, n, \Lambda, \Sigma^-, \Xi^-$). They can be read from the three terms with coefficients $a_1$, $a_2$, $a_3$ in Eq.~(\ref{eq:lagkb}):
\begin{subequations}\label{eq:kbsigma}
\begin{eqnarray}
\Sigma_{Kn}&=&-(a_2+2a_3)(m_u+m_s) = \Sigma_{K\Sigma^-} \ ,\label{eq:kbsigma1} \\
\Sigma_{K\Lambda}&=& -\left(\frac{5}{6}a_1+\frac{5}{6}a_2+2a_3\right)(m_u+m_s) \ , \label{eq:kbsigma2} \\
\Sigma_{Kp}&=&-(a_1+a_2+2a_3)(m_u+m_s) = \Sigma_{K\Xi^-} . \label{eq:kbsigma3}
\end{eqnarray}
\end{subequations}
  (See also Appendix~\ref{subsec:appendixA}.)
  
  In Eq.~(\ref{eq:effbm}), the dimensionless constants $d_b$ ($b=p, n, \Lambda, \Sigma^-, \Xi^-$) of the range terms are related with $d_p\equiv (d_1+d_2)m_K/2$ = $d_{\Xi^-}$, $d_n \equiv d_1m_K/2$ = $d_{\Sigma^-}$, $d_\Lambda=(d_1+5d_2/6)m_K/2$ in terms of $d_1$, $d_2$, and $m_K$. 
The mass difference between the $\Lambda^\ast$ and $p$ is denoted as $\delta M_{\Lambda^\ast p}\equiv M_{\Lambda^\ast}-M_p$,  
and $\gamma_{\Lambda^\ast}$ is introduced as the imaginary part of the inverse propagator for the $\Lambda^\ast$. 
The constants $d_p$, $d_n$, $g_{\Lambda^\ast}$, and $\gamma_{\Lambda^\ast}$ are determined to reproduce the $s$-wave on-shell $K$-$N$ scattering lengths including the pole contribution from $\Lambda^\ast$~\cite{martin1981}. 
\footnote{
There has been a novel improvement on the $\bar K$-$N$ interaction by the SIDDHARTA experiment through the kaonic hydrogen $X$-rays~\cite{siddharta2011}, while there is not a quantitatively large difference of the value of the $s$-wave $\bar K N$ scattering length between Ref.~\cite{martin1981} and \cite{siddharta2011}. Therefore the conventional values in Ref.~\cite{martin1981} are adopted.}
One obtains 
$g_{\Lambda^\ast}$ = 0.583, $\gamma_{\Lambda^\ast}$=12.4~MeV, $d_p$ = $0.351-\Sigma_{Kp}/m_K$, 
$d_n$ = $0.130-\Sigma_{Kn}/m_K$ (see Appendix~\ref{subsec:appendixB}).
As is the case with Ref.~~\cite{fmmt1996}, the range terms and $\Lambda^\ast$-pole term are absorbed into the effective baryon mass as shown in Eq.~(\ref{eq:effbm}). 

It has become established by the use of chiral SU(3) dynamics that the $\Lambda^\ast$ is the $\bar K N$ bound state with the strong couplings such as $\bar KN\leftrightarrows\pi\Sigma$~\cite{ksw1995,or1998,lk2002,ihw2011,ro2000}. Properties of $\bar K$ -$N$ dynamics in the medium was studied in the chiral unitary approach by taking into account the self-consistent $\bar K N$ self-energy and the Pauli-blocking effects for the nucleon in the medium~\cite{ro2000}.
The two resonance pole structure of the $\Lambda^\ast$ has also been advocated~\cite{om2001,joo2003,hj2012}. 
In this paper, one simply takes a conventional single pole picture for the $\Lambda^\ast$, 
since, as shown in Sec.~\ref{subsec:onsetKC}, the $\Lambda^\ast$ contribution to the off-mass shell $K^-$ self-energy for $\rho_{\rm B}\gtrsim \rho_0$  is negligible, irrespective of details of the $\Lambda^\ast$ structure.

\subsection{Estimation of the $K$-$B$ sigma terms}
\label{subsec:sigma}

In line with Ref.~\cite{kn86}, the quark masses $m_i$ are set to be ($m_u, m_d, m_s$) = (6, 12, 240) MeV. 
The parameters $a_1$ and $a_2$ are fixed to be $a_1$ = $-$0.28, $a_2$ = 0.56 so as to reproduce the empirical octet baryon mass splittings, where the octet baryon rest masses are given Eq.~(\ref{eq:fbmass}) in Appendix~\ref{subsec:appendixA} in the leading-order chiral perturbation.  
With this set of parameters, the following relations should be assured. 
For the flavor nonsinglet condensate, 
\begin{eqnarray}
\sigma_0&\equiv&\hat m\langle N|(\bar u u + \bar d d-2\bar s s)|N\rangle \cr 
&=&\Sigma_{\pi N}-(2\hat m/m_s) \sigma_s \cr
&\simeq&\frac{3}{m_s/\hat m-1}~(M_{\Xi}-M_\Lambda) \simeq 25~{\rm MeV} \ , 
\label{eq:sigma0-dmass}
\end{eqnarray}
 where $\hat m\equiv (m_u+m_d)/2$, $\Sigma_{\pi N}$ is the $\pi N$ sigma term defined by 
\begin{equation}
\Sigma_{\pi N} = \hat m \langle N|(\bar uu+\bar dd)|N\rangle = -2\hat m (a_1+2a_3) \ , 
\label{eq:sigma-piN}
\end{equation}
 and $\sigma_s$ ($\equiv m_s\langle N|\bar s s |N\rangle$) is the strangeness condensate in the nucleon. 
For $\sigma_3\equiv\hat m \langle p|(\bar u u -\bar d d )|p\rangle$ (= $-2\hat m a_1$), 
\begin{equation}
\sigma_3\simeq \frac{1}{m_s/\hat m -1}(M_{\Xi}-M_{\Sigma})\simeq 5~{\rm  MeV} \ . 
\label{eq:sigma3-dmass}
\end{equation}
Recent lattice QCD results suggest small $\bar s s $ condensate in the nucleon, i.~e., $y_N\equiv 2\langle N| \bar s s|N\rangle/\langle N|(\bar u u+\bar d d) |N\rangle$ = 0.03 $-$ 0.2~\cite{ohki08,Durr2016,Alex2020}. In particular, for $\sigma_s$ = 0, one can see from Eq.~(\ref{eq:sigma0-dmass}) 
that $\sigma_0=\Sigma_{\pi N}\simeq$ 25~MeV. This value of $\Sigma_{\pi N}$ is too small compared to the phenomenological values (40$-$60) MeV~\cite{gls1991,a2021} which are deduced from the analyses of $\pi$-$N$ scattering and pionic atoms, or lattice QCD results $\approx$ 40 MeV~\cite{a2021}. 
It has been shown that nonlinear effects beyond chiral perturbation can make both the value of $\Sigma_{\pi N}$ and the octet baryon mass splittings
 consistent with experiments with a small strangeness content of the proton~\cite{a2021,jk1987,hk1991}.  
Here one takes into account the nonlinear effect on the quark condensates which originate from the additional rest mass contribution of baryons, $\Delta M(m_s)$ in higher order with respect to $m_s$ (see the Appendix~\ref{subsec:appendixA}). 
Then the $Kb$ sigma terms (\ref{eq:kbsigma}) are modified by the replacement: $a_3\rightarrow \widetilde a_3\equiv a_3-\Delta/4$ with $\Delta\equiv \partial\Delta M/\partial m_s$.   
The nonlinear effect $\Delta$ is absorbed into $\widetilde a_3$. 

\begin{figure}[h]
\begin{center}
\includegraphics[height=0.30\textheight]{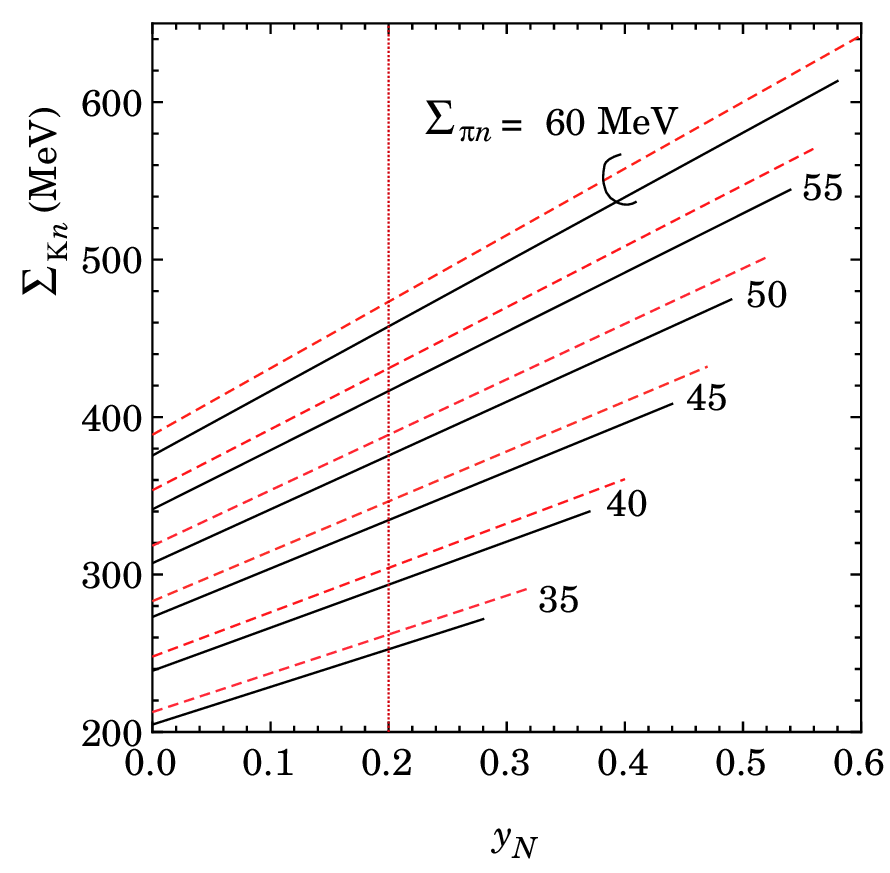}~
\end{center}~
\vspace{0.5cm}~
\caption{The $K$-neutron sigma term $\Sigma_{Kn}$ as a function of $y_N$ ($\equiv 2\langle N| \bar s s|N\rangle/\langle N|(\bar u u+\bar d d) |N\rangle$) for a given value of $\Sigma_{\pi N}$ = (35$-$60) ~MeV. The solid lines (dashed lines) are for the current quark masses ($m_u, m_d, m_s$)=(6, 12, 240)~MeV [($m_u, m_d, m_s$)=(2.2, 4.7, 95)~MeV]. The vertical dotted line denotes the upper value of $y_N$ = 0.2 suggested by the lattice QCD results, taken from~\cite{ohki08,Durr2016, Alex2020}. The right endpoint of each line corresponds to $\Delta$ = 0 (the case of chiral perturbation). See the text for details.\\ }
\label{fig:sigkn}
\end{figure}

To estimate the allowable value of the $\Sigma_{KN}$, one starts with the general relation among $\Sigma_{KN}$, $\Sigma_{\pi N}$, and $\sigma_s$, or $y_N$, 
\begin{subequations}\label{eq:knsigma2}
\begin{eqnarray}
\Sigma_{KN}&=&\frac{m_u+m_s}{2\hat m}\left(\frac{\Sigma_{\pi N}}{1+z_N}+\frac{\hat m}{m_s}\sigma_s\right)  \label{eq:knsigma2-1} \\
&=&\frac{m_u+m_s}{2\hat m}\Sigma_{\pi N}\left(\frac{1}{1+z_N}+\frac{1}{2}y_N\right) \ , \label{eq:knsigma2-2}
\end{eqnarray}
\end{subequations}
where 
\begin{equation}
z_N\equiv \langle N|\bar d d|N\rangle / \langle N|\bar u u|N\rangle \qquad (N=p, n) \ .
\label{eq:zN}
\end{equation}
The parameters $z_N$ and $y_N$ are rewritten specifically as 
\begin{equation}
z_p=\frac{a_3}{a_1+a_3}=1/z_n \ ,  
\label{eq:zN1} 
\end{equation}
\begin{equation}
y_N =\frac{2(a_2+ a_3)-\Delta}{a_1+2 a_3} \qquad (N=p, n) \ . 
\label{eq:yN}
\end{equation}
Once $\Sigma_{\pi N}$ is given, $a_3$ and $z_p$ ($z_n$) are determined from Eqs.~(\ref{eq:sigma-piN}) and (\ref{eq:zN1}), respectively. 
Then $\Sigma_{KN}$ is obtained from Eq.~(\ref{eq:knsigma2-2}) together with $y_N$, which is given as a function of $\Delta$ through Eq.~(\ref{eq:yN}). 
With the nonlinear effect $\Delta$, $\sigma_0$ is represented as $\sigma_0=-2\hat m (a_1-2a_2+\Delta)$. 
In Fig.~\ref{fig:sigkn}, the $K$-neutron sigma term $\Sigma_{Kn}$ as a function of $y_N$ is shown at a fixed value of $\Sigma_{\pi N}$ considering uncertainty ranging from  35~MeV to 60~MeV. The vertical dotted line shows boundaries of the allowable region for $y_N$, taken from~\cite{ohki08,Durr2016, Alex2020}.
The standard value for $\Sigma_{\pi N}$ has been taken to be $\approx$ 45~MeV phenomenologically. Recently the higher values (50$-$60)~MeV are suggested from the phenomenological analyses~\cite{a2021}. 
In view of this, reading off from Fig.~\ref{fig:sigkn}, one takes two cases of $\Sigma_{Kn}$ = 300 MeV with $y_N$ = 0 and 400 MeV with $y_N$ = 0.2 as typical values for $\Sigma_{Kn}$ throughout this paper. The corresponding quantities, $\Sigma_{\pi N}$, $\Delta$, and $\Sigma_{Kb}$ ($b=p, \Lambda, \Sigma^-, \Xi^-$) together with $a_3$, $\widetilde a_3$ are also determined for fixed values of ($m_u$, $m_d$, $m_s$) and ($a_1$, $a_2$). 
The result is listed in Table~\ref{tab:kbsigma}. 
\begin{table*}[!]
\caption{The parameters $a_1$, $a_2$, $a_3$, $\widetilde a_3$ in the chiral-symmetry breaking terms in the effective chiral Lagrangian (\ref{eq:lagkb}), and the quantities in terms of them for two sets of the current quark masses ($m_u$, $m_d$, $m_s$): $y_N\equiv 2\langle N| \bar s s|N\rangle/\langle N|(\bar u u+\bar d d) |N\rangle$, $\Delta$ being a shift of the strangeness content in the nucleon from the value in the leading order chiral perturbation, $\Sigma_{\pi N}$ the $\pi N$ sigma term, and the ``K-baryon sigma terms'' $\Sigma_{Kb}$ ($b=p, n, \Lambda, \Sigma^-, \Xi^-$) adopted in this work. The $K$-neutron sigma term, $\Sigma_{Kn}$, is set to be two typical values 300 MeV and 400 MeV. See the text for details. 
}
\begin{center}
\begin{tabular}{c || c || c  c || c c | c c c c }
\hline
    ($m_u, m_d, m_s$) & ($a_1, a_2$) & $a_3$ & $\Sigma_{\pi N}$ & $y_N$ & $\Delta$ & $\widetilde a_3$ & $\Sigma_{Kn}$ (= $\Sigma_{K\Sigma^-}$)   &  $\Sigma_{Kp}$ (= $\Sigma_{K\Xi^-}$)  & $\Sigma_{K\Lambda}$  \\
(MeV)  & &  & (MeV)  & & &  & (MeV) & (MeV) & (MeV)   \\ \hline\hline
(6, 12, 240)~\cite{kn86} & ($-$0.280, 0.560) & $-$1.22  & 49.0 & 0 & $-$1.32 & $-$0.890 & 300 & 369 & 380   \\
 &  &$-$1.33  & 53.0 & 0.20 & $-$0.955 & $-$1.09   & 400 & 469 & 480   \\\hline
(2.2, 4.7, 95)~\cite{PDG2020} &  ($-$0.697, 1.37) & $-$3.09 & 47.4  & 0 &$-$3.43 & $-$2.23 & 300 & 368 & 379  \\
&  & $-$3.37 & 51.3 & 0.20 & $-$2.51 & $-$2.74  & 400 & 468 & 479  \\\hline
\end{tabular}
\label{tab:kbsigma}
\end{center}
\end{table*} 

Recent lattice simulation results on the current quark masses give rather smaller values, $(m_u, m_d, m_s)$ = (2.2, 4.7, 95) MeV~\cite{PDG2020}, than adopted in this paper. However, the ratio $(m_u+m_s)/2\hat m$ ($\approx$14) is almost the same in both set of quark masses. Therefore, one can see from (\ref{eq:knsigma2}) that the $\Sigma_{KN}$ is little altered by the use of different quark masses once the $\Sigma_{\pi N}$ is given. It can also be shown that the least-square fitting for the empirical octet mass splitting by the use of the lighter quark masses little changes the result of not only $\Sigma_{KN}$ but also $\Sigma_{KY}$, as seen from Table~\ref{tab:kbsigma}. 

It is to be noted that the pion-baryon sigma terms [$\sigma_{bq}=\hat m \langle b|(\bar u u +\bar d d)|b\rangle$] and strangeness sigma terms [$\sigma_{bs}=m_s \langle b|\bar s s|b\rangle$] in the octet baryons ($b$) have been derived from analyses of the lattice QCD simulations for the octet baryon masses~\cite{sty2013,Lutz2014}. The relative ordering of $\sigma_{bq}$, $\sigma_{bs}$ for $b= (N, Y(=\Lambda, \Sigma^-, \Xi^-))$ estimated in the present model agrees well with those in Ref.~\cite{sty2013}, while there is a little difference in absolute values for $\sigma_{Yq}$, $\sigma_{Ys}$ between the present results and those in Fig.~2 in~\cite{sty2013}. 
\vspace{0.3cm}

\section{Baryon interactions}
\label{sec:Bforce}

In the framework in this paper, the baryon interactions are separated into two parts: (I) Two-body $B$-$B$ interaction which may be relevant mainly to the intermediate and long-range regions ($\rho_{\rm B}\lesssim 2\rho_0$), where baryons can be considered to be point-like and the $B$-$B$ interaction is mainly mediated by the scalar ($\sigma$, $\sigma^\ast$) mesons and vector ($\omega$, $\rho$, $\phi$) mesons in the RMF.  
(II) Phenomenological three-body forces. One is the three-nucleon repulsion (TNR) and TNA, both of which are important to reproduce the saturation of the SNM around $\rho_0$. Another is the three-baryon repulsion (TBR), which is an extension of the TNR to high-densities where hyperons as well as nucleons are occupied in matter. TBR is relevant to the short-range region, where quark structure of baryon reveals itself, and the origin of short-range repulsion is connected with the quark confinement mechanism which is spin-flavor independent. 

\subsection{Minimal RMF for baryon-baryon interaction}
\label{subsec:MRMF}

The two-body $B$-$B$ interaction in the MRMF is implemented by the following prescription for the baryon part of the Lagrangian ${\cal L}_B$ [(\ref{eq:lagb})]: (i) The scalar meson ($\sigma$, $\sigma^\ast$)-$B$ couplings are absorbed into the effective baryon mass by the replacement of 
$M_b^\ast$ [(\ref{eq:effbm})] with 
\begin{equation}
M_b^\ast\rightarrow\widetilde M_b^\ast\equiv M_b^\ast - g_{\sigma b}\sigma-g_{\sigma^\ast b}\sigma^\ast \ ,
\label{eq:wtmb}
\end{equation}
where $g_{\sigma b}$, $g_{\sigma^\ast b}$ are the scalar-meson-baryon  coupling constants. 
  (ii) The vector meson ($\omega$, $\rho$, $\phi$)-$B$ couplings are introduced through the replacement of the derivative in (\ref{eq:lagb}) by the covariant derivative, 
  $\partial_\mu\rightarrow D_\mu^{(b)}\equiv \partial_\mu+i g_{\omega b}\omega_\mu+i g_{\rho b} {\hat I}_3^{\ (b)} R_\mu^3 +ig_{\phi b}\phi_\mu$, where the vector meson fields for the $\omega$, $\rho$, $\phi$ mesons
are denoted as $\omega^\mu$, $R_a^\mu$ with the isospin component $a$, and $\phi^\mu$, respectively, $g_{mb}$ is the vector meson-$B$ coupling constant and ${\hat I}_3^{\ (b)}$ is a sign of the third component of the isospin for baryon $b$. Together with the free meson part of the Lagrangian density, one obtains the $B$-$M$ Lagrangian density as 
\begin{eqnarray}
{\cal L}_{B,M}&=&\sum_{b}\overline{\psi}_b \left(i\gamma^\mu D_\mu^{(b)}-\widetilde M_b^\ast \right) \psi_b \cr
&+&\frac{1}{2}\left(\partial^\mu\sigma\partial_\mu\sigma - m_\sigma^2\sigma^2\right) 
+\frac{1}{2}\left(\partial^\mu\sigma^\ast\partial_\mu\sigma^\ast-m_{\sigma^\ast}^2\sigma^{\ast 2}\right) \cr\cr
&-&\frac{1}{4}\omega^{\mu\nu}\omega_{\mu\nu}+\frac{1}{2}m_\omega^2\omega^\mu\omega_\mu 
-\frac{1}{4}R_a^{\mu\nu} R^a_{\mu\nu}+\frac{1}{2}m_\rho^2 R_a^\mu R^a_\mu \cr\cr
&-& \frac{1}{4}\phi^{\mu\nu}\phi_{\mu\nu}+\frac{1}{2}m_\phi^2\phi^\mu\phi_\mu \ ,
\label{eq:lagbm}
\end{eqnarray}
where the kinetic terms of the vector mesons are given in terms of 
$\omega^{\mu\nu}\equiv \partial^\mu \omega^\nu-\partial^\nu \omega^\mu$, $R_a^{\mu\nu}\equiv \partial^\mu R_a^\nu-\partial^\nu R_a^\mu$, and $\phi^{\mu\nu}\equiv \partial^\mu \phi^\nu-\partial^\nu \phi^\mu$. 
Throughout this paper, only the time-components of the vector mean fields, $\omega_0$, $R_0$ ($\equiv R_0^{\rm 3}$), $\phi_0$, are considered for description of the ground state and are taken to be uniform. 
The meson masses are set to be $m_\sigma$ = 400 MeV, $m_{\sigma^\ast}$ = 975 MeV, $m_\omega$ = 783 MeV, 
$m_\rho$ = 769 MeV, and $m_\phi$ = 1020 MeV. 

In Eq.~(\ref{eq:lagbm}), there is no extra term with nonlinear self-interacting meson potentials, which would bring about three-body or many-body baryon interactions. Instead, in the present framework, many-body baryon interactions, which should be relevant to the stiffness of the EOS in high densities, are represented by phenomenological three-body forces. 

\subsection{Three-baryon repulsive force}
\label{subsec:TBR}

It may be naturally understood that the three-body repulsion is qualitatively independent on spin-flavor of baryons, reflecting the confinement mechanisms of quarks at high-density region. Thus it is assumed to work universally between any baryon species. Along with this viewpoint, one adopts a specific model for the UTBR proposed by Tamagaki based on the string-junction model (SJM) ~\cite{t2008,tnt2008}.   
One utilizes the density-dependent effective two-body potential $U_{\rm SJM}(1,2;\rho_{\rm B})$ between baryons 1 and 2, by integrating out variables of the third baryon participating the UTBR: 
\begin{widetext}
\begin{eqnarray}
U_{\rm SJM}(1,2; \rho_{\rm B}) &=&\int d^3 {\bf r}_3 \sum_\gamma^{({\rm occ.})}\langle\phi_\gamma (3)| W({\bf r}_1, {\bf r}_2; {\bf r}_3)|\phi_\gamma(3)\rangle f_{\rm src}^2({\bf r}_1-{\bf r}_3)f_{\rm src}^2({\bf r}_2-{\bf r}_3) \cr
&=&
\rho_{\rm B}\int d^3 {\bf r}_3W({\bf r}_1, {\bf r}_2;{\bf r}_3)f_{\rm src}^2({\bf r}_1-{\bf r}_3)f_{\rm src}^2({\bf r}_2-{\bf r}_3) \ , 
\label{eq:usjm}
\end{eqnarray}
\end{widetext}
where $W({\bf r}_1, {\bf r}_2; {\bf r}_3)$ is the three-body baryon interaction, $f_{\rm src}^2({\bf r}_i-{\bf r}_j)$ is the short-range correlation (s.r.c.)~function squared between $B_i$ and $B_j$, and ``occ.'' stands for the occupied states. 
The three-body baryon interaction is written as  
\begin{equation}
W({\bf r}_1, {\bf r}_2; {\bf r}_3)=W_0 g({\bf r}_1-{\bf r}_3)g({\bf r}_2-{\bf r}_3)
\label{eq:w}
\end{equation}
 with $W_0$ being the strength of the order of $B$-antibaryon ($\bar B$) excitation energy ($\simeq$ 2 GeV)~\cite{t2008}, and $g({\bf r}_i-{\bf r}_j)$ being the wavefunction between $B_i$ and $B_j$. Taking the wavefunction $g({\bf r})$ as the Gaussian form, $g({\bf r})=\exp[-(r/\eta_c)^2]$ with $r=|{\bf r}_i-{\bf r}_j|$ and $\eta_c$ [= (0.45 $-$ 0.50) fm] being the range of the repulsive core for baryon forces, one obtains
\begin{equation}
U_{\rm SJM}(r; \rho_{\rm B})=\frac{\rho_{\rm B}W_0}{2\pi^2}\int_0^\infty dq q^2 j_0(qr)\left(G_{\rm src}(q)\right)^2 \ , 
\label{eq:usjm3}
\end{equation}
where $G_{\rm src}(q)$ is the Fourier transform of the wavefunction $g({\bf r})$ multiplied by the s.r.c. function : 
\begin{equation}
G_{\rm src}(q)
=4\pi\int_0^\infty dr r^2 f_{\rm src}^2({r})\exp[-(r/\eta_c)^2] j_0(qr) 
\label{eq:gsrc}
\end{equation}
with $j_0(qr)=\sin(qr)/(qr)$ being the spherical Bessel function of the 1st kind. 
The $f_{\rm src}(r)$ is used as the statistical weighted average of the $^1{\rm E}$ and $^3{\rm O}$ wave functions as
$f_{\rm src}(r)\equiv [f_{\rm src}(^1{\rm E})(r)+3f_{\rm src}(^3{\rm O})(r)]/4$, where, for simplicity, the result of the reaction matrix calculation in neutron matter by the use of the one-pion exchange Gaussian (OPEG)-A potential is substituted for $f_{\rm src}(r)$, by the assumption that the density-dependence is weak~\cite{t2008,tnt2008}: 
\begin{eqnarray}
 f_{\rm src}^{^1{\rm E}}(r)&=&1-1.2e^{-(r/0.6)^2}+0.4e^{-(r/1.2)^2}  \ , \cr
 f_{\rm src}^{^3{\rm O}}(r)&=&1-0.55e^{-(r/0.7)^2}+0.06e^{-(r/1.6)^2} \ . 
\label{eq:fsrc}
\end{eqnarray}
In the following, the approximate form of $U_{\rm SJM}$ is used as 
\begin{equation}
U_{\rm SJM2}(r; \rho_{\rm B}) = V_r \rho_{\rm B}(1+c_r\rho_{\rm B}/\rho_0)\exp[-(r/\lambda_r)^2)] \ ,
\label{eq:aUTBR}
\end{equation}
 where $V_r$=95 MeV$\cdot$fm$^3$, $c_r$=0.024, and $\lambda_r$=0.86 fm corresponding to $\eta_c$ = 0.50 fm for SJM2~\cite{tnt2008}. The $U_{\rm SJM}$ grows almost linearly with $\rho_{\rm B}$. 
Finally one obtains the effective two-body potential, 
$\widetilde U_{\rm SJM}(r;~\rho_{\rm B})~=f_{\rm src}(r) U_{\rm SJM}(r; \rho_{\rm B})$. 

\subsection{Three-nucleon attractive force}
\label{subsec:TNA}

To simulate the attractive contribution from the TNA to the binding energy for $\rho_{\rm B}\lesssim \rho_0$ , one adopts the density-dependent effective two-body potential by Nishizaki, Takatsuka and Hiura~(NTH~1994)~\cite{nth1994}, which was phenomenologically introduced and the direct term of which agrees with the expression by Lagaris and Pandharipande~(LP1981)~\cite{lp1981} :  
\begin{equation}
U_{\rm TNA}(r; \rho_{\rm B})=V_a\rho_{\rm B} \exp(-\eta_a \rho_{\rm B})\exp[-(r/\lambda_a)^2]
(\vec{\bf\tau}_1\cdot\vec{\bf\tau}_2)^2 \ ,
\label{eq:tna}
\end{equation}
where the range parameter $\lambda_a$ is fixed to be 2.0 fm. 
The $U_{\rm TNA}(r; \rho_{\rm B})$ depends upon not only density but also isospin $\vec \tau_1\cdot\vec \tau_2$ with Pauli matrices $\vec \tau_i$. The parameters $V_a$ and $\eta_a$ are determined together with 
other parameters to reproduce the saturation properties of the SNM for the allowable values of $L$ (see Sec.~\ref{subsec:stiff-SNM}). 

\section{Description of the ground state for the ($Y$+$K$) phase}
\label{sec:ground-state}

The present interaction model in the MRMF consists of the Lagrangian density ${\cal L}_{K}$ +${\cal L}_{B,M}$ [Eqs.~(\ref{eq:lagk}) and (\ref{eq:lagbm})]. In addition, the three-baryon forces UTBR and TNA are taken into account as the effective two-baryon potentials [Eqs.~(\ref{eq:aUTBR}) and (\ref{eq:tna})]. 

\subsection{Energy density expression for the ($Y$+$K$) phase}
\label{subsec:energy}

The energy density ${\cal E}$ for the ($Y$+$K$) phase is separated into the KC part, ${\cal E}_K$, the baryon kinetic part and meson part for two-body baryon interactions, ${\cal E}_{B,M}$, three-body interaction parts, ${\cal E}$~(UTBR)+${\cal E}$~(TNA), and free lepton parts, ${\cal E}_e$ for electrons and ${\cal E}_\mu$ for muons. 
From (\ref{eq:lagk}) and (\ref{eq:lagbm}) one obtains
\begin{equation}
{\cal E}_K=\frac{1}{2}(\mu_K f\sin\theta)^2+f^2m_K^2(1-\cos\theta) \ , 
\label{eq:ekfinal}
\end{equation}
\vspace{-0.5cm}~
\begin{eqnarray}
{\cal E}_{B,M}&=&\sum_b \frac{2}{(2\pi)^3}\int_{|{\bf p}|\leq p_F(b)} d^3|{\bf p}|(|{\bf p}|^2+\widetilde M_b^{\ast 2})^{1/2} \cr
&+&\frac{1}{2}\left(m_\sigma^2\sigma^2+m_{\sigma^\ast}^2\sigma^{\ast 2}\right) \cr\cr
&+& \frac{1}{2}\left(m_\omega^2\omega_0^2+m_\rho^2 R_0^2+m_\phi^2\phi_0^2\right) \ ,  
\label{eq:ebm}
\end{eqnarray}
where baryons ($b$) are occupied within each Fermi sphere with Fermi momentum $p_F(b)$. 
 
The contribution from the UTBR is written in the Hartree approximation as
\begin{eqnarray}
{\cal E}~({\rm UTBR}) &=& 2\pi\rho_{\rm B}^2\int d rr^2 \widetilde U_{\rm SJM2}(r; \rho_{\rm B}) \cr
&=& \frac{\pi^{3/2}}{2}V_r (\widetilde\lambda_r)^3\rho_{\rm B}^3\left(1+c_r\frac{\rho_{\rm B}}{\rho_0}\right) \cr
&=&  \frac{\pi^{3/2}}{2}\rho_{\rm B}^2 U_{\rm SJM2}(r=0; \rho_{\rm B})\cdot (\widetilde\lambda_r)^3 \ ,  
\label{eq:edUTBRtil}
\end{eqnarray}
where $\displaystyle (\widetilde\lambda_r)^3\equiv \frac{4}{\pi^{1/2}}\int_0^\infty dr r^2f_{\rm src}(r)e^{-(r/\lambda_r)^2}$ (=0.589496 $\cdots$fm$^3$) for SJM2. With the use of the spatial average for the s.~r.~c. function $f_{\rm src}(r)$ being denoted as $\overline{f}_{\rm src}$, one can write $(\widetilde\lambda_r)^3\simeq \overline{f}_{\rm src}\cdot (\lambda_r)^3$. Thus $\widetilde\lambda_r$ is interpreted as the range of the effective two-body potential $\widetilde U_{\rm SJM2}(r; \rho_{\rm B})$.
It is to be noted that the exchange term of the energy density contribution from the UTBR is given by
\begin{equation}
{\cal E}^{\rm ex}({\rm UTBR}) = -\int d^3 r \sum_b\vert C_b(r)\vert^2 \widetilde U_{\rm SJM2}(r; \rho_{\rm B}) \ , 
\label{eq:edUTBRex}
\end{equation}
where
\begin{equation}
 C_b(r)\equiv \int_{|{\bf k}_b|\leq p_F(b)}\frac{d^3 |{\bf k}_b|}{(2\pi)^3} e^{-i{\bf k}_b\cdot {\bf r}} 
 = \frac{3}{2}\rho_b\frac{j_1(p_F(b)r)}{p_F(b)r} 
 \label{eq:Cb}
 \end{equation}
with $j_1(x)$ [$\equiv(\sin x-x\cos x)/x^2$] being the spherical Bessel function of the 1st kind, $\rho_b$ and $p_F (b)$ the number density and the Fermi momentum of baryon $b$ ($b=p, n, \Lambda, \Sigma^-, \Xi^-$), respectively. The exchange contribution (\ref{eq:edUTBRex}) to the energy density is shown to be numerically small in comparison with the direct term (\ref{eq:edUTBRtil}) except for $r\approx 0$. Therefore, it is reasonable to neglect the exchange contribution to the energy density in line with the Hartree approximation in the RMF framework.  

Likewise the energy-density contribution from the direct term of the TNA is represented as
\begin{eqnarray}
{\cal E}~({\rm TNA})&=&\frac{1}{2} \int d^3 r V_a\rho_{\rm B} e^{-\eta_a \rho_{\rm B}} e^{-(r/\lambda_a)^2} \cr
&\times&\rho_{\rm B}^2\lbrace 3-2(1-2x_p)^2\rbrace \cr
&=&\gamma_a\rho_{\rm B}^3e^{-\eta_a\rho_{\rm B}}\lbrace 3-2(1-2x_p)^2\rbrace 
\label{eq:edTNA}
\end{eqnarray}
with $\displaystyle\gamma_a\equiv(\pi^{3/2}/2) V_a\lambda_a^3$ and $x_p=\rho_p/\rho_{\rm B}$ the proton-mixing ratio. The exchange term of the TNA has been shown to be very small in comparison with (\ref{eq:edTNA})~\cite{nth1994}, so it is also neglected. 

For leptons, the energy density contribution from the ultra-relativistic electrons is given as
\begin{equation}
{\cal E}_e\simeq\mu_e^4/(4\pi^2) \ . 
\label{eq:ee}
\end{equation}
In case $\mu_\mu > m_\mu$ with $\mu_\mu$ being the muon chemical potential and $m_\mu$ the muon mass (=105.66 MeV), there appear muons in the ground state and participate in charge neutrality and $\beta$ equilibrium conditions. The energy density for muons are given as
\begin{eqnarray}
\hspace{-0.8cm}{\cal E}_\mu&=&\frac{2}{(2\pi)^3}\int_{|{\bf p}|\leq p_F(\mu^-)} d^3|{\bf p}|(|{\bf p}|^2+m_\mu^2)^{1/2} \cr
&=&\frac{m_\mu^4}{8\pi^2}\Big\lbrack r(1+2r^2)\sqrt{r^2+1} 
- \log(r+\sqrt{r^2+1}) \Big\rbrack 
\label{eq:emuon}
\end{eqnarray}
with $r\equiv p_F(\mu^-)/m_\mu$, where 
\begin{equation}
p_F(\mu^-) \equiv\sqrt{\mu^2-m_\mu^2}\cdot\Theta(\mu^2-m_\mu^2)
\label{eq:pFmu}
\end{equation}
 is the muon Fermi momentum with the charge chemical potential $\mu$ and the step function $\Theta(x)$.
 
With Eqs.~(\ref{eq:ekfinal})$-$(\ref{eq:emuon}), the total energy density ${\cal E}$ is given by
\begin{equation}
{\cal E}={\cal E}_K+{\cal E}_{B,M}+{\cal E}({\rm UTBR})+{\cal E}({\rm TNA})+{\cal E}_e+{\cal E}_\mu \ .
\label{eq:total-edensity}
\end{equation} 

\subsection{Classical field equations for kaon condensates and meson mean fields}
\label{subsec:eom}

Throughout this paper, the classical $K^-$ field ($\displaystyle |K^-|=f\theta/\sqrt{2}$) and meson mean fields ($\sigma, \sigma^\ast, \omega, \rho, \phi$) 
are set to be uniform and only depends on total baryon density $\rho_{\rm B}$. 
The equations of motion for these fields are derived from the Lagrangian density ${\cal L}_K+{\cal L}_{B,M}$ in the mean-field approximation. 

The classical kaon field equation follows from 
\[ \partial({\cal L}_K+{\cal L}_{B,M})/\partial\theta=0 \ . \] 
One obtains
\begin{equation}
\mu_K^2Z^{-1}\cos\theta+2X_0\mu_K-m_K^{\ast 2}=0 \ , 
\label{eq:keom2}
\end{equation} 
where 
\begin{equation}
Z^{-1}\equiv 1 + \frac{1}{f^2}\sum_b\rho_b^s\Bigg\lbrace \frac{d_b}{m_K}+\delta_{b,p}\frac{g_{\Lambda^\ast}^2}{2}\frac{\delta M_{\Lambda^\ast p} -\mu_K}{(\delta M_{\Lambda^\ast p} -\mu_K)^2
+\gamma_{\Lambda^\ast}^2}\Bigg\rbrace  \ , 
\label{eq:Z}
\end{equation}
and the effective kaon mass squared is defined by
\begin{equation}
m_K^{\ast 2}\equiv m_K^2-\frac{1}{f^2}\sum_{b=p,n,\Lambda, \Sigma^-, \Xi^-}\rho_b^s\Sigma_{Kb}  
\label{eq:ekm2}
\end{equation}
with $\rho_b^s$ being a scalar density for baryon $b$: 
\begin{equation}
 \rho_b^s=\frac{2}{(2\pi)^3} \int_{|{\bf p}|\leq p_F(b)} d^3|{\bf p}|\frac{\widetilde M_b^\ast}{(|{\bf p}|^2
+\widetilde M_b^{\ast 2})^{1/2}} \ . 
\label{eq:rhobs}
 \end{equation}
For the equations of motion for meson fields, one obtains

\begin{subequations}\label{eq:cieom}
\begin{eqnarray}
m_\sigma^2\sigma&=&\sum_{b=p,n,\Lambda,\Sigma^-,\Xi^-}g_{\sigma b}\rho_b^s \ , \label{eq:cieom1} \\
m_\sigma^{\ast 2}\sigma^\ast&=&\sum_{Y=\Lambda,\Sigma^-,\Xi^-}g_{\sigma^\ast Y}\rho_Y^s \ , \label{eq:cieom2}\\
m_\omega^2\omega_0&=&\sum_{b=p,n,\Lambda,\Sigma^-,\Xi^-} g_{\omega b}\rho_b \ ,  \label{eq:cieom4}\\
m_\rho^2 R_0&=&\sum_{b=p,n,\Lambda,\Sigma^-,\Xi^-} g_{\rho b}{\hat I}_3^{(b)} \rho_b \ , \label{eq:cieom5} \\
m_\phi^2\phi_0 &=&\sum_{Y=\Lambda,\Sigma^-,\Xi^-}g_{\phi Y}\rho_Y \ . \label{eq:cieom6}
\end{eqnarray}
\end{subequations}

\subsection{Ground-state conditions}
\label{subsec:grcond}

The ground state energy for the ($Y+K$) phase is obtained under the charge neutrality, baryon number, and $\beta$-equilibrium conditions. The charge neutrality condition is written as  
\begin{equation}
\rho_Q=\rho_p-\rho_{\Sigma^-}-\rho_{\Xi^-}-\rho_{K^-}-\rho_e-\rho_\mu=0 \ , 
\label{eq:charge}
\end{equation}
where $\rho_Q$ denotes the total negative charge density, $\rho_{K^-}$ is the number density of KC and is given from kaon part of the Lagrangian density (\ref{eq:lagk}) as
\begin{eqnarray}
\rho_{K^-}&=&-iK^-(\partial{\cal L}_K/\partial\dot{K^-})+iK^+(\partial{\cal L}_K/\partial\dot{K^+}) \cr
&=&\mu_K f^2\sin^2\theta+2f^2X_0(1-\cos\theta) \ . 
\label{eq:rhokc}
\end{eqnarray}
In Eq.~(\ref{eq:charge}), $\rho_e$ is the electron number density and is related to the electron chemical potential $\mu_e$ as $\rho_e=\mu_e^3/(3\pi^2)$ in the ultra-relativistic limit. $\rho_\mu$ is the muon number density and is given by $\rho_\mu=[p_F(\mu^-)]^3/(3\pi^2)$. 

The baryon number conservation is given by
\begin{equation}
\rho_p +\rho_n + \rho_\Lambda+\rho_{\Sigma^-}+\rho_{\Xi^-}=\rho_{\rm B} \ .
\label{eq:bn}
\end{equation}
In addition, the following chemical equilibrium conditions for weak processes are imposed: 
 $n\rightleftharpoons p+K^-$, $n\rightleftharpoons p+e^- (+\bar\nu_e)$, $n + e^-\rightleftharpoons \Sigma^-(+\nu_e)$, $\Lambda + e^-\rightleftharpoons \Xi^-(+\nu_e)$, $n\rightleftharpoons \Lambda(+\nu_e\bar\nu_e)$, and those involved in muons in place of $e^-$ if muons are present. 
 These conditions are followed by the relations between the chemical potentials 
\begin{eqnarray}
\mu=\mu_K&=&\mu_e=\mu_\mu=\mu_n-\mu_p \ , \cr
\mu_\Lambda&=&\mu_n , \cr
 \mu_{\Sigma^-}&=&\mu_{\Xi^-}=\mu_n+\mu_e \ ,
\label{eq:chem}
\end{eqnarray}
where $\mu$ and $\mu_i$ (=$\partial{\cal E}/\partial \rho_i$) ($i$= $p$, $n$, $\Lambda$, $\Sigma^-$, $\Xi^-$, $K^-$, $e^-$, $\mu^-$) are the charge chemical 
potential and the chemical potential for each particle species ($i$), respectively, at a given baryon number density $\rho_{\rm B}$. 
For baryon $b$ ($b$ = $p$, $n$, $\Lambda$, $\Sigma^-$, $\Xi^-$ ), the baryon chemical potentials are denoted as
\begin{widetext}
\begin{eqnarray}
\mu_b=\partial{\cal E}/\partial\rho_b =\left(p_F(b)^2+\widetilde M_b^{\ast 2}\right)^{1/2}&+&g_{\omega b}\omega_0+g_{\rho b}{\hat I}_3^{(b)} R_0+g_{\phi b}\phi_0-\mu Q_V^b(1-\cos\theta) \cr
&+&\frac{285}{2}\pi^{3/2}(\widetilde\lambda_r)^3\rho_{\rm B}^2\left(1+0.032\rho_{\rm B}/\rho_0\right)+\delta_{b,N}\Delta\mu_{\rm TNA} \ , 
\label{eq:mub}
\end{eqnarray}
\end{widetext}
where the last two terms on the r.~h.~s. are contributions from the UTBR, $\partial{\cal E}_{\rm UTBR}/\partial\rho_b$,  and from the TNA, $\partial{\cal E}_{\rm TNA}/\partial\rho_b$, respectively, with
\begin{widetext}
\begin{eqnarray}
\Delta\mu_{\rm TNA}=
\begin{cases}
\gamma_a\Big\lbrace(3-\eta_a\rho_{\rm B})(1+8x_p-8x_p^2)+8(1-2x_p)(1-x_p)\Big\rbrace e^{-\eta_a\rho_{\rm B}}\rho_{\rm B}^2 & {\rm for}  \ N=p \ , \\
\gamma_a\Big\lbrace(3-\eta_a\rho_{\rm B})(1+8x_p-8x_p^2)-8(1-2x_p)x_p\Big\rbrace e^{-\eta_a\rho_{\rm B}}\rho_{\rm B}^2 & {\rm for}  \ N=n \ . \\
\end{cases}
\label{eq:muTNA}
\end{eqnarray}
From Eq.~(\ref{eq:mub}), baryon potential $V_b$ is identified with
\begin{eqnarray}
V_b = -g_{\sigma b}\sigma-g_{\sigma^\ast b}\sigma^\ast &+&g_{\omega b}\omega_0+g_{\rho b}{\hat I}_3^{(b)} R_0+g_{\phi b}\phi_0-(\Sigma_{Kb}+\mu Q_V^b)(1-\cos\theta) \cr
&+&\frac{285}{2}\pi^{3/2}(\widetilde\lambda_r)^3\rho_{\rm B}^2\left(1+0.032\rho_{\rm B}/\rho_0\right) + \delta_{b,N}\Delta\mu_{\rm TNA}\ . 
\label{eq:vb}
\end{eqnarray}
\end{widetext}
The above ground-state conditions are implemented in terms of the effective energy density,
\begin{equation}
{\cal E}_{\rm eff}={\cal E} + \mu\rho_Q +\nu\rho_{\rm B}\ .
\label{eq:effE}
\end{equation}
From $\partial{\cal E}_{\rm eff}/\partial \mu=\rho_Q=0$, $\partial{\cal E}_{\rm eff}/\partial \nu=\rho_{\rm B}$, one obtains the charge neutrality and baryon number conservation, respectively. 
From $ \partial{\cal E}_{\rm eff}/\partial \rho_i=0$ \ ($i$= $p$, $n$, $\Lambda$, $\Sigma^-$, $\Xi^-$, $K^-$, $e^-$, $\mu^-$), one obtains the chemical equilibrium conditions for weak processes. 

Classical field equation for kaon condensates, (\ref{eq:keom2}), and equations of motion for meson mean fields,(\ref{eq:cieom}), are also derived from $\partial{\cal E}_{\rm eff}/\partial \theta=0$ and $\partial{\cal E}_{\rm eff}/\partial \varphi_m= 0$ ($\varphi_m =\sigma,\sigma^\ast,\omega_0,R_0, \phi_0$), respectively.

\section{Properties of symmetric nuclear matter and pure neutron matter}
\label{sec:SNM}

Here the ground state properties of SNM in the (MRMF+UTBR+TNA) model are elucidated, and the effects of the three-nucleon forces, UTBR and TNA, on the stiffness of the EOS of SNM at densities around and beyond $\rho_0$ are discussed. The EOS of pure neutron matter (PNM) is also briefly mentioned.

\subsection{Meson-nucleon coupling constants from saturation properties in SNM}
\label{subsec:saturation}

With Eqs.~(\ref{eq:ebm})$-$(\ref{eq:edTNA}), the total energy per nucleon, $E$(total) in SNM is written as 
$E$(total)=$E$(two-body)+$E$(TNR)+$E$(TNA) = [${\cal E}$(two-body)+${\cal E}$(UTBR)+${\cal E}$(TNA)]/$\rho_{\rm B}$, where
\begin{eqnarray}
{\cal E} ({\rm two-}{\rm body}) &=& \sum_{N=p,n} \frac{2}{(2\pi)^3}\int_{|{\bf p}|\leq p_F} d^3|{\bf p}|(|{\bf p}|^2+M_N^{\ast 2})^{1/2} \cr
&+&\frac{1}{2}m_\sigma^2\sigma^2+\frac{1}{2}m_\omega^2\omega_0^2  
\label{eq:edSNM}
\end{eqnarray}
with $p_F$ being the Fermi momentum of the nucleon in SNM, $ p_F=\left(3\pi^2 \rho_{\rm B}/2\right)^{1/3}$, and $M_N^{\ast }$ (=$M_N-g_{\sigma N}\sigma$) the effective nucleon mass. 
The saturation conditions are given by 
\begin{eqnarray}
E~({\rm total})\vert_{\rho_{\rm B}=\rho_0}-M_N&=&-B_0 \ , \cr
\partial E({\rm total})/\partial\rho_{\rm B}\vert_{\rho_{\rm B}=\rho_0}&=& 0 
\label{eq:saturation}
\end{eqnarray}
with $B_0$ (=16.3 MeV) being the binding energy. The meson mean-fields $\langle\sigma\rangle_0$ and $\langle\omega_0\rangle_0$ are obtained by
\begin{eqnarray}
m_\sigma^2\langle\sigma\rangle_0&=&g_{\sigma N}\rho_N^s\vert_{\rho_{\rm B}=\rho_0} \ , \cr
m_\omega^2\langle\omega_0\rangle_0&=&g_{\omega N}\rho_0 \ , 
\label{eq:eomSNM}
\end{eqnarray}
where $\rho_N^s$ (=$\rho_p^s+\rho_n^s$) is the nuclear scalar density. 
The meson-nucleon coupling constants, $g_{\sigma N}$, $g_{\omega N}$, $g_{\rho N}$, the meson mean-fields at $\rho_0$, $\langle\sigma\rangle_0$, $\langle\omega_0\rangle_0$, and the parameters $\gamma_a$, $\eta_a$ in TNA associated with isospin symmetry are obtained from  Eqs.~(\ref{eq:saturation}) and (\ref{eq:eomSNM}), together with the empirical values of the incompressibility $K_0$=240 MeV~\cite{GC2018}, the symmetry energy $S_0$ (=31.5 MeV) at $\rho_0$~\cite{lh2013}, and the slope $L$ of the symmetry energy, being defined as  
$L$$\equiv 3\rho_0\left(d S(\rho_{\rm B}) / d\rho_B\right)_{\rho_B=\rho_0}=3 P_{\rm PNM}(\rho_0) / \rho_0 $, where $S(\rho_{\rm B})$ is the symmetry energy and $P_{\rm PNM}(\rho_{\rm B})$ is the pressure for pure neutron matter (PNM); 
both are given as functions of baryon density. 
The expressions for $K_0$, $S_0$, and $L$ in the present model are given by Eqs.~(\ref{eq:K0}), (\ref{eq:s0}), and (\ref{eq:L}), respectively in Appendix~\ref{subsec:appendixB}. 
Allowable values of $S_0$ and $L$ have been constrained from analyses of terrestrial experiments such as measurements of giant dipole resonances, dipole polarizabilities, neutron skins of neutron-rich nuclei, and Heavy-Ion Collisions, etc.  
Up to now, the empirical value of the $L$ has a large uncertainty, ranging from 30 MeV to 90 MeV~\cite{lp2016,oertel2017,xia2021}. 
Recent measurements of neutron skin thickness from scattering of the spin-polarized electrons on $^{208}$Pb by the PREX-II experiments~\cite{adhikari2021} and on $^{48}$C by the CREX~\cite{adhikari2022} lead to controversy about the value of $L$~\cite{reed2021,Reinhard2021,reed2024}.  
Another experiment is the measurement of the spectra of charged pions produced by colliding rare isotope Sn beams with isotopically enriched Sn targets. The deduced value of $L$ has been shown to be 42 $< L <$ 117 MeV~\cite{estee2021}.
Also there are attempts to get information on $L$ from the astronomical observations associated with $X$-ray bursters~\cite{sotani2018}. 
 In the present work, the parameters, $V_r$, $c_r$, and $\lambda_r$ in $U_{\rm SJM2}$ [(\ref{eq:aUTBR})] are fixed to be those used in SJM2 model in Ref.~\cite{tnt2008}, so that the $E$~(TNR) is commonly given irrespective of choice of the other parameters.  Allowable values of the $L$ are chosen so that the overall density-dependence of the energy contributions, $E$~(two-body), $E$~(TNA), and $E$~(total) around $\rho_0$ in SNM are similar to those obtained by the standard variational calculation by the use of the $v_{14}$ two-body potential with addition of the phenomenological TNR and TNA by Lagaris-Pandahripande~[LP (1981)]~\cite{lp1981}. It turns out that $E$~(two-body) in the present model is not bound for $L \lesssim 55$ MeV and that its density-dependence is quite different from that of LP~(1981). For $L \gtrsim75$ MeV, an onset density of KC exceeds a central density of neutron star with maximum mass. Hence the typical three values $L$ = (60, 65, 70) MeV are taken throughout this paper.
In Table~\ref{tab:para1}, the relevant quantities associated with the (MRMF+UTBR+TNA) model are listed for three cases of $L$ = (60, 65, 70) MeV. 
\begin{table*}[!]
\caption{The parameters $\gamma_a$, $\eta_a$ for TNA, the coupling constants, $g_{\sigma N}$, $g_{\omega N}$, $g_{\rho N}$, the meson mean-fields, $\langle\sigma\rangle_0$, $\langle\omega_0\rangle_0$, and the effective mass ratio for the nucleon, 
$(M_N^\ast/M_N)_0$, in SNM at $\rho_{\rm B}$ = $\rho_0$, obtained in the (MRMF+ UTBR+TNA) model
 in case of $L$=60, 65, and 70 MeV. The  $\sigma$-$Y$ coupling constants ($Y$=$\Lambda$, $\Sigma^-$, $\Xi^-$) determined from the potential depths for $Y$ in SNM [(\ref{eq:ypot})]  are also listed. }~
\begin{center}
\begin{tabular}{ c || c | c | c | c | c || c | c || c || c | c | c }
\hline
 & $\gamma_a$ & $\eta_a$  &  $g_{\sigma N}$ & $g_{\omega N}$  & $g_{\rho N}$ & $\langle\sigma\rangle_0$ & $\langle\omega_0\rangle_0$  & $(M_N^\ast/M_N)_0$ & $g_{\sigma\Lambda}$ & $g_{\sigma\Sigma^-}$ & $g_{\sigma\Xi^-}$ \\ 
  & (MeV$\cdot$fm$^6$) & (fm$^3$) &          &         &   &  (MeV) &  (MeV)  &  &  &  &  \\     \hline\hline
SJM2+TNA-L60    & $-$1662.63     &  17.18 &  5.27 & 8.16  & 3.29 & 39.06 & 16.37  & 0.78 & 3.29 & 2.00 & 1.82  \\
SJM2+TNA-L65    & $-$1597.67     &  18.25 &  5.71 & 9.07  & 3.35 & 42.16 & 18.18  & 0.74 & 3.54 & 2.34 & 1.93   \\
SJM2+TNA-L70    & $-$1585.48     &  19.82 &  6.07 & 9.77  & 3.41 & 44.62 & 19.59  & 0.71 & 3.74 & 2.61 & 2.02   \\
\hline
\end{tabular}
\label{tab:para1}
\end{center}
\end{table*}

\subsection{Role of three-body repulsion and attraction on the equation of state in SNM}
\label{subsec:stiff-SNM}

\begin{figure}[h]
\begin{center}
\includegraphics[height=0.38\textwidth]{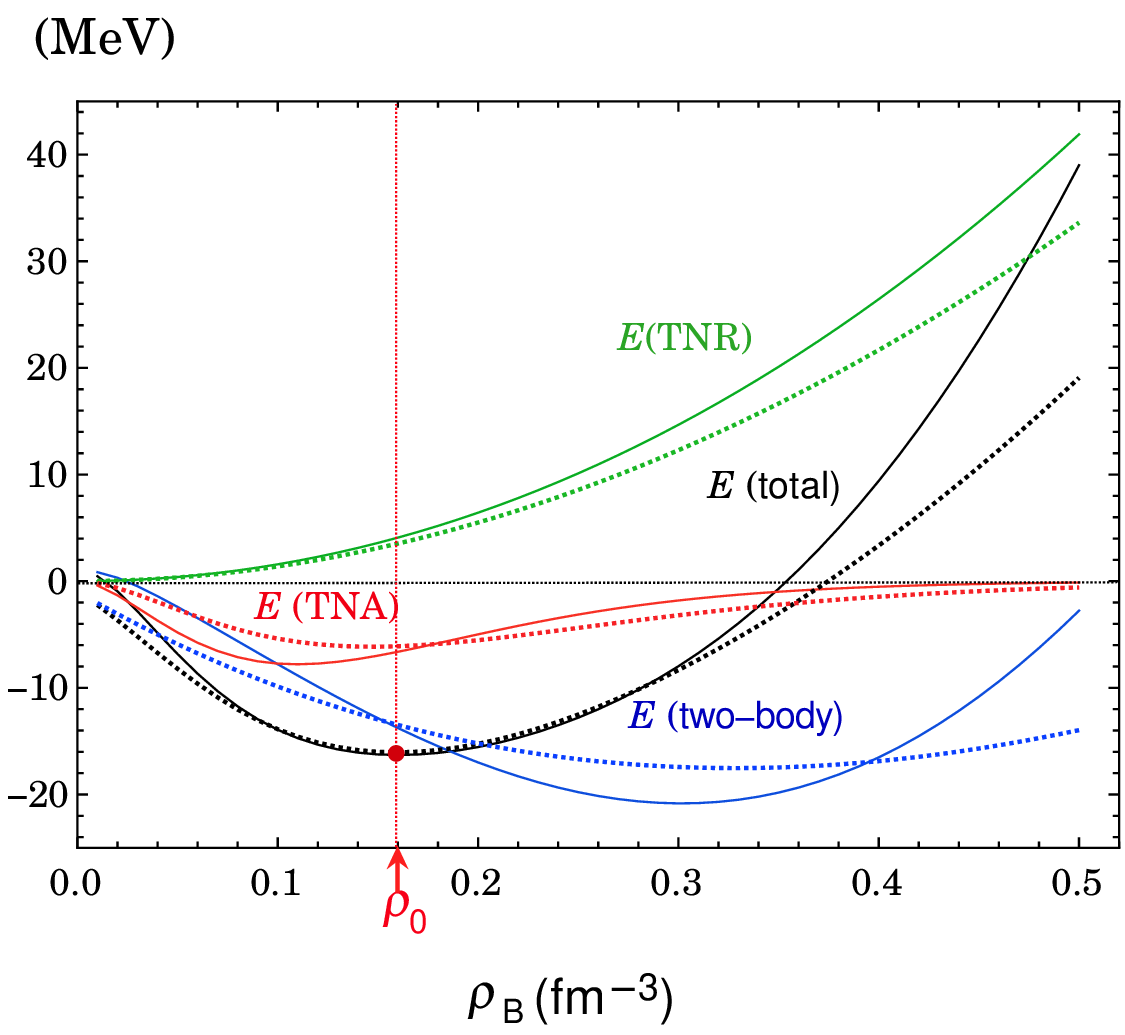}~
\end{center}~
\caption{The total energy per nucleon, $E$ (total) $(={\cal E}/\rho_{\rm B}$), and energy contributions as functions of $\rho_{\rm B}$ in SNM (solid lines), obtained by the (MRMF+UTBR+TNA) model in the case of the slope $L$ = 65 MeV.  $E$ (TNR) (=${\cal E}$~(UTBR)/$\rho_{\rm B}$) is the contribution from the three-nucleon-repulsion, $E$ (TNA) (=${\cal E}$(TNA)/$\rho_{\rm B}$) from the three-nucleon attraction, and $E$(two-body) (=${\cal E}_{B,M}/\rho_{\rm B}$) the sum of kinetic and two-body interaction energies. For comparison, those obtained from LP~(1981)\cite{lp1981} are shown by the dotted lines. }
\label{fig:esnm}
\end{figure}

The total energy per nucleon, $E$~(total) $(={\cal E}/\rho_{\rm B}$), in SNM is shown as a function of baryon number density $\rho_{\rm B}$ by the solid line in Fig.~\ref{fig:esnm} for a typical value of the slope parameter $L$ = 65 MeV. 
The energy contributions from the three-nucleon-repulsion [$E$~(TNR)], the three-nucleon attraction [$E$~(TNA)], and the sum of kinetic and two-body interaction energies [$E$~(two-body)] in SNM are also shown by the solid lines. For comparison, those for LP~(1981) are depicted by the dotted lines. In the case of LP~(1981), the numerical value of $E$~(two-body) are read off from Fig.~2, Tables~4 and 5 in~\cite{lp1981}. The parameter $\lambda_a$~(LP) in TNA is fixed to be 2.0 fm, and the remaining two parameters $\gamma_a$~(LP) and $\eta_a$~(LP) are set to be $\gamma_a$~(LP) =$-$700 MeV$\cdot$fm$^6$ and $\eta_a$~(LP) = 13.6~fm$^3$, respectively~\cite{lp1981}. For $E$~(TNR) in LP~(1981),  one uses the following expression by NTH~1994
\begin{eqnarray}
\hspace{-1.0cm}~E ({\rm TNR})~({\rm LP})&=&2\pi\rho_{\rm B}\int_0^\infty dr r^2U_{\rm TNR}(r; \rho_{\rm B})~({\rm LP}) \cr
&=&\frac{\pi^{3/2}}{2}V_r\lambda_r^3\rho_{\rm B}(1-e^{-\eta_r\rho_{\rm B}}) \cr
&=& \frac{\pi^{3/2}}{2}\rho_{\rm B} U_{\rm TNR}(r=0; \rho_{\rm B}) ({\rm LP})\cdot\lambda_r^3 
\label{eq:ETNR-LP}
\end{eqnarray}
by the use of the density-dependent effective two-body potential 
\begin{equation}
U_{\rm TNR}(r; \rho_{\rm B})~({\rm LP})=V_r \left(1-\exp(-\eta_r \rho_{\rm B})\right)\exp[-(r/\lambda_r)^2] \label{eq:UTNR-LP}
\end{equation}
with $\lambda_r$=1.40 fm, $\eta_r$=0.15 fm$^3$\cite{nth1994}. 

As seen from Fig.~\ref{fig:esnm}, both the TNR and TNA play important roles to locate the total energy minimum at the empirical saturation point. Indeed, it is necessary to include both the TNR and TNA in the total energy $E$ (total) in addition to the nuclear two-body interaction within the MRMF, in order to reproduce the empirical saturation property  (\ref{eq:saturation}) and incompressibility (= 240 MeV) for the SNM. The TNR  (the TNA) pushes up (pushes down) the $E$ (two-body) curve for $\rho_{\rm B}\gtrsim \rho_0$ ($\rho_{\rm B}\lesssim \rho_0$). 

As compared with LP~(1981), the energy contributions from the TNR and TNA around $\rho_0$ are quantitatively close to those in LP~(1981) : $E$ (TNR)= 4.1 MeV and $E$ (TNA)=$-$6.6 MeV at $\rho_0$, while $E$ (TNR)~(LP)= 3.5 MeV and $E$ (TNA)~(LP)=$-$6.1 MeV. 
It is to be noted that the energy contribution from the TNR is determined by the repulsive-core height of the effective two-body potential times its range volume as seen in Eqs.~(\ref{eq:edUTBRtil}) and (\ref{eq:ETNR-LP}). 
Both the height and range of the effective two-body potentials are different between SJM2 and LP~(1981). Specifically $U_{\rm SJM2}(0;~\rho_0)$ =15.6 MeV and $\widetilde\lambda_r$~(SJM2)=0.86 fm, while $U_{\rm TNR}(0;~\rho_0)$~(LP)= 2.89 MeV and $\lambda_r$~(LP) = 1.40 fm~\cite{t2008,mmt2021}.  
Nevertheless one has almost the same amount of the $E$~(TNR) at $\rho_{\rm B}$=$\rho_0$ in two cases. 

The subtle difference between the $E$~(TNR) obtained from SJM2 and that from LP(1981) at $\rho_{\rm B}=\rho_0$ leads to substantial difference of repulsive energy at high densities $\rho_{\rm B}\gtrsim$ 0.40~fm$^{-3}$ due to the nonlinear density-dependence of the $E$ (TNR) and $E$ (TNR)~(LP) ($\propto\rho_{\rm B}^2$). 
There is also a rapid increase in $E$ (two-body) over the density $\rho_{\rm B}\approx$ 0.40~fm$^{-3}$ as compared with the case of LP~(1981). This is due to the fact that, in a meson-exchange picture like the MRMF model, attraction brought about by the $\sigma$-meson exchange is saturated at high densities, while repulsion by the $\omega$-meson exchange steadily  increases. Such a relativistic effect further leads to stiffer EOS than the case of LP~(1981).  

For $\rho_{\rm B}\lesssim \rho_0$, the $E$~(two-body) in the present case is less attractive than that of LP~(1981) as seen in Fig.~\ref{fig:esnm}. The deficit attraction in the $E$~(two-body) is compensated with more attractive TNA than the case of LP~(1981). For $\rho_{\rm B}\gtrsim \rho_0$, the $E$ (TNA) in both cases decrease in magnitude and become negligible for $\rho_{\rm B}\gtrsim $ 0.4~fm$^{-3}$. 

\subsection{Dependence of the EOS for SNM and PNM on the slope $L$}
\label{subsec:L-EOS-SNM}

\begin{figure}[h]
\begin{center}
\includegraphics[height=0.38\textwidth]{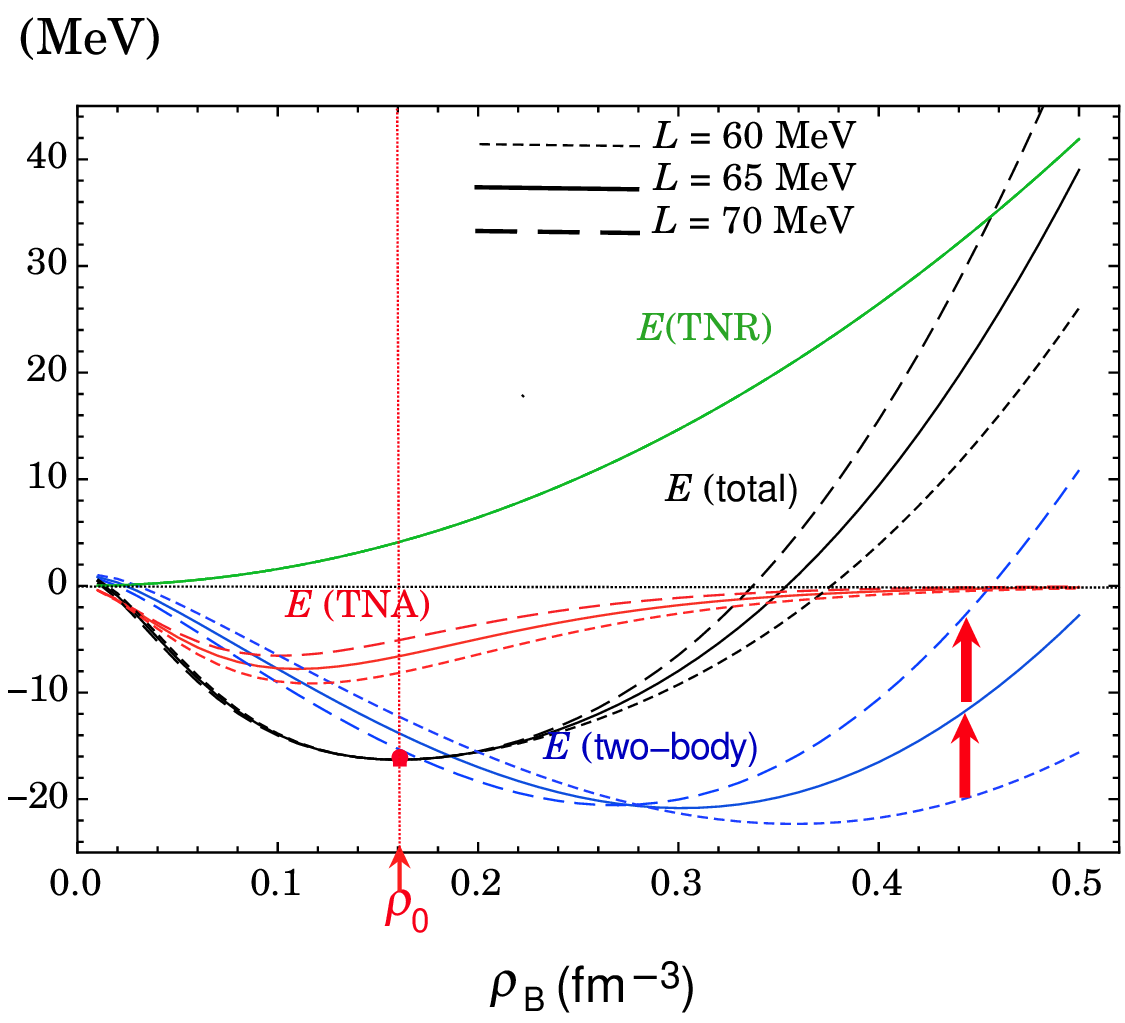}~
\end{center}~
\caption{The total energy per baryon, $E$~(total) $(={\cal E}/\rho_{\rm B}$), and each energy contribution from the three-nucleon-repulsion [$E$~(TNR)], the three-nucleon attraction [$E$~(TNA)], and the sum of kinetic and two-body interaction energies [$E$~(two-body)] in SNM are shown as functions of $\rho_{\rm B}$ by the dotted, solid, and dashed lines for $L$=60, 65, and 70 MeV, respectively. The $E$~(TNR) is commonly taken from the SJM2~\cite{t2008} for the cases of $L$=(60, 65, 70) MeV and is represented by the single green solid line. }
\label{fig:esnmL60-70}
\end{figure}

Next one considers dependence of the EOS for SNM on the slope $L$. In Fig.~\ref{fig:esnmL60-70}, the total energy per baryon as well as each energy contribution in SNM are shown as functions of $\rho_{\rm B}$ by the dotted, solid, and dashed lines for $L$=60, 65, and 70 MeV, respectively. The $E$ (TNR) is commonly taken from the SJM2~\cite{t2008} for the cases of $L$=(60, 65, 70) MeV and is represented by the single green solid line.

The amount of the attractive energy from the TNA around $\rho_{\rm B}=\rho_0$ depends on the slope $L$ through the isospin-dependence of the TNA [Eqs.~(\ref{eq:tna}) and (\ref{eq:edTNA})]. 
The contribution to $L$ from the TNA, $L^{(\rm TNA)}$, is given by 
$L^{(\rm TNA)}$=$6\gamma_a\rho_0^2 (\eta_a\rho_0 - 2)e^{-\eta_a\rho_0}$ and has a negative contribution to $L$ since $\gamma_a<0$ and $\eta_a\rho_0 > 2$ (see Table~\ref{tab:para1}). 
For a larger $L$, one has a smaller $|\gamma_a|$, so does the contribution to the binding energy from the TNA at $\rho_{\rm B}=\rho_0$. 
To compensate the deficit of the attractive binding energy contribution from the TNA, 
the contribution to the binding energy from the two-body $B$-$B$ interaction through the $\sigma$ and $\omega$ meson-exchange in the RMF framework gets larger for a larger $L$. 
As a result, both the coupling constants $g_{\sigma N}$, $g_{\omega N}$, $g_{\rho N}$ and meson mean-fields at $\rho_0$, $\langle\sigma\rangle_0$, $\langle\omega_0\rangle_0$ are shifted to enlarged values to keep the binding energy and saturation properties of the SNM as empirical ones. 
As a result, the effective mass ratio for the nucleon, 
$(M_N^\ast/M_N)_0$, in SNM at nuclear saturation density is smaller for larger case of $L$, since both $g_{\sigma N}$ and $\langle\sigma\rangle_0$ become larger. Nevertheless it lies within the allowable range determined from experimental analyses of neutron scattering from lead nuclei, 0.7$-$0.8~\cite{glendenning2000} for all the cases of $L$~(see Table \ref{tab:para1}). 

At high densities beyond $\rho_0$, the attraction from the $\sigma$-exchange is saturated, so that the remaining repulsion from the $\omega$-exchange is more remarkable for larger $\omega$-mean field. 
The increase in repulsive energy from the $\omega$-meson exchange with the increase in $L$ from 60 MeV to 70 MeV, as shown by the arrows for the $E$ (two-body) in Fig.~\ref{fig:esnmL60-70}, plays a decisive role on stiffening the EOS in SNM. Inversely, as $L$ gets smaller than $L\lesssim 55$ MeV, the repulsive energy from the $\omega$-meson exchange is not large enough to compensate the attractive energy by the $\sigma$-meson exchange beyond $\rho_0$, so that the system becomes unbound with the two-body interaction only in the present interaction model. Thus, the strength of the binding energy contribution by the TNA at $\rho_0$ and the density profile of the two-body $B$-$B$ interaction via the $\sigma$ and $\omega$ exchanges show subtle 
correlation through the choice of $L$ in the (MRMF+UTBR+TNA) model. 

In contrast, as for the (MRMF+NLSI) model, where many-baryon forces are given by  the NLSI terms in place of the UTBR and TNA, the two-body $B$-$B$ interaction via the $\sigma$ and $\omega$ exchanges are determined irrespectively of the choice of $L$, since the total energy does not have isospin-dependent terms like the TNA in the case of the SNM, as is shown in Appendix~\ref{subsec:a4}. 

\begin{figure}[h]
\begin{center}~
\includegraphics[height=0.35\textheight]{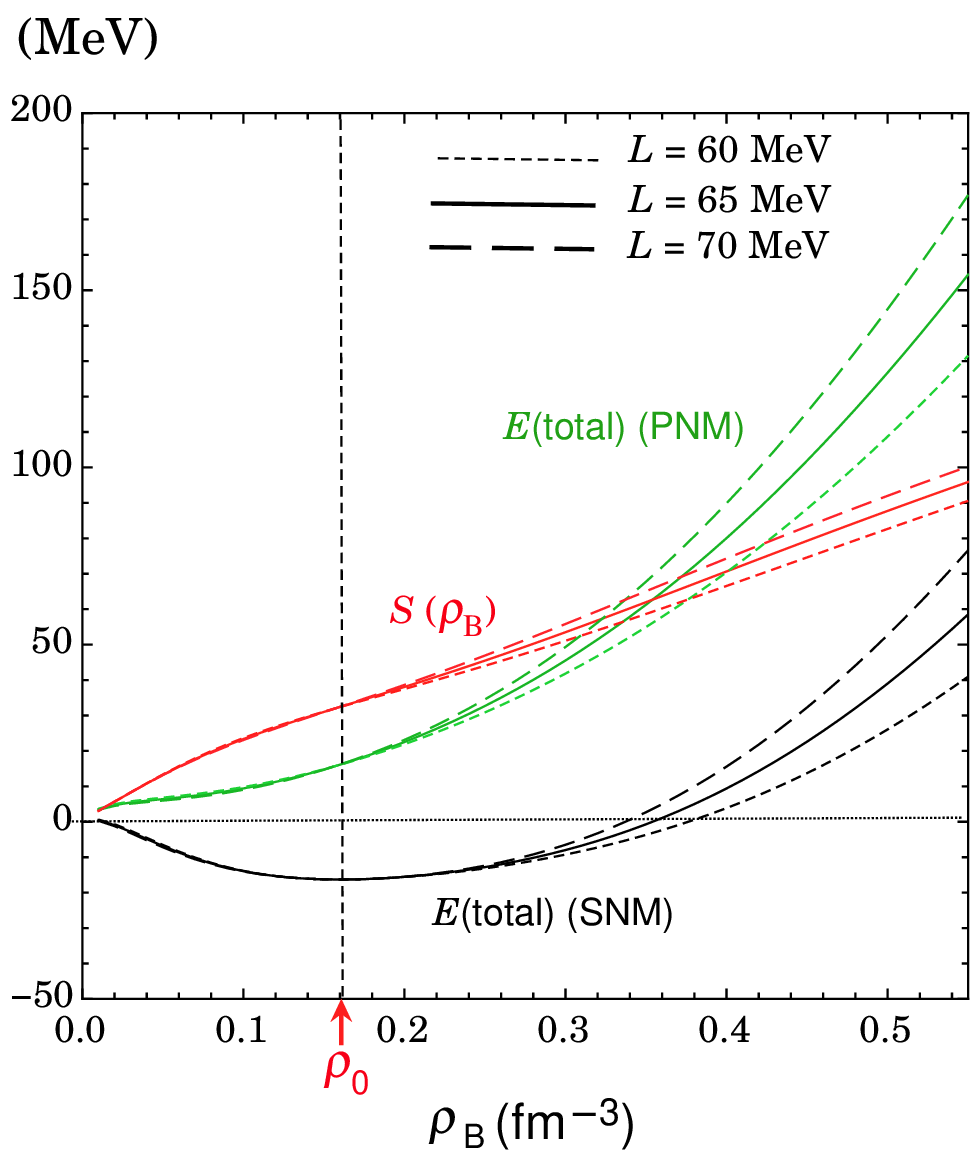}~
\end{center}~
\caption{The total energy per nucleon, $E$ (total), in SNM and in PNM, as a function of $\rho_{\rm B}$ for $L$=(60, 65, 70) MeV. The density-dependence of the symmetry energy $S(\rho_{\rm B})$, defined by $S(\rho_{\rm B})$=$E$ (total)(PNM)$-$$E$ (total)(SNM), is also shown.  
See the text for the details. \\}
\label{fig:symenergy}
\end{figure}

\begin{figure}[h]
\begin{center}
\vspace{-0.5cm}~
\includegraphics[height=0.36\textwidth]{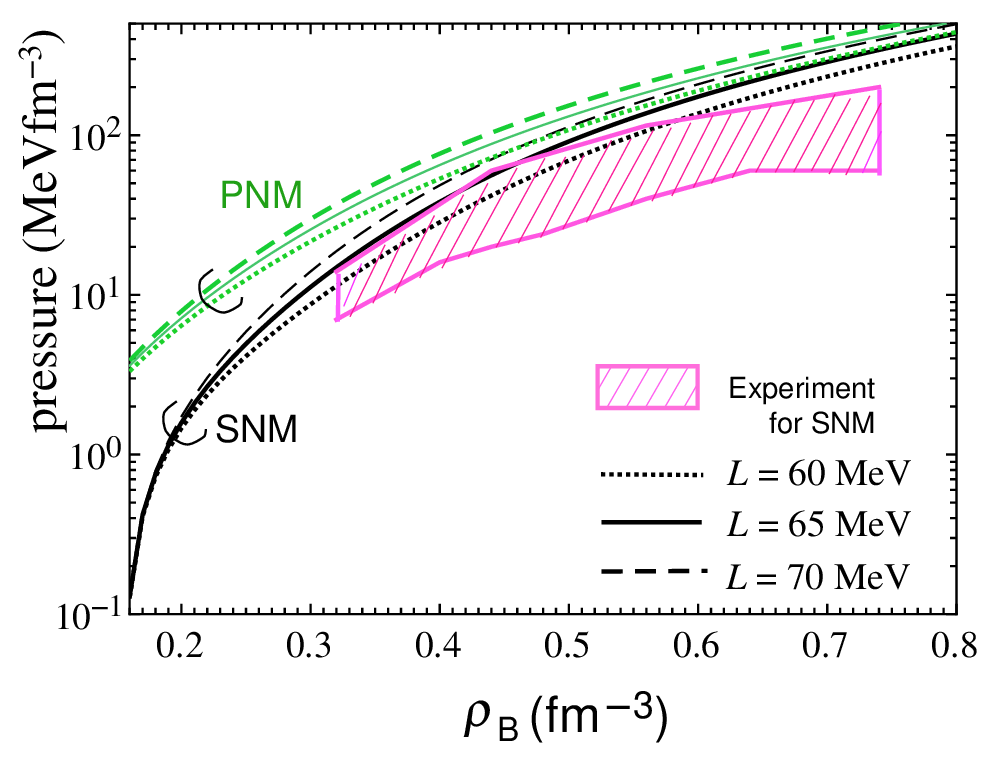}~
\end{center}~
\vspace{0.5cm}~
\caption{The pressure in SNM (black curves) and in PNM (green curves) as a function of $\rho_{\rm B}$ for $L$=(60, 65, 70) MeV. The shaded region, corresponding to the region of pressures consistent with the experimental flow data in the case of SNM, is taken from Ref.~\cite{danielewicz2002}. \\ }
\label{fig:psnm}
\end{figure}

To compare the $L$-dependence of the EOS between the PNM and SNM, the total energy per nucleon, $E$ (total), is shown in Fig.~\ref{fig:symenergy} and the pressure $p$ is shown in Fig.~\ref{fig:psnm}, as a function of $\rho_{\rm B}$ in the case of $L$=(60, 65, 70) MeV, for both PNM and SNM cases.  
The shaded region in Fig.~\ref{fig:psnm}, corresponding to the region of pressures consistent with the experimental flow data in the case of SNM, is taken from Ref.~\cite{danielewicz2002}.
One can see the EOS gets stiffer for larger $L$ over the densities $\rho_{\rm B}=(1-5)\rho_0$ for PNM as well as the case of SNM. 
Therefore, in the present interaction model, not only the TNR but also the slope $L$ controls the stiffness of the EOS at high densities. This feature will be shown to be commonly applied to hadronic matter with the ($Y$+$K$) phase. (see Sec.~\ref{sec:KC-EOS}. ) 
~

In~\cite{danielewicz2002}, the allowable region in the pressure ($p$)-density ($\rho_{\rm B}$) plane for the EOS in SNM and PNM has been obtained from the analyses of transverse and elliptic flow data in nuclear collisions [The shaded region in Fig.~\ref{fig:psnm} in the case of SNM]. Quantitatively, the result for $L\lesssim$ 65 MeV lie within the upper bound of the experimentally consistent region for $\rho_{\rm B}$ = (2$-$4)~$\rho_0$, while those for $L\gtrsim$70 MeV slightly above it. For the latter case, some extra softening effect might be responsible in order to be consistent with the experimental implications. 

In Fig.~\ref{fig:symenergy}, the density-dependence of the symmetry energy, $S(\rho_{\rm B})$, defined by $S(\rho_{\rm B})$=$E$ (total)(PNM)$-$$E$ (total)(SNM), is shown in the case of $L$=(60, 65, 70) MeV by the red solid lines. $S(\rho_{\rm B})$ increases monotonically with density for a given $L$. Since there is a cancellation of the $L$-dependent part of the total energy between the cases of PNM and SNM, the slope $L$-dependence of $S(\rho_{\rm B})$ is not remarkable even at high densities.

\section{Kaon properties in hyperon-mixed matter and onset of KC}
\label{sec:KP}

Here kaon properties in hyperon-mixed matter are considered, and onset density of KC with the interaction model (ChL+MRMF+UTBR+TNA) is obtained. 

\subsection{Meson-hyperon coupling constants for description of hyperon-mixed matter}
\label{subsec:YMM}

For the description of hyperon-mixed matter, one sets the values of the meson-hyperon coupling constants to obtain the hyperon-nucleon and hyperon-hyperon interactions in the MRMF. 

The vector meson couplings for hyperons ($Y$) are obtained from the vector-nucleon couplings $g_{\omega N}$, $g_{\rho N}$, $g_{\phi N}$ through the SU(6) symmetry relations~\cite{sdg94} : 
 \begin{subequations}
\begin{eqnarray}\label{eq:gmY}
g_{\omega\Lambda}&=&g_{\omega\Sigma^-}=2g_{\omega \Xi^-}=(2/3) g_{\omega N} \ , 
\label{eq:gmY1} \\
g_{\rho \Lambda}&=& 0 \ , g_{\rho\Sigma^-}=2g_{\rho\Xi^-}=2g_{\rho N} \ , \label{eq:gmY2} \\ 
 g_{\phi\Lambda}&=& g_{\phi\Sigma^-}=(1/2) g_{\phi\Xi^-}=-(\sqrt{2}/3) g_{\omega N} \ . 
\end{eqnarray}
\end{subequations}
The scalar ($\sigma$, $\sigma^\ast$) meson-hyperon couplings are determined from the phenomenological analyses of recent hypernuclear experiments. 
The $\sigma$-$Y$ coupling constant, $g_{\sigma Y}$, is related with the potential depth of the hyperon $Y$ ($Y=\Lambda$, $\Sigma^-$, $\Xi^-$) at $\rho_B=\rho_0$ in SNM, $V_Y^N$, which is read off from Eq.~(\ref{eq:vb}) with $\langle \sigma^\ast\rangle_0$ = $\langle R_0\rangle_0$ = $\langle \phi_0\rangle_0$ = 0 as  
\begin{equation}
V_Y^N=-g_{\sigma Y}\langle\sigma\rangle_0 +g_{\omega Y}\langle\omega_0\rangle_0 
+\partial{\cal E}_{\rm UTBR}/\partial \rho_Y \ ,
\label{eq:ypot}
\end{equation}
where $\langle m\rangle_0$ ($m$=$\sigma$, $\sigma^\ast$, $\omega_0$, $R_0$, $\phi_0$) are the meson mean fields at $\rho_B=\rho_0$ in SNM, and the last term, which is equal to the second term from the last on the r.~h.~s. of Eq.~(\ref{eq:vb}), comes from the energy density contribution from the UTBR. 
From recent theoretical and experimental analyses on hypernuclear experiments~\cite{ghm2016}, the values of the $V_Y^N$ in Eq.~(\ref{eq:ypot}) are set to be $V_\Lambda^N=-27$ MeV, $V_{\Sigma^-}^N$ = 23.5 MeV, and $V_{\Xi^-}^N$ = $-14$ MeV, and one obtains $g_{\sigma\Lambda}$, $g_{\sigma\Sigma^-}$, and $g_{\sigma\Xi^-}$. These coupling constants are obtained for each case of $L$ and listed in Table~\ref{tab:para1}.
 
The $\sigma^\ast$-$Y$ coupling constants $g_{\sigma^\ast Y}$ are relevant to the $Y$-$Y$ interaction as well as binding energy of hypernuclei. The previous model with $B$-$B$ interactions within the RMF including the nonlinear self-interacting $\sigma$ potential was extended to finite systems of hypernuclei~\cite{mmt2009,mmt2014} in view of the density functional theory. With this model, the separation energy $B_{\Lambda\Lambda}$($^{\ \ 11}_{\Lambda\Lambda}$Be) was calculated and 
$g_{\sigma^\ast \Lambda}$ was determined to be 7.2 so as to reproduce the empirical values of the 
$B_{\Lambda\Lambda}$($^{\ \ 11}_{\Lambda\Lambda}$Be). 
Also $g_{\sigma^\ast\Xi^-}$ was taken to be 4.0, for which one obtains the theoretical values of the separation energies $B^{\rm th}_{\Xi}(^{ \ \ 15}_{\ \Xi(s)}$C) = 8.1 MeV and $B^{\rm th}_{\Xi}(^{\ \ 12}_{\ \Xi(s)}$Be) = 5.1 MeV, which are consistent with the empirical values deduced from the ``Kiso'' event, $\Xi^-$ + $^{14}$N $\rightarrow$ $^{15}_{\ \Xi}$C $\rightarrow$ $^{10}_{\ \Lambda}$Be + $^5_\Lambda$He~\cite{n15,hayakawa2021}. The remaining unknown coupling constant, $g_{\sigma^\ast \Sigma^-}$, is simply set to be zero since there is little experimental information on $\Sigma$ hypernuclei. For the details of estimating the $g_{\sigma Y}$ and $g_{\sigma^\ast Y}$, 
see Ref.~\cite{mmt2022}. 

\subsection{Onset of $\Lambda$-mixing and composition of matter below the onset density of KC}
\label{subsec:onset-Lambda}

As is shown below, the $\Lambda$ hyperons always appear at a lower density than KC for various cases of $L$ and $\Sigma_{Kn}$. Therefore first the onset of the $\Lambda$-mixing and composition of matter are considered prior to KC. 

\begin{figure*}[!]
\begin{minipage}[l]{0.50\textwidth}
\begin{center}
\includegraphics[height=0.32\textheight]{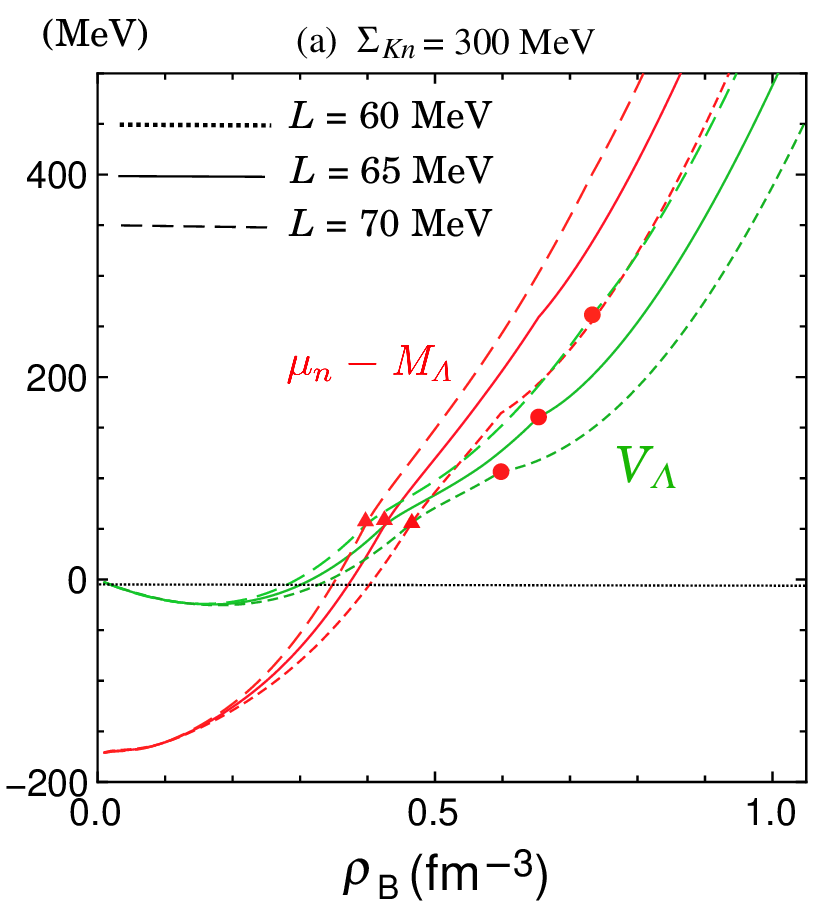}~
\end{center}
\end{minipage}~\vspace{-0.5cm}
\begin{minipage}[r]{0.50\textwidth}
\begin{center}
\includegraphics[height=0.32\textheight]{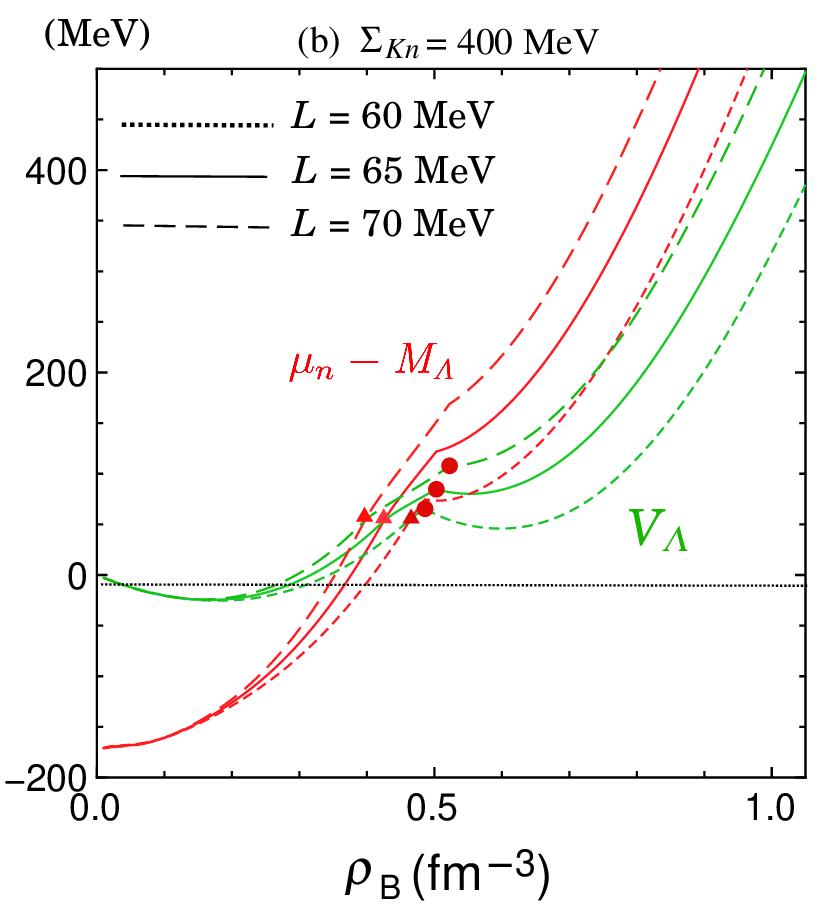}
\end{center}
\end{minipage}
\vspace{1.5cm}~
\caption{(a) \ The $\Lambda$ potential $V_\Lambda$ (green curves) and ($\mu_n-M_\Lambda$) (red curves) as functions of baryon number density $\rho_{\rm B}$ for $\Sigma_{Kn}$ = 300 MeV in the case of $L$=60 MeV (dotted lines), 65 MeV (solid lines), and 70 MeV (dashed lines). The filled triangle denotes the onset density of $\Lambda$-mixing, $\rho_{\rm B}^c (\Lambda)$, in $\beta$-equilibrated nuclear matter, and the filled circle denotes the onset density of KC, $\rho_{\rm B}^c(K^-)$, in $\beta$-equilibrated hyperon-mixed matter. (b) The same as (a), but for $\Sigma_{Kn}$ = 400 MeV. The profiles for $\Sigma_{Kn}$ = 300 MeV and 400 MeV are the same before appearance of KC. See the text for details.
}
\label{fig:vlam}
\end{figure*}

In Fig.~\ref{fig:vlam}, the $\Lambda$ potential $V_\Lambda$ (green curves) and the difference between the neutron chemical potential $\mu_n$ and the $\Lambda$ rest mass, ($\mu_n-M_\Lambda$), (red curves) are shown as functions of baryon number density $\rho_{\rm B}$ in the case of $L$=60 MeV (dotted lines), 65 MeV (solid lines), and 70 MeV (dashed lines). (a) and (b) is for $\Sigma_{Kn}$ = 300 MeV and 400 MeV, respectively. Note that the profiles for $\Sigma_{Kn}$ = 300 MeV and 400 MeV are the same before appearance of KC. 
The onset density of the $\Lambda$ hyperons, $\rho_{\rm B}^c (\Lambda)$, in $\beta$-equilibrated neutron-star matter is given by the intersection of the curves for the $\mu_n-M_\Lambda$ and $V_\Lambda$, denoted as the filled triangle. The $\Lambda$ hyperons are mixed over the densities where $\mu_n-M_\Lambda > V_\Lambda$. 
Numerically $\rho_{\rm B}^c(\Lambda)$ is weakly dependent on $L$: $\rho_{\rm B}^c(\Lambda)$ = (0.47, 0.43, 0.40) fm$^{-3}$ for $L$ = (60, 65, 70) MeV.

For the larger value of the slope $L$, the neutron chemical potential $\mu_n$  rises up with density more rapidly than the  $\Lambda$ potential $V_\Lambda$ in the vicinity of $\rho_{\rm B}^c(\Lambda)$, due to the enhanced repulsive interaction through the vector-meson exchange for the neutron as compared to the $\Lambda$. Therefore, for larger $L$, the onset condition of the $\Lambda$-mixing is met at the lower density, leading to smaller $\rho_{\rm B}^c(\Lambda)$. 

Once the $\Lambda$ hyperons appear, the neutron fraction as well as the proton fraction is diminished through the weak process, $n\rightarrow \Lambda (+\bar\nu_l + \nu_l)$ ($l$ : leptons), and $p+l^-\rightarrow \Lambda + \nu_l$. [See also Fig.~\ref{fig:fractions} in Sec.~\ref{subsec:onsetKC}.]

 \begin{table*}[!]
\caption{The onset densities at which hyperon-mixing starts and those of KC in the (ChL+MRMF+ UTBR+TNA) model for $\Sigma_{Kn}$ = 300 MeV and 400 MeV in case of $L$=60, 65, and 70 MeV. The $\rho_{\rm B}^c(\Lambda)$ is the onset density of 
$\Lambda$ hyperons in the normal neutron-star matter, $\rho_{\rm B}^c(\Xi^- \ {\rm in} \ \Lambda )$ the one of $\Xi^-$ hyperons in the $\Lambda$-mixed matter, $\rho_{\rm B}^c~(K^-)$ the one of KC in the hyperon ($\Lambda$ and/or $\Xi^-$)-mixed matter, and $\rho_{\rm B}^c(\Xi^- \ {\rm in} \ K^- \Lambda )$ the one of the $\Xi^-$ hyperons in the KC phase in the $\Lambda$-mixed matter. For reference, the central density of the maximum mass star, $\rho_{\rm B, center}(M_{\rm max})$, obtained after solving the T.O.V equation, and the depth of the $K^-$ optical potential $U_K$, estimated at $\rho_{\rm B}$ = $\rho_0$ in the SNM, are listed for each case of $L$ and $\Sigma_{Kn}$. }
\begin{center}
\begin{tabular}{ c | c | c || c | c | c | c || c}
\hline
 $L$ & $\Sigma_{Kn}$  & $U_K$ & $\rho_{\rm B}^c(\Lambda)$ & $\rho_{\rm B}^c(\Xi^-~{\rm in}~\Lambda)$  &  $\rho_{\rm B}^c(K^-)$ &  $\rho_{\rm B}^c(\Xi^-~{\rm in}~{K^-\Lambda})$ & $\rho_{\rm B,center} (M_{\rm max})$ \\
 (MeV) & (MeV)  & (MeV) & (fm$^{-3}$) &  (fm$^{-3}$)   &  (fm$^{-3}$)   & (fm$^{-3}$) &(fm$^{-3}$) \\ \hline\hline
60       & 300     & $-$111 &  0.466     &  $-$            & 0.598         & 1.04   & 1.22     \\
           & 400    & $-$132  &  0.466     &  $-$            & 0.486        &  0.994   & 1.34   \\ \hline
65       & 300     & $-$111 &  0.425     & 0.568         & 0.653         & $-$   & 1.07     \\
           & 400    & $-$131  &  0.425     & $-$             & 0.503         & 0.900     & 1.16    \\ \hline
70       & 300     & $-$111 &  0.397    &  0.516         & 0.733         & $-$   & 0.966       \\
           & 400    & $-$131  &  0.397    &  (0.516)         &  0.523        & 0.790    & 1.01       \\ \hline
\hline
\end{tabular}
\label{tab:onset}
\end{center}
\end{table*}

\subsection{Onset density of kaon condensation in hyperon-mixed matter}
\label{subsec:onsetKC}

The lowest kaon energy $\omega_K(\rho_B)$ decreases as a function of $\rho_{\rm B}$ due to the $K$-$B$ scalar and vector attraction. In the case of the $s$-wave kaon condensation, the $\omega_K(\rho_B)$ intersects with the kaon chemical potential $\mu_K$, which is equal to the charge chemical potential $\mu$ in Eq.~(\ref{eq:chem}). 
The onset density $\rho_B^c (K^-)$ for the $s$-wave kaon condensation as a continuous phase transition is given by the condition~\cite{mt92}
\begin{equation}
\omega_K (\rho_B^c(K^-))=\mu  \ .
\label{eq:onsetk}
\end{equation}
Above the onset density, $\rho_{\rm B}\geq\rho_{\rm B}^c(K^-)$, the condensed kaons spontaneously appear in the ground state through the weak reaction processes, $n+N\rightarrow p+N+K^-$, $l \rightarrow K^- +\nu_l $ ($l = e^-, \mu^-$), and strong reaction processes, 
$\Lambda\rightarrow p+K^-$, $\cdots$ in the presence of hyperons. The total strangeness-rich matter accompanying  the ($Y$+$K$) phase is eventually realized, where the relaxation time is governed by the weak processes~\cite{mti2000}.  

The $\omega_K(\rho_B)$ is given as a pole of the kaon propagator at $\rho_B$, i.e., $D_K^{-1}(\omega_K; \rho_{B})=0$. The kaon inverse propagator, $D_K^{-1}(\omega_K; \rho_{B})$, is obtained through expansion of the effective energy density ${\cal E}_{\rm eff}$~(\ref{eq:effE}) with respect to the classical kaon field, 
\begin{equation}
{\cal E}_{\rm eff}(\theta)={\cal E}_{\rm eff}(0)-\frac{f^2}{2}D_K^{-1}(\mu; \rho_{\rm B})\theta^2+O(\theta^4)  \ .
\label{eq:dkinv1}
\end{equation}
With the use of Eqs.~(\ref{eq:ekfinal}), (\ref{eq:ebm}), (\ref{eq:charge}), (\ref{eq:rhokc}), and 
by setting $\mu_K\rightarrow \omega_K$, $\theta\rightarrow 0$, one obtains 
\begin{equation}
D_K^{-1}(\omega_K; \rho_{B})
=\omega_K^2-m_K^2-\Pi_K(\omega_K; \rho_B) \ , 
\label{eq:dkinv}
\end{equation}
where $\Pi_K(\omega_K; \rho_B)$ is the self-energy of $K^-$ mesons: 
\begin{widetext}
\begin{eqnarray}
\Pi_K(\omega_K; \rho_B)
=
-\frac{1}{f^2}\Bigg\lbrack &\rho_p^s& \Bigg\lbrace\Sigma_{Kp}+\left(\frac{\omega_K}{m_K}\right)^2\left(d_p m_K+\frac{g_{\Lambda\ast}^2}{2} m_K^2 \frac{M_{\Lambda^\ast}-M_p-\omega_K}{(M_{\Lambda^\ast}-M_p-\omega_K)^2+\gamma_{\Lambda^\ast}^2}\right)\Bigg\rbrace \cr\cr
+&\rho_n^s&\Bigg\lbrace\Sigma_{Kn}+\left(\frac{\omega_K}{m_K}\right)^2 d_n m_K \Bigg\rbrace 
+\sum_{Y=\Lambda, \Sigma^-, \Xi^-}\rho_Y^s\Bigg\lbrace\Sigma_{KY}+\left(\frac{\omega_K}{m_K}\right)^2 d_Y m_K\Bigg\rbrace \Bigg\rbrack -2X_0\omega_K\ .
\label{eq:selfk}
\end{eqnarray}
\end{widetext}
With the coefficients $d_p$ and $d_n$, which are determined so as to reproduce the empirical values of the on-shell $s$-wave $K$-$N$ scattering lengths (Appendix~\ref{subsec:appendixB}), one can see from Eq.~(\ref{eq:selfk}) that for on-shell $K^-$ mesons ($\omega_K=m_K$), the $s$-wave $K$-$N$ scalar attraction simulated by $\Sigma_{Kb}$ terms is canceled out by each corresponding range term. Also the $\Lambda^\ast$-pole, which lies about 30 MeV below the $KN$ threshold, works repulsively to the on-shell $s$-wave $K^-$$p$ scattering length. The range terms and $\Lambda^\ast$-pole term thus have sizable contributions to the $K^-$ self-energy in free space. However, as baryon density increases such that $\rho_{\rm B}\gtrsim\rho_0$, the $K^-$ energy $\omega_K$ decreases due to the $s$-wave $K$-$N$ attractive interactions, getting $\omega_K/m_K \ll 1$. In such a case, the $\Sigma_{Kb}$ terms and $s$-wave $K$-$N$ vector interaction, $-2X_0\omega_K$,  become dominant over the range terms and the $\Lambda^\ast$-pole term\footnote{In Ref.~\cite{fmmt1996}, the same result has been obtained in the second-order perturbation with respect to the axial-vector current (the second-order effect) in the framework of current algebra and PCAC~\cite{t88}. The present result on the lowest $K^-$ energy naturally agrees with those obtained by other works based on the chiral perturbation theory~\cite{lbm95,tw1995}.}, and the $K^-$ self-energy is well described without the range terms and $\Lambda^\ast$ term as
\begin{equation}
\Pi_K(\omega_K; \rho_B)
 = -\frac{1}{f^2}\sum_{b=p,n,\Lambda, \Sigma^-, \Xi^-}\left(\rho_b^s\Sigma_{Kb}+\omega_K\rho_bQ_V^b\right) \ , 
\label{eq:selfk2}
\end{equation}
which one uses for actual numerical calculations throughout this paper. 
The term leading to the $s$-wave $K$-$B$ scalar interaction in the self-energy (\ref{eq:selfk2}) is proportional to $\rho_b^s$, which has a complicated $\rho_{\rm B}$-dependence:
\begin{equation}
\rho_b^s=\rho_b\Bigg\lbrace 1-\frac{3}{5}\frac{p_F(b)^2}{2M_b^{\ast 2}}+\frac{9}{56}\left(\frac{p_F(b)}{M_b^\ast}\right)^4+O(p_F(b)^6) \Bigg\rbrace 
\label{eq:exprhobs}
\end{equation}
beyond the linear density approximation in the nonrelativistic limit. (See Fig.~\ref{fig:rhobs} in Sec.~\ref{subsec:onsetKC}.)

A scale of the $s$-wave $K$-$N$ attraction is characterized by the $K^-$ optical potential $U_K$ in the SNM, which is defined in terms of the $K^-$ self-energy [(\ref{eq:selfk2})] as $U_K=\Pi_K(\omega_K; \rho_{\rm B}) /(2\omega_K)\vert_{\rho_{\rm B}=\rho_0}$. In Table~\ref{tab:onset}, the $U_K$ is listed for each case of $L$ and $\Sigma_{Kn}$. 
The value of $U_K$ has sensitive dependence on $\Sigma_{Kn}$, while it depends little on the slope $L$. 
The deduced value of the depth $|U_K|$ (110~MeV$-$130~MeV) is larger than the theoretical values in the chiral unitary approach~\cite{ro2000} and the recent optimal value of the real part of the $K^-$ optical potential depth $|V_0|$ = 80 MeV with the imaginary part $W_0$ = $-$40 MeV, which was deduced from the measured spectrum shape of the inclusive missing-mass spectrum of $^{12}$C~($K^-, p$) reactions in the J-PARC~E05 experiment~\cite{ichikawa2020}. Nevertheless, the result is within the allowable range, considering that a much larger value of $|V_0|$ ($\gtrsim$ 150~MeV) is confronted with the difficulty of reproducing the spectrum~\cite{ichikawa2020}. 

\begin{figure*}[!]
\begin{minipage}[l]{0.50\textwidth}
\begin{center}
\includegraphics[height=0.32\textheight]{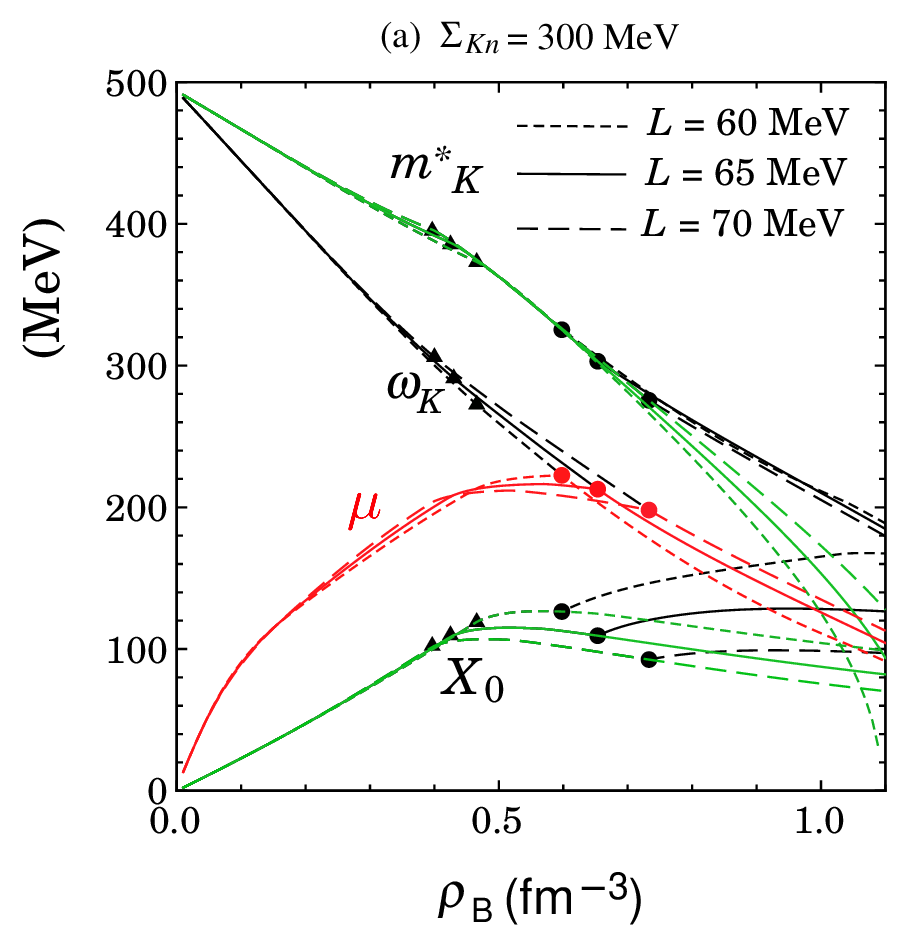}~
\end{center}
\end{minipage}~\vspace{-0.5cm}
\begin{minipage}[r]{0.50\textwidth}
\begin{center}
\includegraphics[height=0.32\textheight]{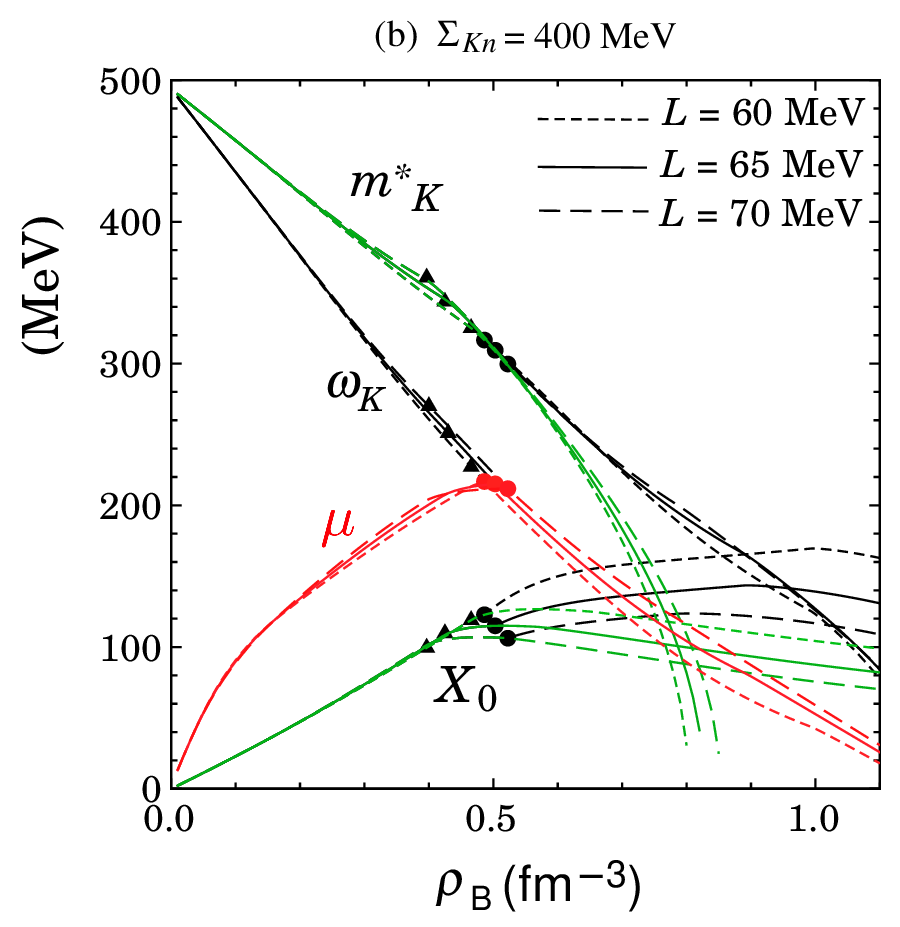}
\end{center}
\end{minipage}
\vspace{1.5cm}~
\caption{(a) The lowest $K^-$ energy $\omega_K$, the effective mass of $K^-$ meson, $m_K^{\ast}$ [ Eq.~(\ref{eq:ekm2})], and the $X_0$ [Eq.~(\ref{eq:x0})] as functions of baryon number density $\rho_{\rm B}$ 
for $\Sigma_{Kn}$ = 300 MeV in the case of $L$ = (60, 65, 70) MeV. The dependence of the charge chemical potential $\mu$ (=$\mu_e=\mu_\mu$ if muons are present) on $\rho_{\rm B}$ is also shown by the red lines for $L$ = (60, 65, 70) MeV. The filled triangle (filled circle) denotes the onset density of $\Lambda$ hyperon-mixing, $\rho_{\rm B}^c(\Lambda)$, (onset density of KC, $\rho_{\rm B}^c(K^-)$)  in each case of $L$. The $\omega_K$ is equal to the charge chemical potential $\mu$ in the ($Y$+$K$) phase for $\rho_{\rm B}\geq\rho_{\rm B}^c(K^-)$. For comparison, the density-dependence of $m_K^{\ast}$ and $X_0$ in pure hyperon-mixed matter, where $\theta$ is set to be zero,  is also shown by the green lines in each case of $L$.  \\ (b) The same as (a), but for $\Sigma_{Kn}$ = 400 MeV. 
The filled triangle corresponds to the same onset density of $\Lambda$ as in (a), See the text for details.
}
\label{fig:wk}
\end{figure*}

 In Fig.~\ref{fig:wk}, the lowest $K^-$ energy $\omega_K$ as a function of $\rho_{\rm B}$ is shown for (a) $\Sigma_{Kn}$=300 MeV and (b) $\Sigma_{Kn}$=400 MeV in the case of $L$ = (60, 65, 70) MeV. The dependence of the charge chemical potential $\mu$ (=$\mu_e$ and $\mu=\mu_\mu$ if muons are present)  on $\rho_{\rm B}$ is also shown by the red lines for $L$=(60, 65, 70) MeV. The filled triangle denotes the onset density of $\Lambda$ hyperon-mixing, $\rho_{\rm B}^c(\Lambda)$, at which hyperon ($\Lambda$)-mixing starts in the normal neutron-star matter. The filled circle denotes the onset density of KC, $\rho_{\rm B}^c(K^-)$,  realized from hyperon ($\Lambda$ and/or $\Xi^-$)-mixed matter in each case of $L$. 
The onset density of KC is read as  $ \rho_{\rm B}^c~(K^-)$ = 
(0.60$-$0.73)~fm$^{-3}$ [(3.7$-$4.6)~$\rho_0$] for $\Sigma_{Kn}$ = 300 MeV and 
 $\rho_{\rm B}^c~(K^-)$ = (0.49$-$0.52)~fm$^{-3}$ [(3.0$-$3.3)~$\rho_0$] for $\Sigma_{Kn}$ = 400 MeV, 
 within the range of the slope $L$ = (60$-$70) MeV. In comparison with the case where only the electrons are included for leptons~\cite{mmt2021}, the onset density for KC is slightly shifted to a high density, since the charge chemical potential $\mu$ becomes small at a given density due to the inclusion of the muons as well as the electrons.  For $\Sigma_{Kn}$ = 400 MeV, the $\omega_K$ is smaller at a given density than the case of $\Sigma_{Kn}$ = 300 MeV due to the stronger $s$-wave $K$-$B$ scalar attraction, so that the $\rho_{\rm B}^c~(K^-)$ for $\Sigma_{Kn}$ = 400 MeV is lower than the case of $\Sigma_{Kn}$ = 300 MeV. 
 In Table~\ref{tab:onset}, the onset densities $\rho_{\rm B}^c(\Lambda)$ and $\rho_{\rm B}^c(K^-)$ in the (ChL+MRMF+ UTBR+TNA) model for $\Sigma_{Kn}$ = 300 MeV and 400 MeV in case of $L$ = (60, 65, 70) MeV are listed. For all the cases of $L$ and $\Sigma_{Kn}$, the onset of $\Lambda$-mixing always precedes the onset of KC. 
 
It can be seen from Fig.~\ref{fig:wk} and Table~\ref{tab:onset} that the larger value of $L$ pushes up the $\rho_{\rm B}^c~(K^-)$ to a higher density for both $\Sigma_{Kn}$ = 300~MeV and 400~MeV. This result comes from considering (i) the $L$-dependence of the density profile of the lowest $K^-$ energy $\omega_K$ and (ii) that of the charge chemical potential $\mu$, in $\beta$-equilibrated $\Lambda$-mixed matter in the range, $\rho_{\rm B}^c(\Lambda)\lesssim \rho_{\rm B}\lesssim \rho_{\rm B}^c(K^-)$.   

\begin{figure*}[!]
\begin{minipage}[l]{0.50\textwidth}
\begin{center}~
\includegraphics[height=.31\textheight]{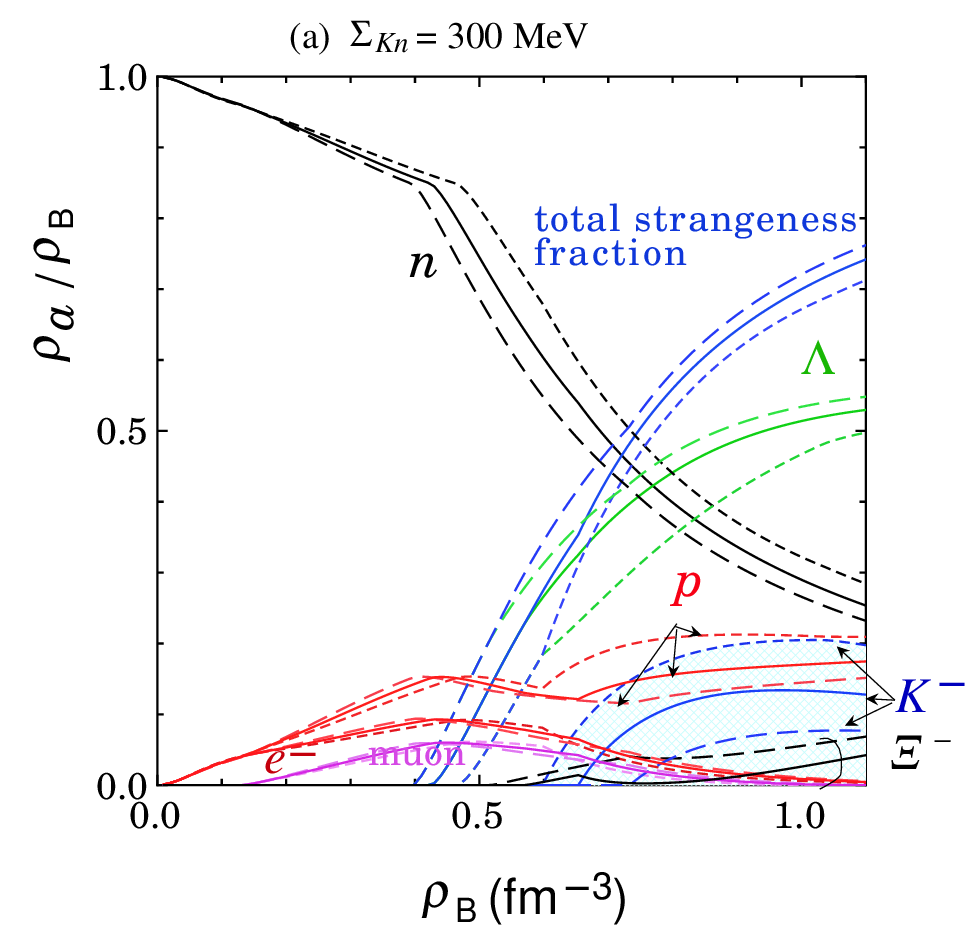}
\end{center}
\end{minipage}~
\begin{minipage}[r]{0.50\textwidth}
\begin{center}~
\includegraphics[height=.31\textheight]{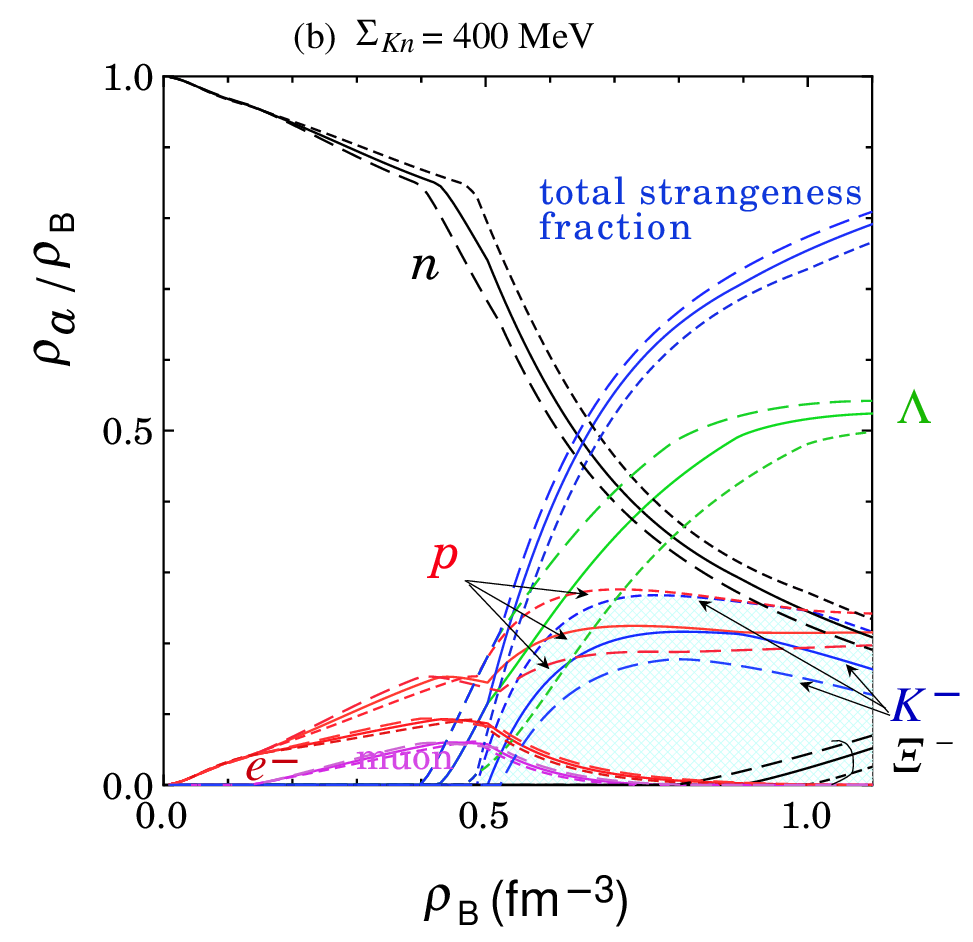}
\end{center}
\end{minipage}~\vspace{2.0cm}~
\caption{(a) The particle fractions in the ($Y$+$K$) phase as functions of the baryon number density $\rho_{\rm B}$ for $\Sigma_{kn}$=300 MeV in the case of $L$ = (60, 65, 70) MeV. The total strangeness fraction is given by $(\rho_{K^-}+\rho_\Lambda+2\rho_{\Xi^-})/\rho_{\rm B}$. (b) The same as (a) but for  $\Sigma_{kn}$ = 400 MeV. }
\label{fig:fractions}
\end{figure*}

\begin{figure*}[!]
\begin{minipage}[l]{0.50\textwidth}
\begin{center}~
\includegraphics[height=.31\textheight]{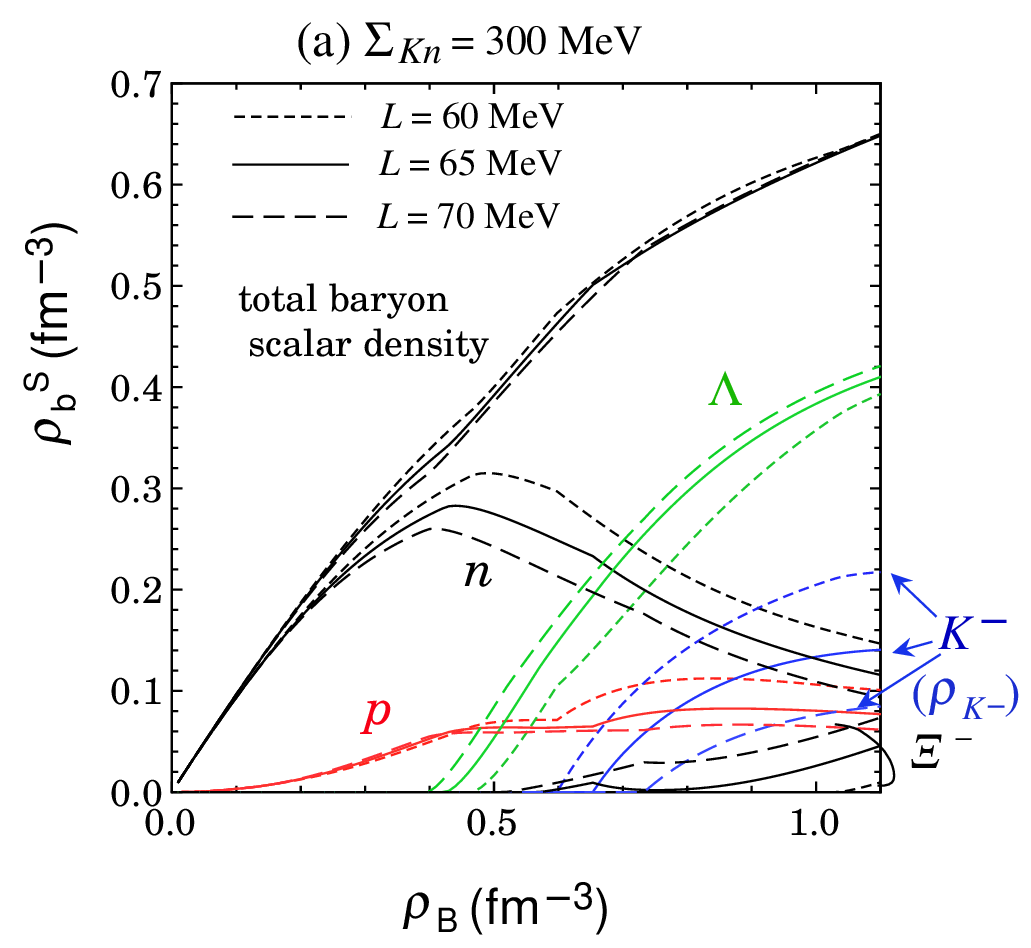}
\end{center}
\end{minipage}~
\begin{minipage}[r]{0.50\textwidth}
\begin{center}~
\includegraphics[height=.31\textheight]{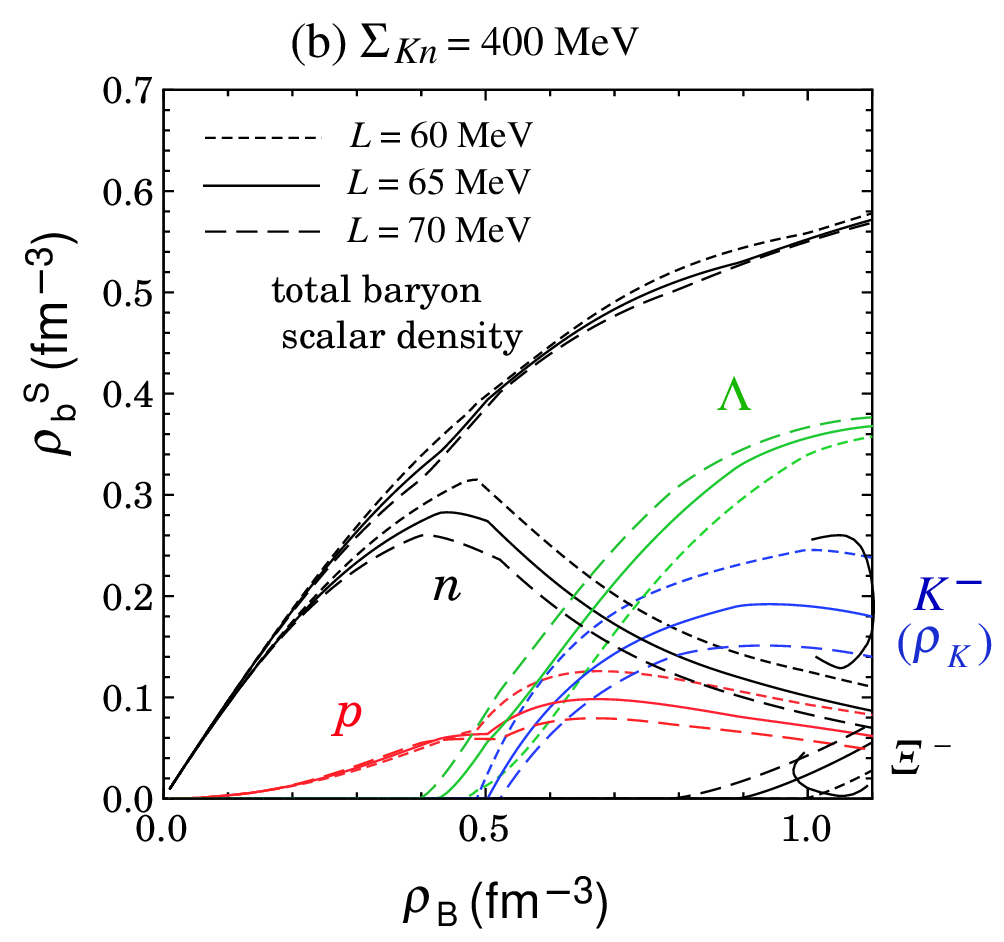}
\end{center}
\end{minipage}~\vspace{2.0cm}~
\caption{(a) The scalar densities for baryons $\rho_b^s$ ($b$=$p, n$, $\Lambda$, $\Xi^-$) including the ($Y$+$K$) phase as functions of the baryon number density $\rho_{\rm B}$ for $\Sigma_{kn}$ = 300 MeV in the case of $L$ = (60, 65, 70) MeV. The total baryon scalar density is given by $\displaystyle\rho_{\rm B}^s=\sum_{b=p,n,\Lambda, \Xi^-}\rho_{\rm b}^s$. For reference, the number density of KC, $\rho_{K^-}$ [(\ref{eq:rhokc})], is shown for $\rho_{\rm B}\gtrsim\rho_{\rm B}^c(K^-)$. (b) The same as (a) but for $\Sigma_{kn}$ = 400 MeV. }
\label{fig:rhobs}
\end{figure*}

Going into details of (i), one shows the density-dependence of the $X_0$ [Eq.~(\ref{eq:x0})] and that of the ``effective mass'' $m_K^\ast$ of the $K^-$ meson [Eq.~(\ref{eq:ekm2})] in Fig.~\ref{fig:wk}.
For reference, the particle fractions $\rho_a/\rho_{\rm B}$ and baryon scalar densities $\rho_b^s$ including the ($Y$+$K$) phase are also shown as functions of $\rho_{\rm B}$ in Fig.~\ref{fig:fractions} and Fig.~\ref{fig:rhobs}, respectively. 
It is reminded that 
both nucleon (neutron and proton) fractions and nucleon scalar densities are suppressed due to the appearance of $\Lambda$ hyperons and baryon number conservation (Sec.~\ref{subsec:onset-Lambda}). From Figs.~\ref{fig:fractions} and \ref{fig:rhobs} one can see that this suppression begins at a lower density and is more remarkable for larger $L$. 
On one hand,  the decrease in nucleon fractions leads to suppression of $K$-$N$ attractive vector interaction simulated by $X_0$ [$\propto\left(\rho_p+\rho_n/2-\rho_{\Xi^-}\right)/(2f^2)$ with $Q_V^\Lambda$ = 0  in Eq.~(\ref{eq:x0})]. Note that the effect of the $\Xi^-$-mixing is negligible since $\rho_{\Xi^-}$ is tiny, even if the $\Xi^-$ hyperons are mixed in the range $\rho_{\rm B}^c(\Lambda)\lesssim \rho_{\rm B}\lesssim \rho_{\rm B}^c(K^-)$. As seen in Fig.~\ref{fig:wk}, this suppression of $X_0$ begins at a lower density and is more remarkable for larger $L$, in accordance with the $L$-dependence of nucleon fractions beyond $\rho_{\rm B}^c(\Lambda)$~[Fig.~\ref{fig:fractions}].

On the other hand, $m_K^\ast$, where the energy contribution from the $s$-wave $K$-$B$ scalar attraction is roughly proportional to the total baryon scalar density, $\rho_{\rm B}^s$ (=$\rho_p^s+\rho_n^s+\rho_\Lambda^s+\rho_{\Xi^-}^s$), is little dependent on the slope $L$ for $\rho_{\rm B}^c(\Lambda)\lesssim \rho_{\rm B}\lesssim \rho_{\rm B}^c(K^-)$, as seen in Fig.~\ref{fig:wk}. This is because the increase in the $\Lambda$ scalar density makes up with the decrease in nucleon scalar densities, irrespective of the value of $L$, as seen in Fig.~\ref{fig:rhobs}.  
Therefore, the $L$-dependence of $\omega_K$ stems mainly from that of the $X_0$, and softening of the $\omega_K$ as $
\rho_{\rm B}$ increases is weakened, corresponding to more marked suppression of $X_0$ for larger $L$. 

With regard to (ii), the $L$-dependence of the density profile of the charge chemical potential $\mu$, one can see from Fig.~\ref{fig:wk} that $\mu$ turns to decrease through the process, $p+l^- \rightarrow \Lambda +\nu_l$, at the lower density for the larger $L$. As a consequence of the combining effect of (i) and (ii), the intersection point of $\omega_K$ and $\mu$, i.~e., the onset density $\rho_{\rm B}^c(K^-)$ shifts to a higher density for the larger $L$. Nevertheless the difference of $\rho_{\rm B}^c (K^-)$ due to different values of $L$ is tiny in the case of $\Sigma_{Kn}$ = 400 MeV. 
 
In Refs.~\cite{panda1995,carlson2000} the effect of $K$-$N$ and $N$-$N$ correlations was taken into account in the study of kaons in dense matter, and it was shown that the correlations tend to push up the $\omega_K$ over the relevant densities. Subsequently the effect of the short-range correlations was discussed in detail and was shown to be moderate~\cite{wrw1997,ww1997} as compared to the preceding results~\cite{panda1995,carlson2000}.
 The present result on the $U_K$ is similar to that of Refs.~\cite{wrw1997,ww1997} with inclusion of the short-range correlations  ($U_K \approx -120$MeV), 
 whereas  the $\omega_K$ in the model is slightly smaller than that of Refs.~\cite{wrw1997,ww1997}.

\section{Ground state properties of the ($Y$+$K$) phase and equation of state}
\label{sec:KC-EOS}

\subsection{Composition in the ($Y$+$K$) phase}
\label{subsec:compositon}

As seen in Fig.~\ref{fig:fractions} and Table~\ref{tab:onset}, the $\Lambda$-mixing starts at a lower density than that of KC or $\Xi^-$ hyperons. Subsequently, the fraction of $\Lambda$ hyperons monotonically increases with density even after KC or $\Xi^-$ hyperons appear. 

  In case KC sets in at a lower density than the $\Xi^-$ hyperons (for $L$ = 60 MeV with $\Sigma_{Kn}$ = 300 MeV, or for $\Sigma_{Kn}$ = 400 MeV), mixing of the $\Xi^-$ hyperons is pushed up to high densities, or even does not occur over the relevant densities, as seen from Figs.~\ref{fig:fractions},~\ref{fig:rhobs}, and Table~\ref{tab:onset}. (Note that for the case of $L$ = 70~MeV with $\Sigma_{Kn}$ = 400~MeV, the $\Xi^-$ hyperons precede KC, but their fraction soon vanishes, until they reappear at high density, $\rho_{\rm B}$ = 0.79~fm$^{-3}$.)
 On the contrary, for $L$ = 65 MeV and 70 MeV with $\Sigma_{Kn}$ = 300 MeV,  $\Xi^-$ hyperons appear at a lower density than KC, and the onset of KC is pushed up to high densities. 
Due to the repulsive $K$-$\Xi^-$ vector interaction term in $X_0$, the form of which is specified by chiral symmetry, KC and $\Xi^-$ hyperons compete against each other. This competitive relationship can also be seen from the interaction part of the $K^-$ number density $\rho_{K^-}$ [Eq.~(\ref{eq:rhokc})].
Thus the existence of KC strongly affects onset of the $\Xi^-$ hyperons and subsequent particle composition at high densities. 
In Table~\ref{tab:onset},  the onset densities at which hyperon ($\Lambda$, $\Xi^-$)-mixing starts and those of KC are summarized. 
It is to be noted that the $\Sigma^-$ hyperons are not mixed over the relevant densities due to the strong repulsion of the $V_{\Sigma^-}^N$ in the model.

The development of KC with increase in baryon density leads to enhancement of the proton fraction
 so that the positive charge carried by protons compensates for the negative charge by KC, keeping charge neutrality. 
On the other hand, the lepton ($e^-$, $\mu^-$) fractions are suppressed after the appearance of KC as well as $\Lambda$ hyperons, since the negative charge carried by leptons is replaced by that of KC,  avoiding a cost of degenerate energy of leptons. The ($Y$+$K$) phase becomes almost lepton-less at high densities (see Fig.~\ref{fig:fractions}). 
As a consequence, the charge chemical potential $\mu$ [=$(3\pi^2\rho_e)^{1/3}$ ] decreases steadily as density increases after the onset of kaon condensation and has the value with $\mu\lesssim O(m_\pi)$ (see Fig.~\ref{fig:wk}). 
These features concerning proton and lepton fractions and the charge chemical potential are characteristic of the hadron phase in the presence of KC.
The total strangeness is carried mainly by $\Lambda$ hyperons and KC in the ($Y$+$K$) phase with a minor fraction of $\Xi^-$ hyperons at high densities. 

The slope $L$-dependence of each matter composition is summarized as follows. As has been shown in Sec.~\ref{subsec:onset-Lambda}~[see Fig.~\ref{fig:vlam}], mixing of the $\Lambda$ hyperons starts at a lower baryon density $\rho_{\rm B}$ for larger $L$ within the range $L$ = (60$-$70)~MeV. Moreover, for larger $L$, the $\Lambda$ hyperons become more abundant at the same $\rho_{\rm B}$ beyond $\rho_{\rm B}^c(\Lambda)$ and even in the ($Y$+$K$) phase, as seen in Fig.~\ref{fig:fractions}. 
More abundant $\Lambda$ hyperons lead to less nucleon ($p$, $n$) fractions due to baryon number conservation. As a result, $K$-nucleon vector attraction simulated by $X_0$ becomes suppressed for large $L$, and so do kaon condensates. Along with the $L$-dependence of KC, the proton fraction (the lepton fraction) becomes suppressed (enhanced) in the case of large $L$. Due to the competing effect with KC, mixing of the $\Xi^-$ hyperons is enhanced for large $L$. (see  also Fig.~\ref{fig:fractions}.)
The $L$-dependence of the total strangeness fraction is controlled mainly by that of the $\Lambda$ hyperons, since the decrease in the KC fraction by the increase in $L$ is almost cancelled with the increase in the strangeness fraction carried by the $\Xi^-$ hyperons. 

\subsection{Self-suppression of the $s$-wave $K$-$B$ scalar attraction in the presence of KC}
\label{subsec:self-suppression}
 
Beyond the onset density of KC, the $s$-wave $K$-$B$ scalar attraction is suppressed in the presence of KC as compared with the pure hyperon-mixed case, as seen from the difference of $m_K^\ast$ between the cases of the ($Y$+$K$) phase (black lines) and pure hyperon-mixed matter (green lines) in Fig.~\ref{fig:wk}. Such suppression of the $s$-wave $K$-$B$ scalar attraction in the presence of KC is explained as follows; once KC appears, 
the effective baryon mass $\widetilde M_b^\ast$ decreases as density increases following Eq.~(\ref{eq:wtmb}) with $M_{\rm b}^\ast$ being Eq.~(\ref{eq:effbm}), and the decrease in $\widetilde M_b^\ast$ leads to suppression of the baryon scalar density, $\rho_b^s$, at higher densities, 
which, in turn, results in the suppression of the $K$-$B$ scalar attraction, through the term proportional to $\rho_b^s $ in $m_K^\ast$ [Eq.~(\ref{eq:ekm2})]. Thus one can see {\it self-suppression mechanism} of the $s$-wave $K$-$B$ scalar interaction in the ($Y$+$K$) phase, which was originally discussed for KC in neutron-star matter without hyperon-mixing as an effect unique to relativistic framework~\cite{fmmt1996}.

\subsection{Energy per baryon and pressure}
\label{subsec:EOS}

\begin{figure*}[!]
\begin{minipage}[l]{0.50\textwidth}
\begin{center}
\includegraphics[height=0.43\textheight]{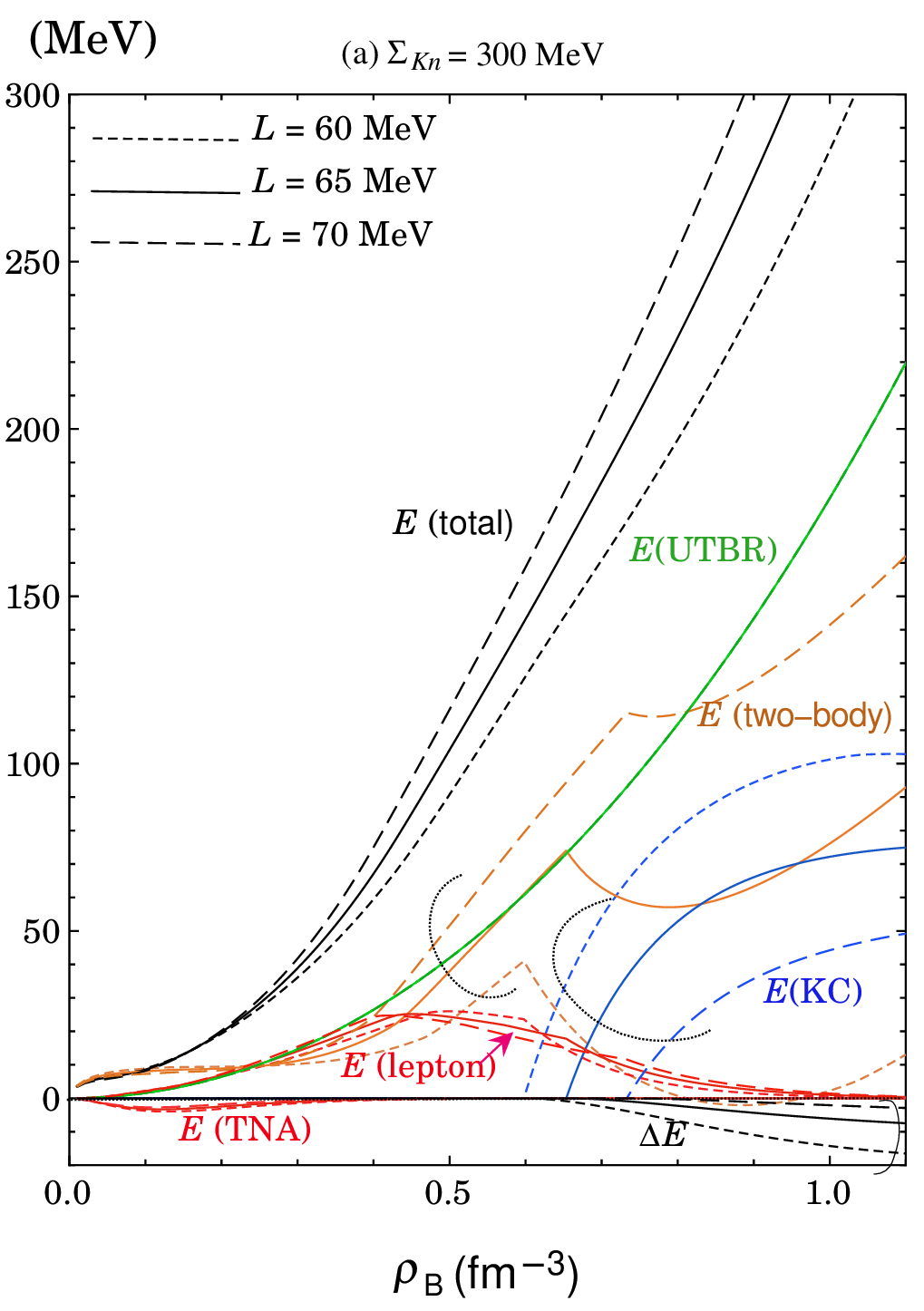}~
\end{center}
\end{minipage}~\vspace{-0.5cm}
\begin{minipage}[r]{0.50\textwidth}
\begin{center}
\includegraphics[height=.43\textheight]{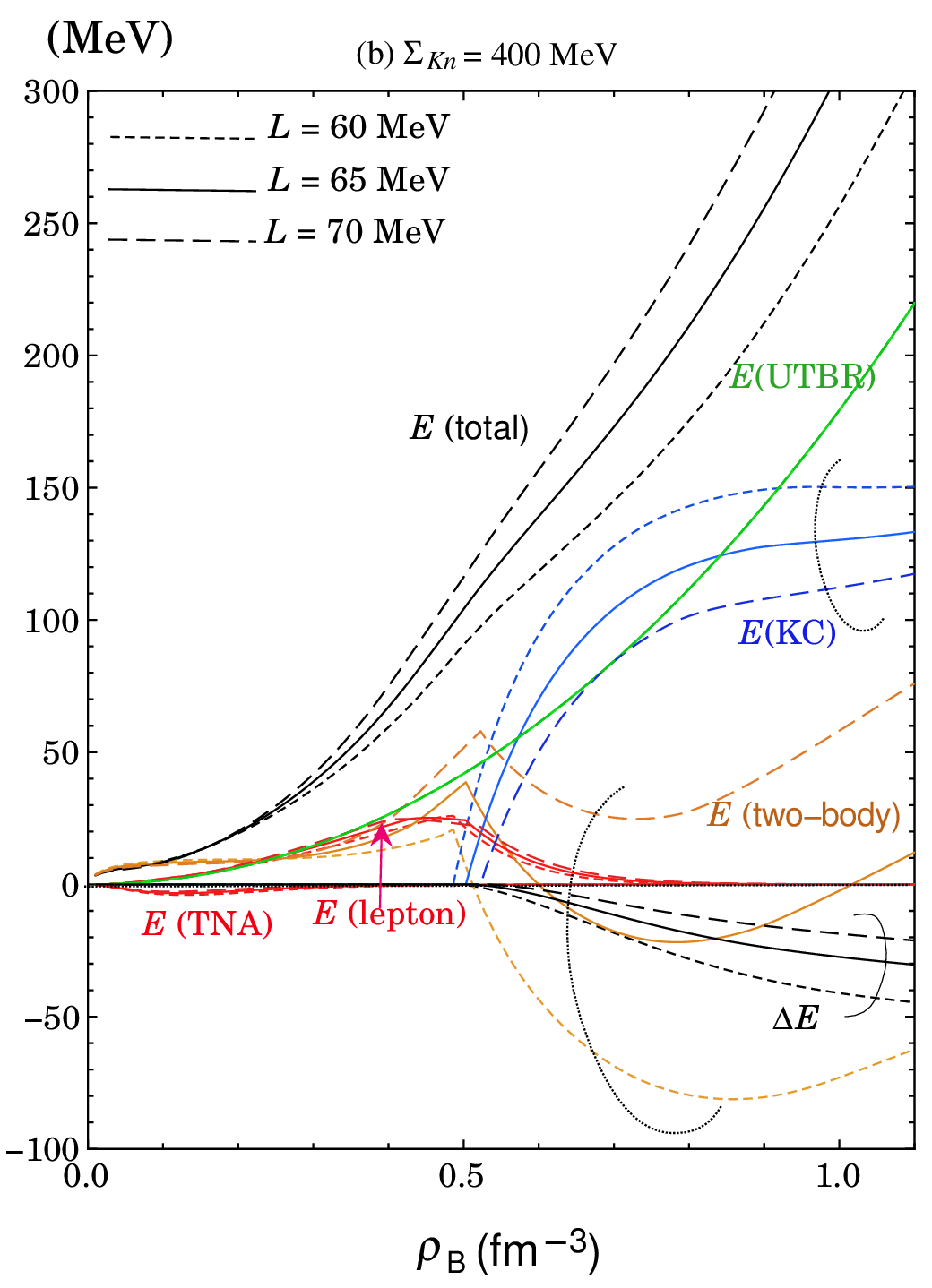}
\end{center}
\end{minipage}
\vspace{1.5cm}~
\caption{(a) The total energy per unit of baryon $E$~(total) [$= {\cal E}/\rho_{\rm B}$], measured from the nucleon rest mass, and each energy contribution, $E$~(KC) [$= {\cal E}_K/\rho_{\rm B}$], $E$~(two-body) [$= {\cal E}_{B,M}/\rho_{\rm B}$], $E$~(UTBR) [$= {\cal E}~({\rm UTBR})/\rho_{\rm B}$], $E$~(TNA) [$= {\cal E}~({\rm TNA})/\rho_{\rm B}$], $E$~(lepton) [$= \left({\cal E}_e+{\cal E}_\mu\right)/\rho_{\rm B}$],  as functions of baryon number density $\rho_{\rm B}$ for $\Sigma_{Kn}$ = 300 MeV in the case of $L$=(60, 65, 70) MeV. 
(b) The same as (a) but for $\Sigma_{Kn}$ = 400 MeV. For reference, the energy difference per baryon between the ($Y$+$K$) phase and pure hyperon-mixed matter, $\Delta E$ ($\leq 0$), is also shown as functions of $\rho_{\rm B}$. 
}
\label{fig:energy}
\end{figure*}

In Fig.~\ref{fig:energy}, the total energy per unit of baryon $E$~(total) [$= {\cal E}/\rho_{\rm B}$], measured from the nucleon rest mass, and each energy contribution are shown as functions of $\rho_{\rm B}$ for (a) $\Sigma_{Kn}$ = 300 MeV and (b) $\Sigma_{Kn}$ = 400 MeV in the case of $L$=(60, 65, 70) MeV. For reference, the energy difference per unit of baryon between the ($Y$+$K$) phase and pure hyperon-mixed matter, $\Delta E$ ($\leq 0$), is also shown as functions of $\rho_{\rm B}$. 
The $E$~(UTBR) [$= {\cal E}~({\rm UTBR})/\rho_{\rm B}$], which is roughly proportional to $\rho_{\rm B}^2$, has a sizable contribution to the total energy and results in stiffening of the EOS at high densities. 
The $E$~(two-body) [$= {\cal E}_{B,M}/\rho_{\rm B}$] also brings about repulsive energy as large as $E$~(UTBR) until the onset of KC. 
For larger $L$, hyperons start to be mixed at lower densities and occupy bigger fractions. Nevertheless the attractive energy contribution from the scalar ($\sigma$)-meson exchange through the reduction of the baryon effective mass [the first term of Eq.~(\ref{eq:vb})] to the $E$~(two-body) is saturated at lower density, while the repulsive energy contribution from vector ($\omega$)-meson exchange increases more rapidly proportional to $\rho_{\rm B}$. Thus the portion of the repulsive energy $E$~(two-body) in $E$~(total) becomes more marked for larger $L$ even in the hyperon-mixed matter. 

In contrast, the repulsive energy stemming from the two-body $B$-$B$ interaction becomes always dominant over the repulsive contribution from the NLSI terms at high densities regardless of the choice of $L$ in the (MRMF+NLSI) model. This is because the many-baryon repulsion tends to diminish the meson mean-fields at high densities, leading to an energetically stable state. [See Fig.~\ref{fig:eNLSI} in Appendix~\ref{subsec:a4} in the case of $L$ = 65 MeV for the SNM, and also Fig.~6 in Ref.~\cite{mmt2022} for hyperon-mixed matter.]

Beyond the onset density of KC, the $E$~(two-body) turns to decrease with density until it increases again at higher density except for $\Sigma_{Kn}$ = 300~MeV and $L$ = 70~MeV, due to the attraction from the $s$-wave $K$-$B$ interaction for both cases of $\Sigma_{Kn}$.  
On the other hand, the $E$~(KC) [$= {\cal E}_K/\rho_{\rm B}$], composed of kinetic and mass terms of KC, increases with baryon density. 
The sum of $E$~(two-body) and $E$~(KC) results in positive energy which increases with baryon density and works to stiffen the EOS as much as $E$~(UTBR). 

In general, mixing of hyperons, especially $\Lambda$ hyperons, tend to diminish more the scale of KC for larger $L$ as a result of competing effect with hyperons. 
Thus the absolute value of the energy difference $|\Delta E|$ becomes small for larger $L$ for both cases of $\Sigma_{Kn}$ = 300~MeV and 400~MeV, as seen in Fig.~\ref{fig:energy}. There is a clear difference in energy for $\Sigma_{Kn}$ = 400 MeV from the case of the pure $Y$-mixed matter, while 
the difference is tiny for $\Sigma_{Kn}$ = 300 MeV, in particular, in the case of $L$ = 70 MeV. 

The attraction from $E$~(TNA) [$= {\cal E}~({\rm TNA})/\rho_{\rm B}$] is responsible only in the vicinity of the saturation density $\rho_0$, and the contribution from $E$~(TNA) gets negligible beyond $\rho_0$. The energy by leptons ($e^-$ and $\mu^-$), $E$~(lepton), has a minor contribution to the total energy, in particular, in the ($Y$+$K$) phase, reflecting the lepton-less nature of kaon-condensed phase. 
 
\begin{figure*}[!]
\begin{minipage}[l]{0.50\textwidth}
\begin{center}~
\includegraphics[height=.32\textheight]{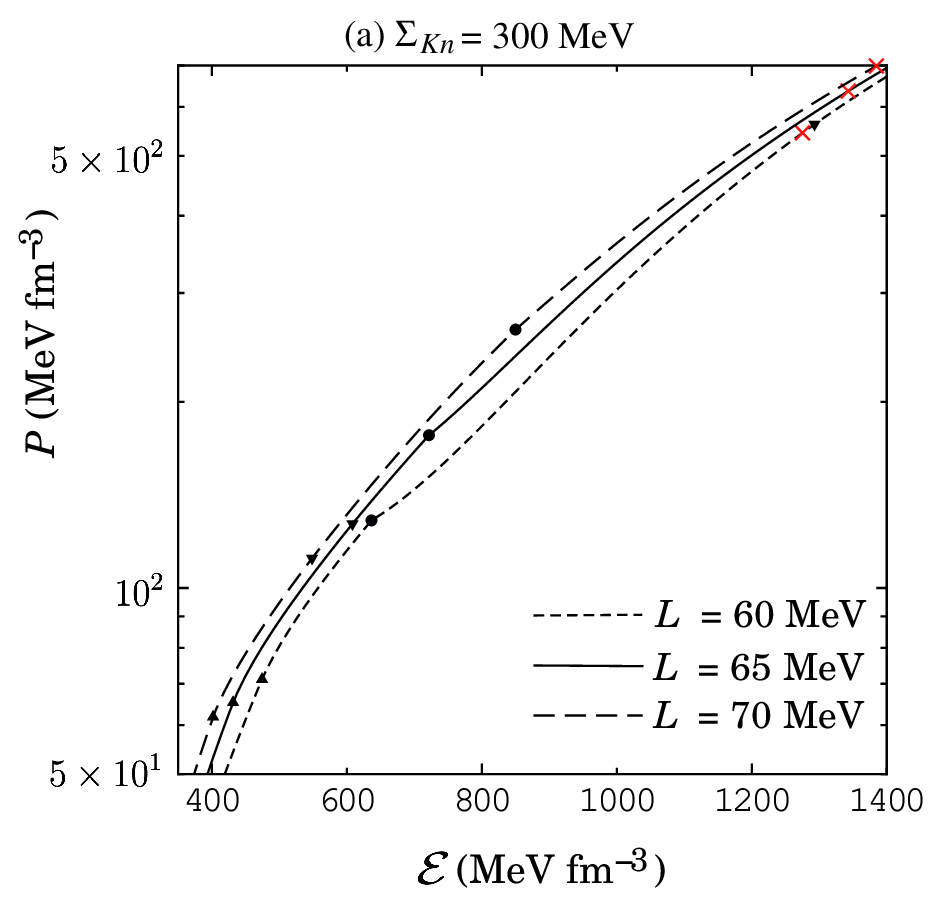}
\end{center}
\end{minipage}~
\begin{minipage}[r]{0.50\textwidth}
\begin{center}
\includegraphics[height=.32\textheight]{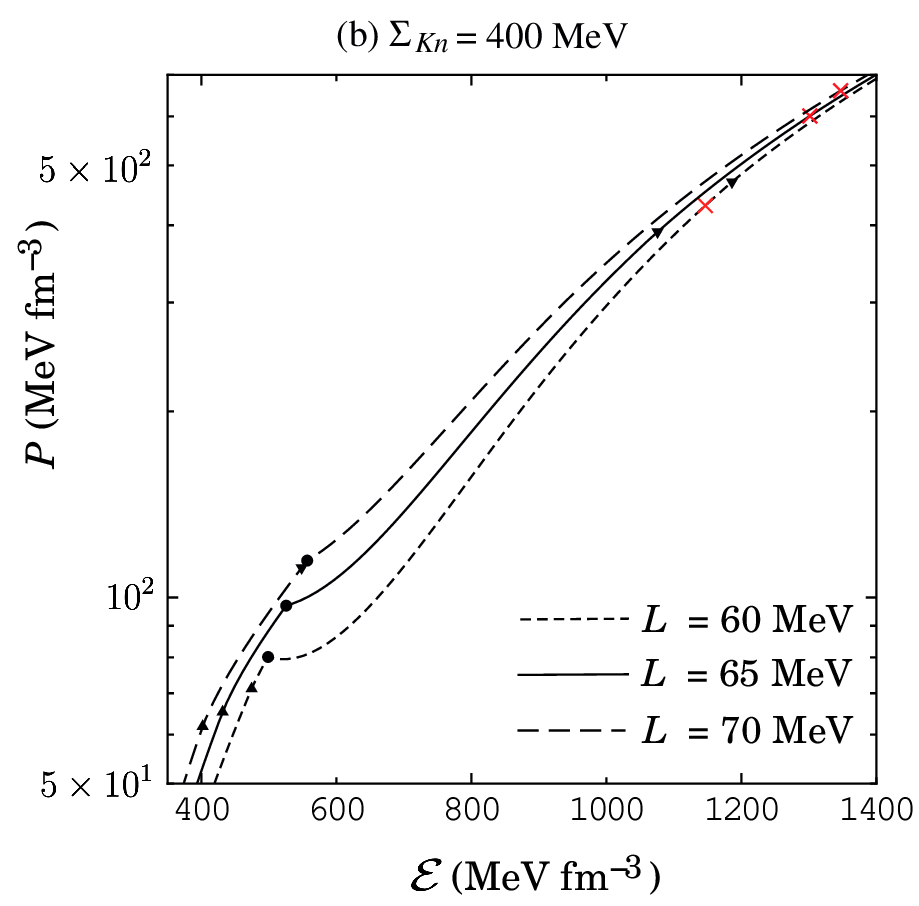}
\end{center}
\end{minipage}
~\vspace{1.0cm}~
\caption{(a) The pressure, $P$ [$\equiv\rho_{\rm B}^2\partial\left({\cal E}/\rho_{\rm B}\right)/\partial\rho_{\rm B}$ ], for the ($Y+K$) phase as a function of energy density ${\cal E}$ [$\equiv (E~({\rm total})+M_N)\rho_{\rm B}$] for $\Sigma_{Kn}$ = 300 MeV. The filled triangles, filled circles, and filled inverted triangles denote the onset densities of $\Lambda$ hyperon, KC, and $\Xi^-$ hyperon, respectively, in each case of $L$. The cross points stand for the causality limit, where the sound speed exceed the speed of light.  (b) The same as (a) but for $\Sigma_{Kn}$ = 400 MeV. 
The filled triangle in each case of $L$ and the filled inverted triangle for $L$ = 70~MeV correspond to  the same onset densities as in (a). }
\label{fig:pres}
\end{figure*}

\begin{figure*}[!]
\begin{minipage}[l]{0.50\textwidth}
\begin{center}~
\includegraphics[height=.32\textheight]{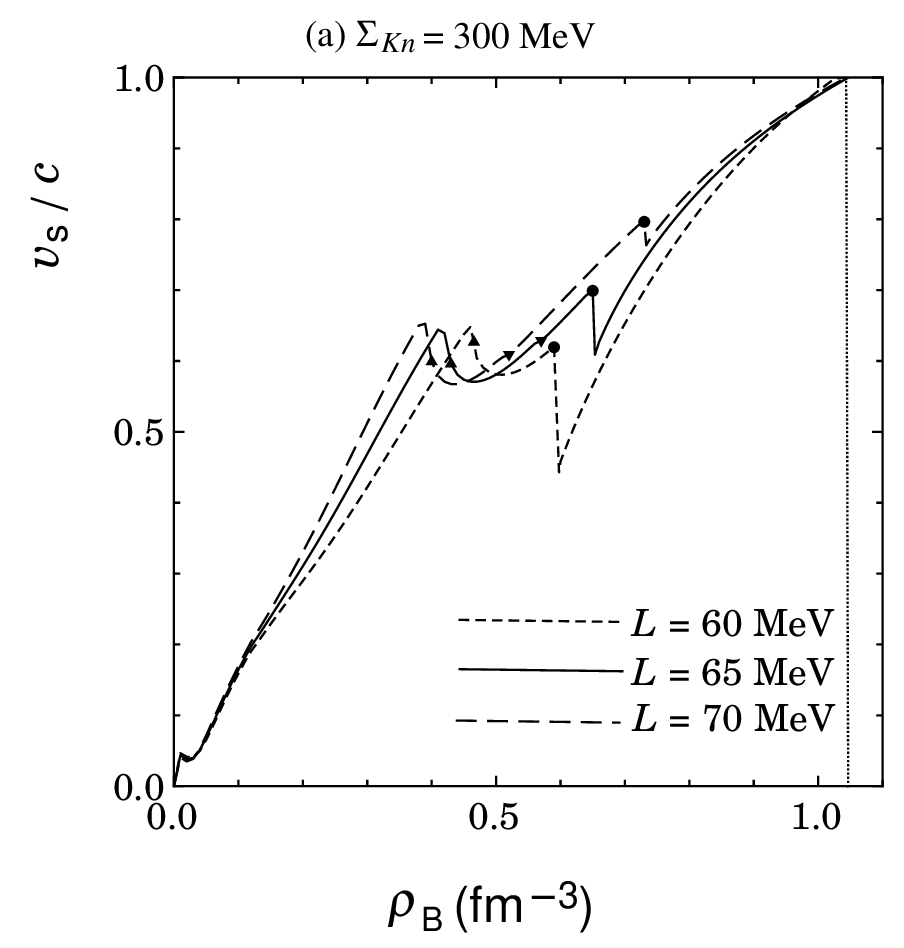}
\end{center}
\end{minipage}~
\begin{minipage}[r]{0.50\textwidth}
\begin{center}
\includegraphics[height=.32\textheight]{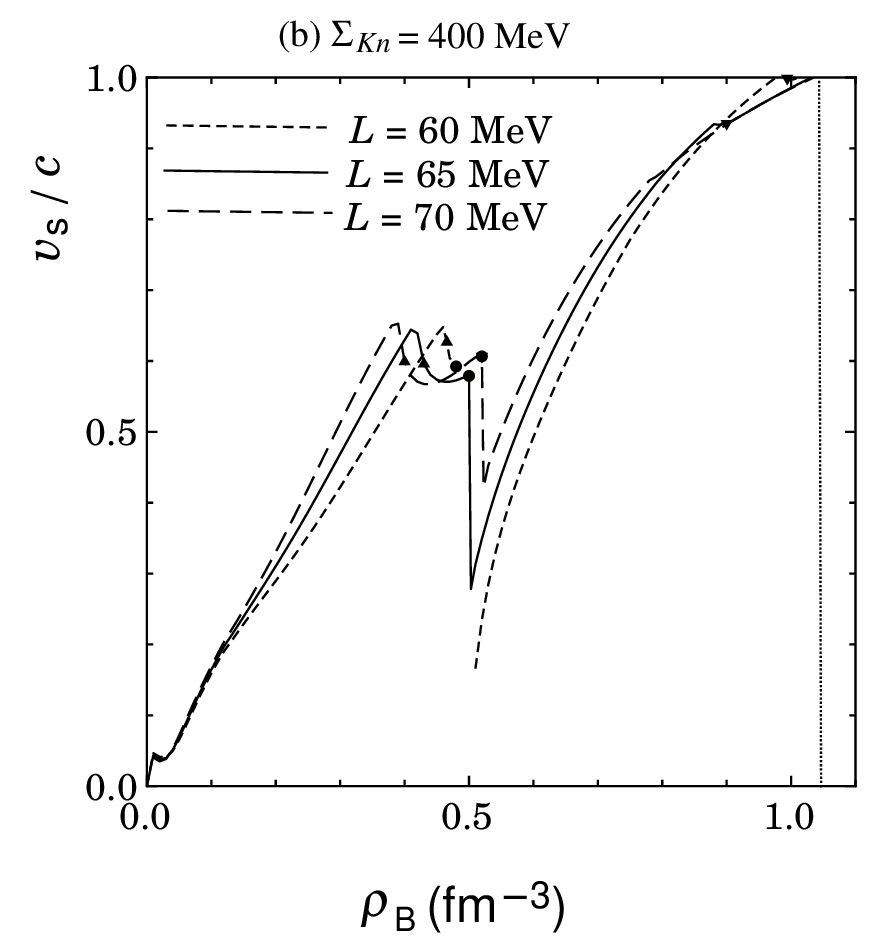}
\end{center}
\end{minipage}
~\vspace{1.0cm}~
\caption{(a) The sound velocity $v_s$ [$\equiv \left(\partial P/\partial {\cal E}\right)^{1/2}$ ] in the unit of the speed of light $c$ as functions of baryon number density $\rho_{\rm B}$ for $\Sigma_{Kn}$ = 300 MeV in the case of $L$=(60, 65, 70) MeV. The symbols are the same as in Fig.~\ref{fig:pres}.  (b) The same as (a) but for $\Sigma_{Kn}$ = 400 MeV.  }
\label{fig:vs}
\end{figure*}

In Fig.~\ref{fig:pres} the pressure, $P$ [$\equiv\rho_{\rm B}^2\partial\left({\cal E}/\rho_{\rm B}\right)/\partial\rho_{\rm B}$ ] is shown as functions of energy density ${\cal E}$ [$\equiv (E~({\rm total})+M_N)\rho_{\rm B}$] for (a) $\Sigma_{Kn}$ = 300 MeV and (b) $\Sigma_{Kn}$ = 400~MeV in the case of $L$=(60, 65, 70) MeV. The sound velocity $v_s$ [$\equiv \left(\partial P/\partial {\cal E}\right)^{1/2}$ ] in the unit of the speed of light $c$ is also shown as functions of baryon number density in Fig.~\ref{fig:vs}. 

One can see from Fig.~\ref{fig:pres} that the EOS becomes stiffer for larger case of $L$ over the whole density region, in conformity with the $L$-dependence of the total energy per unit of baryon shown in Fig.~\ref{fig:energy}. After the onset density of KC $\rho_{\rm B}^c~(K^-)$, the pressure increases monotonically with energy density ${\cal E}$ except for the case of ($L$, $\Sigma_{Kn}$) =(60~MeV, 400~MeV), where there is an unstable region, $dP/d{\cal E}<0$, for the small density interval, 
$\rho_{\rm B}^c~(K^-)\leq \rho_{\rm B} \leq \rho_{\rm B}^c~(K^-)+{\it \Delta} \rho_{\rm B}$ with ${\it \Delta} \rho_{\rm B}\simeq$ 0.03~fm$^{-3}$. In this density region, the sound velocity $v_s$ is not well defined, as seen in Fig.~\ref{fig:vs}. For the other cases of ($L$, $\Sigma_{Kn}$), the $v_s$ continuously changes with $\rho_{\rm B}$ but abruptly decreases during the small density interval. The abrupt change of the $v_s$ can also be seen across the onset density of the $\Lambda$ hyperon-mixing, but the scale of change is tiny as compared to the case of KC.
Recently, responses to radial and nonradial oscillations of compact stars with kaon-condensed phase have been considered~\cite{kheto2023}. The existence of the rapid change of the sound velocity in the vicinity of the onset density of KC may affect chracteristic features of fundamental and higher-order oscillation modes.

It should be mentioned here that there is a causal limit, beyond which the sound velocity exceeds the speed of light, in the present model for each case of $L$, specifically beyond $\rho_{\rm B}\approx$ 1.03~fm$^{-3}$. 
As indicated in Fig.~\ref{fig:MR}, 
the causality condition is fulfilled for a larger value of $L$ = 70 MeV until the mass reaches the maximum mass, while it is violated for $L$ = 60 MeV and 65 MeV before the mass reaches the maximum mass. 
[See also Table~\ref{tab:onset} for the central density of the maximum mass star, $\rho_{\rm B, center}(M_{\rm max})$.]
The many-baryon repulsion more than two-body force eventually leads to violation of the causality condition for sufficiently high baryon densities. 
In order to improve the EOS so as to meet the causality condition over the whole densities in compact stars, one  
adopted the Lorentz scalar form of the many-baryon repulsive force and tried to obtain the EOS with the ($Y$+$K$) phase which is sufficient for the fulfillment of causality condition~\cite{muto2024}. 
It is a common feature that  the sound velocity $v_s$ in hadronic matter becomes large as density increases and that it approaches the light velocity $c$ due to the many-baryon repulsive forces. 
The result should be compared with the case of hadron-quark crossover~\cite{baym2019,kojo2021,fujimoto2022}.  

\section{Structure of neutron stars with the ($Y$+$K$) phase}
\label{sec:MR}
\subsection{Gravitational mass to radius and central density relations}
\label{subsec:MR}

\begin{figure*}[!]
\begin{minipage}[l]{0.50\textwidth}
\begin{center}~
\includegraphics[height=.37\textheight]{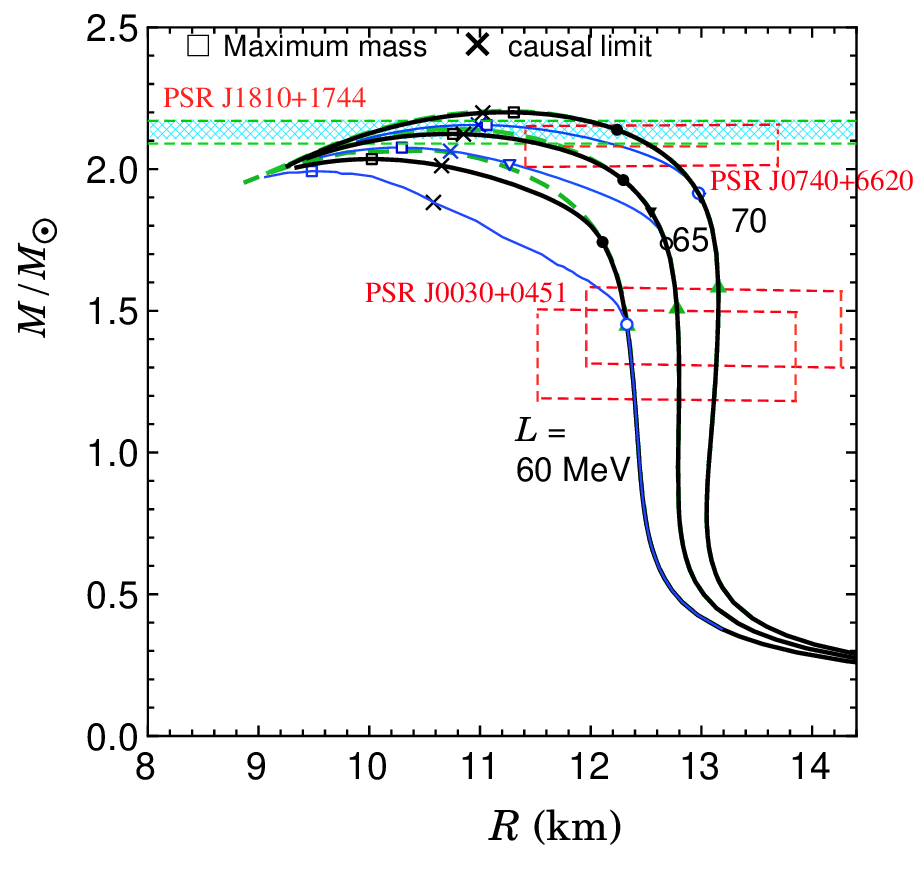}
\caption{The gravitational mass~$M$ to radius~$R$ relations after solving the Tolman-Oppenheimer-Volkoff  equation for $L$ = (60, 65, 70) MeV obtained with the (ChL+MRMF+UTBR(SJM2)+TNA) model. The branches including KC in the core are denoted as the black bold solid lines (blue thin solid lines) for $\Sigma_{Kn}$ = 300 MeV (400 MeV).  For comparison, the branch with pure hyperon-mixed matter, where KC is switched off by setting $\theta=0$, is shown by the green dashed line for each case of $L$. See the text for details. \\ }
\label{fig:MR}
\end{center}
\end{minipage}~
\begin{minipage}[r]{0.50\textwidth}
\begin{center}~
\includegraphics[height=.37\textheight]{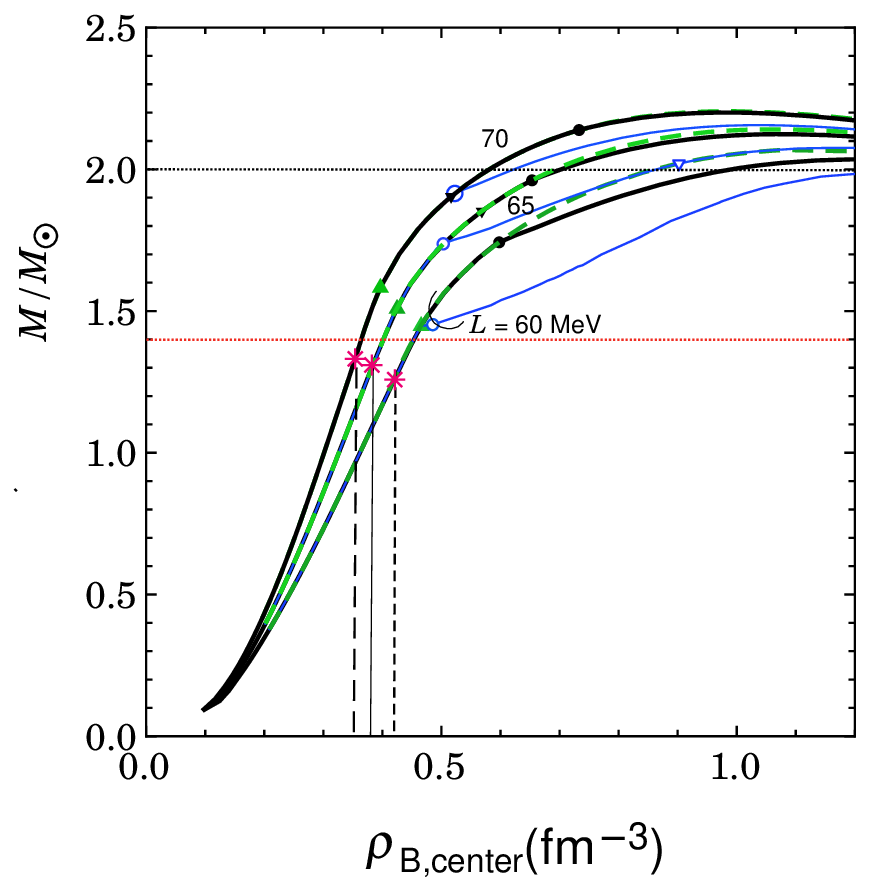}
\caption{The gravitational mass~$M$ to baryon number density at the center of the star, $\rho_{\rm B, {\rm center}}$,  for $L$ = (60, 65, 70) MeV obtained with the (ChL+MRMF+UTBR+TNA) model. The meanings of the curves and attached symbols are the same as in Fig.~\ref{fig:MR} except for the asterisk, at which the density at the center of the star reaches the onset of the direct Urca process. See the text for details. \\ }
\label{fig:Mrhoc}
\end{center}
\end{minipage}~
\end{figure*}

Here the effects of KC on the static properties of compact stars are discussed.  
In Fig.~\ref{fig:MR}, the gravitational mass $M$ - radius $R$ relations are shown for $L$ = (60, 65, 70) MeV. 
In Fig.~\ref{fig:Mrhoc}, the gravitational mass $M$ is shown as functions of baryon number density at the center of the star, $\rho_{\rm B, center}$. 
They are obtained by solving the Tolman-Oppenheimer-Volkoff equation with the EOS including the ($Y$+$K$) phase. For low density region $\rho_{\rm B} < $ 0.10~fm$^{-3}$ below the density of uniform matter, one utilizes the EOS of Ref.~\cite{BPS1971} and combine with the EOS obtained in the model for $\rho_{\rm B}\geq$ 0.10~fm$^{-3}$. 
The branches including KC in the core are denoted as the bold solid lines (thin solid lines) for $\Sigma_{Kn}$ = 300 MeV (400 MeV).  For comparison, the branch including pure hyperon-mixed matter, where KC is switched off by setting $\theta=0$, is shown by the green dashed line for each case of $L$. The filled triangle [$\blacktriangle$] stands for the branch point where the $\Lambda$ hyperons appear from nuclear matter in the center of the star. The branch point at which KC appears in the center of the star is indicated by the filled circle [$\bullet$] (open circle [$\circ$]) in the case of $\Sigma_{Kn}$ = 300 MeV (400 MeV). 
In Fig.~\ref{fig:MR}, the location of the maximum mass point for each branch including the ($Y$+$K$) phase is indicated by the open square [$\square$]. The cross point [$\times$] corresponds to the causal limit beyond which the sound velocity exceeds the speed of light. 
Figure~\ref{fig:MR} shows that the maximum mass and its radius shift to larger values for larger $L$. Also the radius of neutron stars for a given mass in the stable branch increases with $L$. 
These properties reflect that the EOS becomes stiffer at high densities for larger $L$, due to the fact that the repulsive contribution from the ``two-body''  baryon interaction gets large for large $L$ (\ref{subsec:EOS}). 
One can also see the $L$-dependence of the stiffness of the EOS from Fig.~\ref{fig:Mrhoc}. At a given $M$, the density at the center of the star, $\rho_{\rm B, center}$, is smaller for larger $L$. 
For reference, the central density of the maximum mass star, $\rho_{\rm B, center}(M_{\rm max})$, is listed in Table~\ref{tab:onset}.

In Table~\ref{tab:MR}, some critical gravitational masses and their radii are listed for $\Sigma_{Kn}$ = 300 MeV and 400 MeV in the case of $L$=(60, 65, 70) MeV. 
$M^c(\Lambda)$ and $R^c(\Lambda)$ [$M^c(K^-)$ and $R^c(K^-)$] is the mass and radius of the neutron star where the central density attains the onset density of the $\Lambda$-hyperons, $\rho_{\rm B}^c(\Lambda)$ [the onset density of KC, $\rho_{\rm B}^c(K^-)$]. $M_{\rm max}$ and $R(M_{\rm max})$ are the maximum mass of the neutron star and its radius. 
\begin{table*}[!]
\caption{Some critical gravitational masses in the unit of the solar mass $M_\odot$ and their radii of neutron stars for $\Sigma_{Kn}$ = 300 MeV and 400 MeV in the case of $L$=(60, 65, 70) MeV, obtained with the (ChL+MRMF+UTBR+TNA) model. The $M^c(\Lambda)$ and $R^c(\Lambda)$ [$M^c(K^-)$ and $R^c(K^-)$] are the mass and radius of the neutron star where the central density reaches the onset density of the $\Lambda$-hyperons, $\rho_{\rm B}^c(\Lambda)$ [the onset density of KC, $\rho_{\rm B}^c(K^-)$]. 
$M_{\rm max}$ and $R(M_{\rm max})$ are the maximum mass of the neutron star and its radius.}
\begin{center}
\begin{tabular}{  c | c || c | c || c | c || c | c }
\hline
 $L$ & $\Sigma_{Kn}$ & $M^c(\Lambda)/M_\odot $ & $R^c(\Lambda)$ & $M^c(K^-)/M_\odot $ & $R^c(K^-)$ & $M_{\rm max}/M_\odot $ & $R(M_{\rm max})$  \\
 (MeV) & (MeV)          &                                           & (km)                   &                               & (km)           
&                         & (km)                      \\
\hline\hline
 60 & \begin{tabular}{c} 
 300   \\
 400   \\
\end{tabular}             & 1.448                                    &  12.33                & 
\begin{tabular}{c} 
1.742     \\
1.452     \\
\end{tabular}  
& \begin{tabular}{c}
12.11     \\
12.33     \\
\end{tabular}
&  \begin{tabular}{c}
2.035       \\
1.993       \\
\end{tabular}
& \begin{tabular}{c} 
10.02    \\
 9.48  \\
\end{tabular}  \\\hline\hline
 65 & \begin{tabular}{c}
 300     \\
 400    \\
 \end{tabular}
& 1.508                                    &  12.78                
& \begin{tabular}{c}
1.961      \\
1.737      \\
\end{tabular}    
& \begin{tabular}{c}
12.29     \\
12.68     \\
\end{tabular}
& \begin{tabular}{c}
2.124      \\ 
2.076     \\
\end{tabular}
& \begin{tabular}{c}
10.76  \\
10.29  \\
\end{tabular} \\\hline\hline      
 70 & \begin{tabular}{c}
 300     \\
 400     \\
 \end{tabular}
& 1.582      &  13.15              
& \begin{tabular}{c}
2.139      \\
1.915  \\
\end{tabular}
&   \begin{tabular}{c}
12.24      \\ 
12.97    \\
\end{tabular} 
&\begin{tabular}{c}
 2.200        \\
2.155         \\
\end{tabular}
& \begin{tabular}{c}
11.31       \\
 11.06      \\
\end{tabular} \\ \hline
 \end{tabular}
\label{tab:MR}
\end{center}
\end{table*}
The mass of the neutron star where the central density reaches the onset density $\rho_{\rm B}^c(K^-)$ is (1.74, 1.96, 2.14)~$M_\odot$ for $L$ = (60, 65, 70) MeV in the case of $\Sigma_{Kn}$ = 300 MeV, 
and (1.45, 1.74, 1.92)$M_\odot$ for $L$ = (60, 65, 70) MeV in the case of $\Sigma_{Kn}$ = 400 MeV, as is also seen from Figs.~\ref{fig:MR} and \ref{fig:Mrhoc}.  

 Constraints on mass and radius from recent observations of some neutron stars are indicated in Fig.~\ref{fig:MR}. 
 From NICER observations of PSR~J0740+6620, $M_{\rm obs.}$ = 2.08~$M_\odot$ and $R_{\rm obs.}$ = (12.35$\pm$0.75) km~\cite{miller2021} with $M_{\rm obs.}$ and $R_{\rm obs.}$ being the observed values of mass and radius, 
and $M_{\rm obs.}$ = (2.072~$^{+0.067}_{-0.066}$) $M_\odot$ and $R_{\rm obs.}$ = (12.39$^{+1.30}_{-0.98}$)~km~\cite{riley2021} (upper rectangle region bounded by the dashed lines). 
 For PSR~J1810+1744, $M_{\rm obs.}$=(2.13$\pm$0.04)$M_\odot$ \cite{romani2021} (hatched region). 
 From NICER observations of PSR~J0030+0451, $M_{\rm obs.}$ = (1.34~$^{+0.15}_{-0.16}$) $M_\odot$ and $R_{\rm obs.}$ = (12.71$^{+1.14}_{-1.19}$)~km~\cite{riley2019}, and $M_{\rm obs.}$ = (1.44~$^{+0.15}_{-0.14}$) $M_\odot$ and $R_{\rm obs.}$ = (13.02$^{+1.24}_{-1.06}$)~km~\cite{miller2019} (two rectangle regions bounded by the dashed lines). 

The curves of $M$-$R$ relations based on the EOS in the case of $L$ =(65, 70) MeV pass through the above constrained regions. In particular, the maximum masses with the ($Y$+$K$) phase in the core are consistent with recent observations of massive neutron stars in both cases of $\Sigma_{Kn}$ = 300 MeV and 400 MeV for $L$ =(65, 70) MeV.  However, the masses within the causal limit for $\Sigma_{Kn}$ = 400 MeV and $L$ = 60 MeV do not reach the range allowable from the observations of most massive neutron stars. 
In this model, the larger values of the slope $L > $ 60~MeV are preferred in order to obtain observed massive neutron stars.  
The radius of PSR~J0740+6620 has recently been improved by the use of the updated NICER data and the $X$-ray Multi-Mirror (XMM-Newton) observation~\cite{salmi2024,dittmann2024}. The result for $L$ =(65, 70) MeV is still consistent with this new constraint. 

The radius $R$ in the stable branches is also in conformity with observational constraints from gravitational waves of the binary neutron star mergers GW170817~\cite{abbott2018,horowitz2018}. 
However, recent NICER observation of the millisecond pulsar PSR J0437$-$4715 reported 
the equatorial radius $R_{\rm obs.}$ = (11.36$^{+0.95}_{-0.63}$)~km for $M_{\rm obs.}=(1.418\pm 0.037)M_\odot$~\cite{reardon2024,choudhury2024}, which, if it is confirmed, suggests that the EOS for $\rho_{\rm B}\lesssim 0.40$~fm$^{-3}$ (= 2.5~$\rho_0$) should be softer than the present case of $L$ =(65, 70) MeV in the model, as seen from Figs.~\ref{fig:MR}. [see also Sec.~\ref{subsec:L-EOS-SNM} for the stiffness of the EOS in PNM and SNM. ]

KC does not directly suffer from the repulsive effects of the UTBR, so that softening of the EOS with the ($Y$+$K$) phase proceeds steadily as density increases due to the energy decrease coming from the $s$-wave $K$-$B$ attraction~\cite{mmt2021} in addition to avoiding effect of $N$-$N$ repulsion by hyperon-mixing~\cite{nyt02}. As a result, the $M$-$R$ branch with the ($Y$+$K$) phase becomes flat in the vicinity of maximum mass as compared with the case for normal neutron stars without including phase transition. Such feature about the $M$-$R$ branch is unique for multi-strangeness phase, and may be detectable by accumulated information on $M$ and $R$ for various massive neutron stars. 

In this paper, muons are taken into account as leptons as well as electrons to obtain the ground state of the $\beta$-equilibrated matter, while only the electrons have been included as leptons in the previous result~\cite{mmt2021}. 
Above the threshold density of muons, a part of the electron fraction is imparted to the muon fraction to form each Fermi sphere in $\beta$ equilibrium, so that the electron Fermi momentum $p_F(e^-)$, which is equal to the charge chemical potential $\mu$, becomes small in the presence of muons as compared to the case where only the electrons are present at the same baryon density~\cite{mmt2021}. 
As a result, the onset density of KC, $\rho_{\rm B}(K^-)$, shifts to a larger density than the $e^-$-only case by about 5$\%$ for all the cases of $L$. Further, the maximum mass and its radius shift to larger values by at most 0.3~$\%$ and 2~$\%$, respectively for $L$=(65, 70)~MeV. Thus the effects of muons on the onset of KC and structure of compact stars are small.

\subsection{Density distribution in the core for typical masses 2.0~$M_\odot$ and 1.4~$M_\odot$}
\label{subsec:density-distribution}

\begin{figure*}[!]
\begin{minipage}[l]{0.50\textwidth}
\begin{center}~
\vspace{1.5cm}~
\includegraphics[height=.31\textheight]{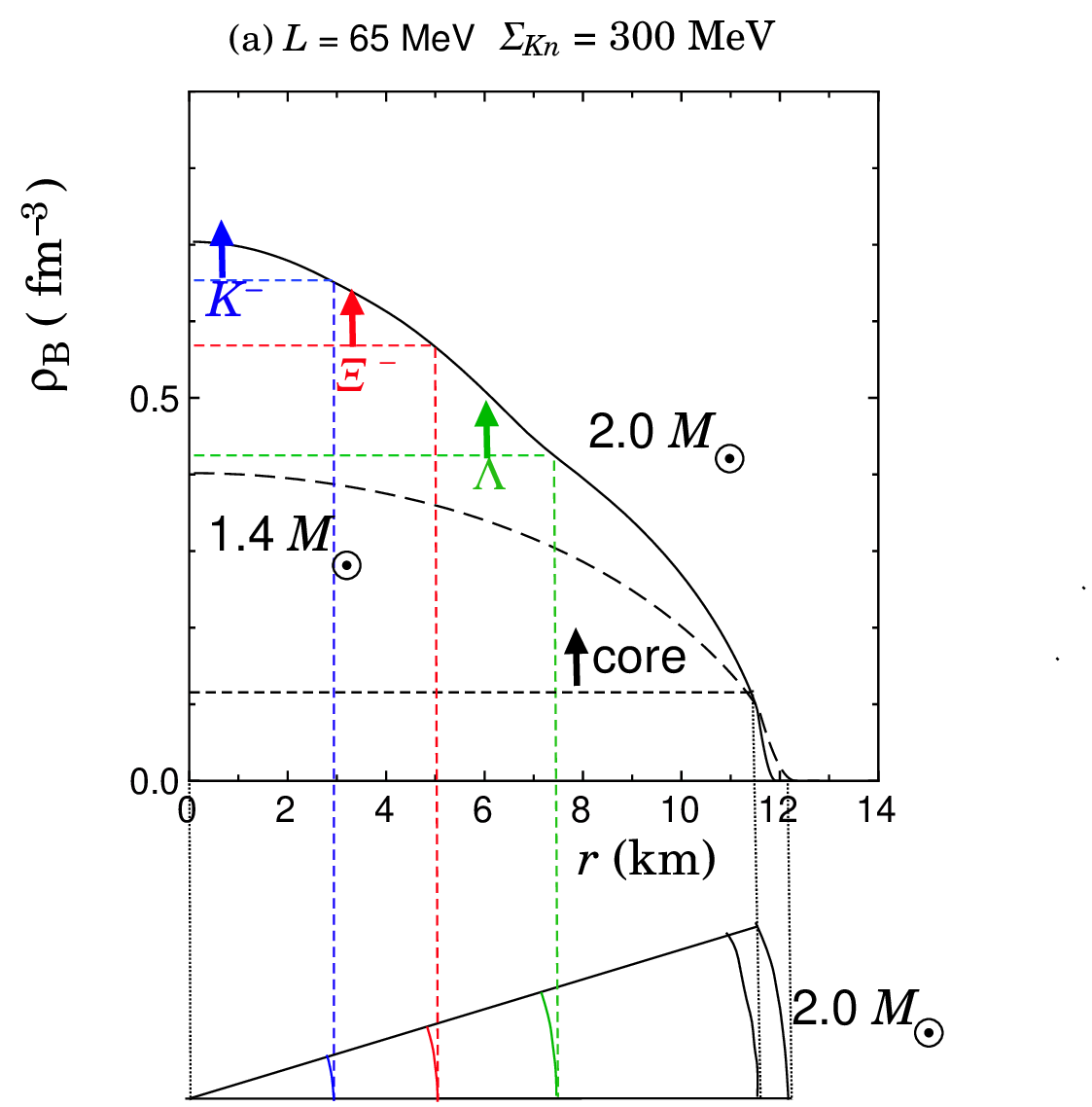}
\end{center}
\end{minipage}~
\begin{minipage}[r]{0.50\textwidth}
\begin{center}~
\vspace{1.5cm}~
\includegraphics[height=.31\textheight]{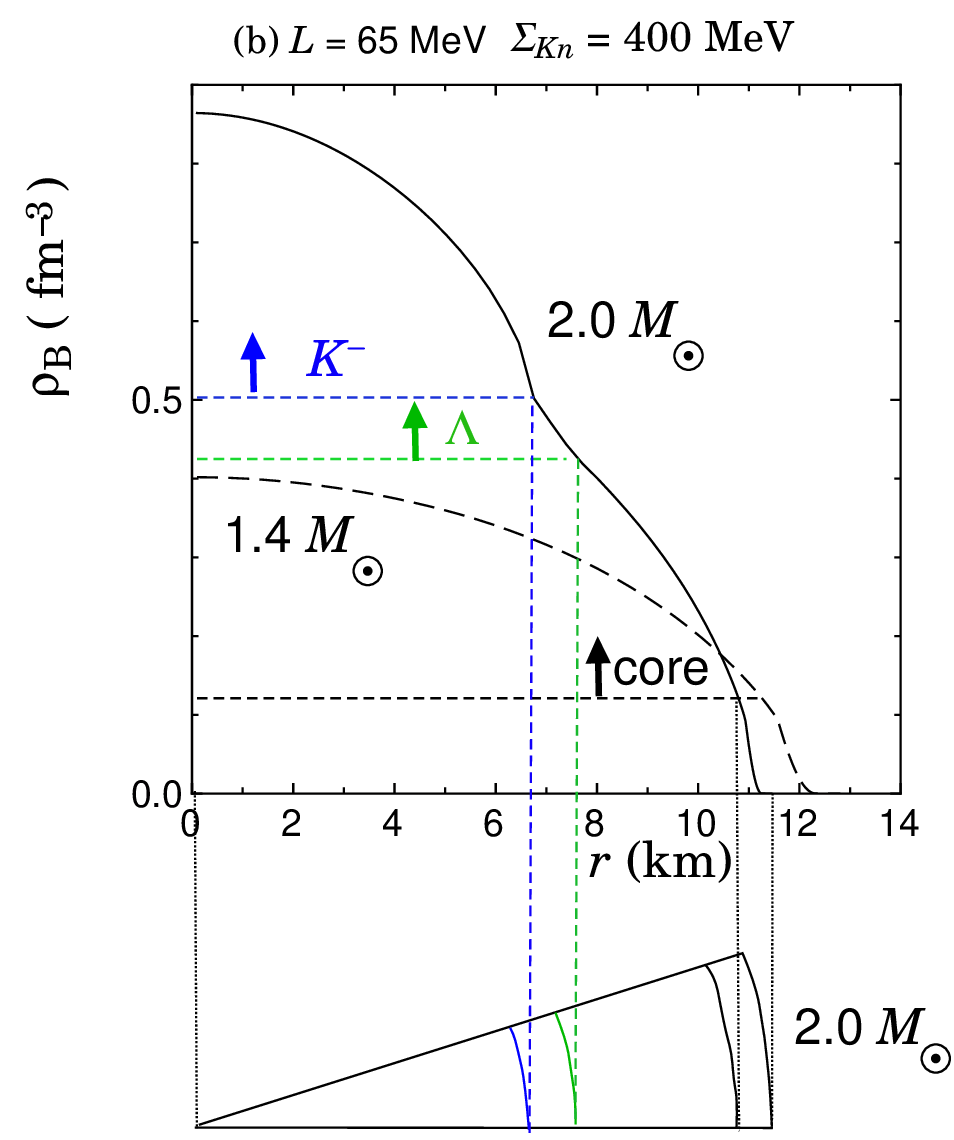}
\end{center}
\end{minipage}~
\caption{(a) The density profiles of a compact star with $M$ = 2.0~$M_\odot$ and $M$ = 1.4~$M_\odot$ for $L$ = 65~MeV and $\Sigma_{Kn}$ = 300~MeV as a function of the radial distance $r$ from the center. The corresponding cross section of the star with $M$ = 2.0~$M_\odot$ is also shown in the lower figure.  (b) The same as (a) but for $\Sigma_{kn}$ = 400 MeV. }
\label{fig:csL65}
\end{figure*}

\begin{figure*}[!]
\begin{minipage}[l]{0.50\textwidth}
\begin{center}~
\vspace{1.5cm}~
\includegraphics[height=.31\textheight]{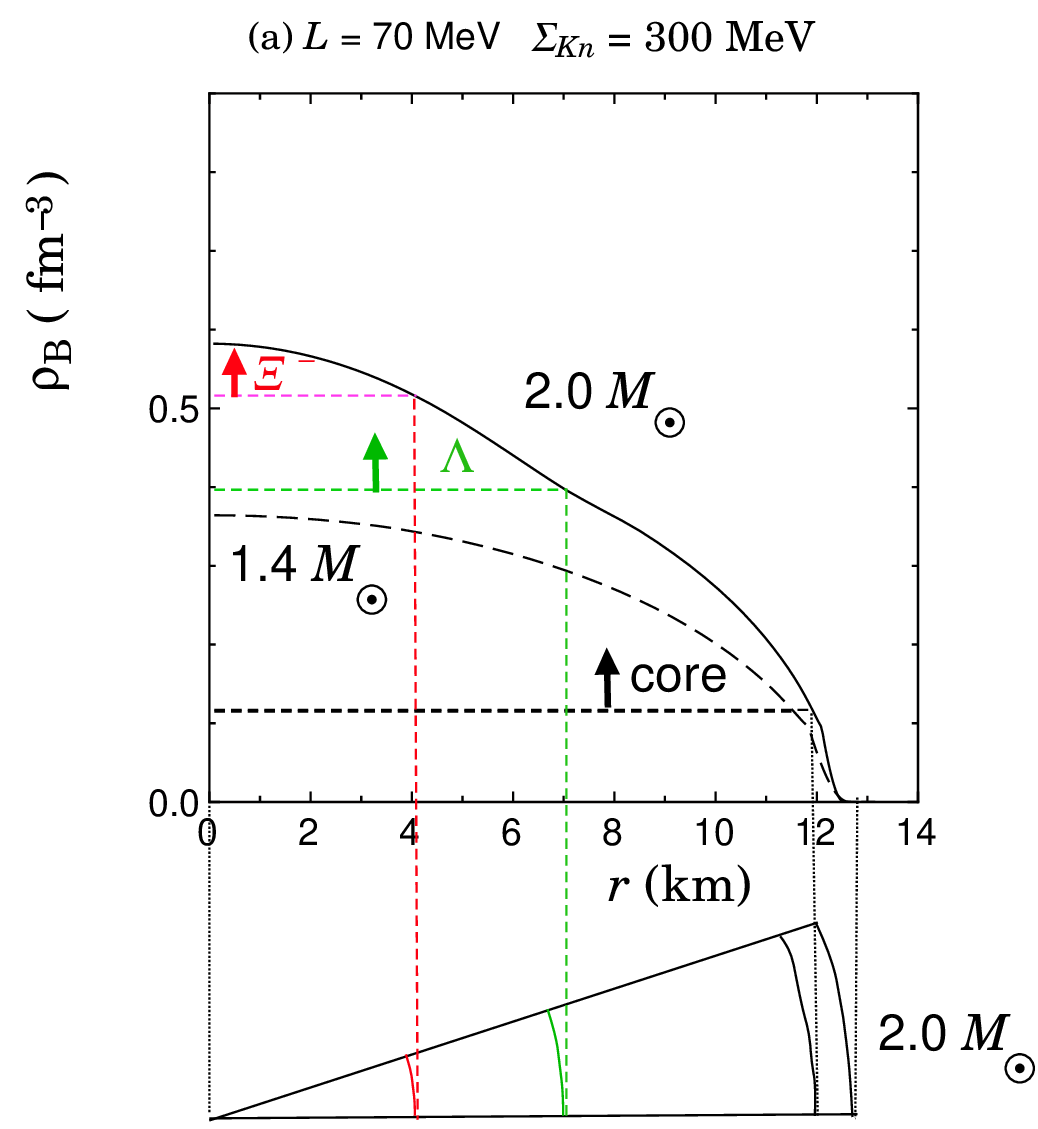}
\end{center}
\end{minipage}~
\begin{minipage}[r]{0.50\textwidth}
\begin{center}~
\vspace{1.5cm}~
\includegraphics[height=.31\textheight]{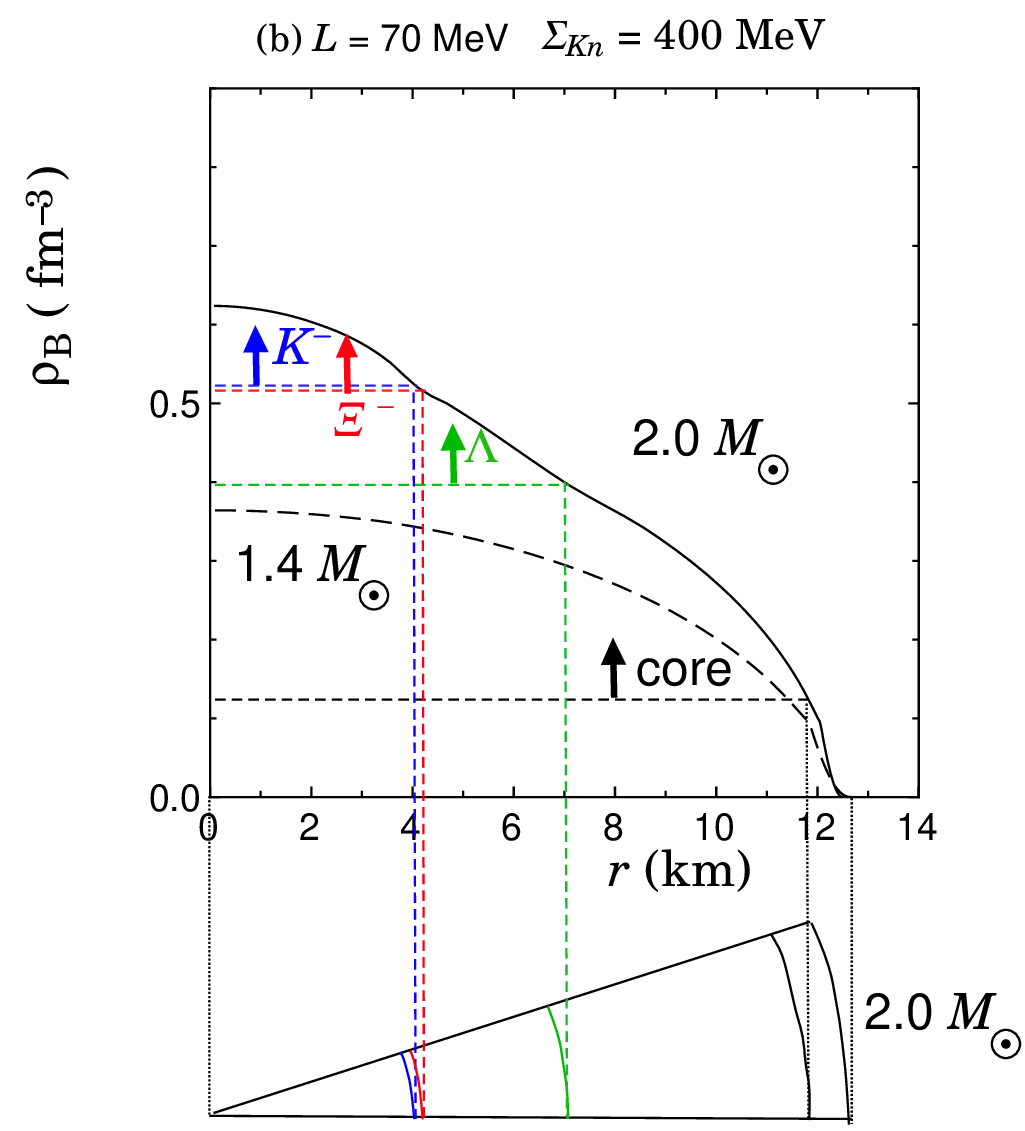}
\end{center}
\end{minipage}~
\caption{(a) The same as Fig.~\ref{fig:csL65}~(a) but for $L$ = 70~MeV. 
(b) The same as Fig.~\ref{fig:csL65}~(b) but for $L$ = 70~MeV. }
\label{fig:csL70}
\end{figure*}

The density profiles of a compact star with $M$ = 2.0~$M_\odot$ and $M$ = 1.4~$M_\odot$, as a function of the radial distance $r$ from the center, are shown in Figs.~\ref{fig:csL65} and \ref{fig:csL70}. The former is for $L$ = 65~MeV with (a) $\Sigma_{Kn}$ = 300~MeV and (b) $\Sigma_{Kn}$ = 400~MeV. Figure~\ref{fig:csL70} is the same as Fig.~\ref{fig:csL65}, but for $L$ = 70~MeV. The corresponding cross section of the star with $M$ = 2.0~$M_\odot$ is also shown in the lower part of each figure. 

For neutron stars with $M$ = 2.0 $M_\odot$, 
substantial portion of the core may be occupied with the ($Y$+$K$) phase except for the case of $L$ = 70~MeV and $\Sigma_{Kn}$ = 300 MeV, where the central density cannot reach the $\rho_{\rm B}^c(K^-)$. 
 In the case of $L$ = 65 MeV and $\Sigma_{Kn}$ = 300 MeV ($L$ = 65 MeV and $\Sigma_{Kn}$ = 400 MeV), one has the ($Y$+$K$) core within the region of radius $r$ =  3.0~km (6.8~km).  In the case of $L$ = 70 MeV and $\Sigma_{Kn}$ = 400 MeV, one has the ($Y$+$K$) core with $r$=4.0~km. Combined with Fig.~\ref{fig:fractions}, one can see that the hyperons are occupied mainly by the $\Lambda$ and that the $\Xi^-$ fractions are tiny or negligible within the ($Y$+$K$) phase for all the choices of $L$ and $\Sigma_{Kn}$. 

For neutron stars with $M\lesssim$1.4 $M_\odot$, the central density does not reach $\rho_{\rm B}^c (\Lambda)\approx 0.4$~fm$^{-3}$ for $L$=(65, 70)~MeV, and the ground state in the core consists of only $n$, $p$, and leptons ($e^-$, $\mu^-$). 

In relation to a neutron star cooling, one shows, with the asterisk in Fig.~\ref{fig:Mrhoc}, the threshold density at the center of the star for the direct Urca process ($np$-DU), $n\rightarrow p+l +\bar\nu_l$, $p+l\rightarrow  n + \nu_l$ ($l=e^-, \mu^-$)~\cite{lattimer1991}, and corresponding gravitational mass of the whole star. 
For a star with canonical mass 1.4 $M_\odot$, the central density $\rho_{\rm B, center}$ slightly exceeds the threshold density for the $np$-DU process for $L$ = (60$-$70) MeV.

For a heavy neutron star with $\approx$ 2.0~$M_\odot$, one can expect the ($Y$+$K$) phase in the core of the star. 
In the presence of KC, extra rapid cooling processes by neutrino and antineutrino emissions,
 i.~e.~the kaon-induced Urca (KU) process for nucleon ($N=p, n$), 
$ N + \langle K^-\rangle\rightarrow N + l +\bar\nu_l $, $N + l \rightarrow N + \langle K^- \rangle + \nu_l $, can occur with $\langle K^- \rangle$ being the classical $K^-$ field~\cite{t88,bkpp1988}.  
In the ($Y$+$K$) phase, the kaon-induced Urca process for hyperons (abbreviated to $Y$-KU) may also be suggested, while the dominant processes may be the hyperon ($\Lambda$) direct Urca process~\cite{plp1992} from the region surrounding the ($Y$+$K$) core.  
These weak processes are important to investigate the possible existence of the ($Y$+$K$) phase. 

\section{Summary and concluding remarks}
\label{sec:summary}

One has investigated the coexistent phase of kaon condensates and hyperons [($Y$+$K$) phase] by the use of the interaction model based on the effective chiral Lagrangian (ChL) for $K$-$B$ and $K$-$K$ interactions combined with the minimal relativistic mean-field theory (MRMF) for two-body baryon interaction, taking into account the universal three-baryon repulsion (UTBR) and the phenomenological three-nucleon attraction (TNA). 
 Interplay between KC and hyperons and resulting onset mechanisms of KC in hyperon-mixed matter and the EOS with the 
($Y$+$K$) phase have been clarified within the (ChL+MRMF+UTBR+TNA) model. 
While the $s$-wave $K$-$B$ scalar attraction simulated by the $K$-$B$ sigma terms favors coexistence of KC and hyperons, they tend to compete with each other by the $s$-wave $K$-$B$ vector repulsion as well as diminishing the $K$-$N$ ($N = p, n$) vector attraction as a result of mixing of hyperons (especially $\Lambda$ hyperons).  
In addition, the former attraction gets moderate at high densities because of suppression of the total baryon scalar density 
as the relativistic effect. Further, the $K$-$B$ scalar attraction is suppressed more along with the growth of KC, 
which is called as self-suppression mechanism. 
As a result, softening of the EOS due to KC becomes moderate in the presence of hyperons at high densities. 

Both the UTBR and two-baryon repulsion (Two-BR) render the stiff EOS at high densities. 
The repulsive energy contribution from the latter stems from the vector-meson exchange within the RMF and is specified by 
the slope $L$ of the symmetry energy. Also, the value of $L$ is partly related with the TNA at the saturation density $\rho_0$ through the isospin-dependence of the TNA. Therefore, not only the UTBR but also the TNA is responsible for the stiffness of the EOS at high densities. 
 A larger value of $L$ leads to stiffer EOS over the relevant densities 
and favors mixing of hyperons at the lower density, which, in turn, pushes up onset density of KC to higher density 
and leads to less energy gain by KC as a result of the competing effect between KC and hyperons.  

The EOS and the resulting mass and radius of compact stars within hadronic picture accompanying the ($Y$+$K$) phase are consistent with recent observations of massive neutron stars. 
In the interaction model (ChL+MRMF+UTBR+TNA), the higher value of the slope $L$ ($\gtrsim$ 65 MeV) is preferred, 
since, for $L < $ 60MeV, considerable softening of the EOS due to the appearance of hyperons and/or KC at high densities ($\rho_{\rm B}\gtrsim 2.5\rho_0$) cannot be compensated with the stiffening effect due to the Two-BR. 

One has assumed that the UTBR works between different kinds of baryons on an equal footing 
on the ground that such repulsion at high densities originates from confinement of quarks in the baryons and is independent of spin-flavor of baryons. Validity of the universality of the UTBR should be examined by comparing with other results in chiral perturbation theory applied to baryons~\cite{petschauer2020}, quark models including the quark Pauli effects~\cite{oka2012,nakamoto2016} and lattice QCD results~\cite{inoue2019}. 
 
 It has been suggested from the transverse and elliptic flow data in nuclear collisions that some extra softening may be needed for the EOS over baryon densities (2$-$4)$\rho_0$~\cite{danielewicz2002}. 
 One of the feasible solutions is appearance of pion condensation (PC) driven by the $p$-wave $\pi NN$ attractive interaction. The onset density of charged pion ($\pi^c$) condensation with realistic medium effects is estimated to be $\rho_{\rm B}\approx 2 \rho_0$~\cite{tatsumi1982}, so that the combined condensation of $\pi^c$ and kaons ($\pi$-$K$ condensation) at high densities should be elaborated within the relativistic framework. In the combined $\pi$-$K$ condensation, not only PC but also KC have a spatial momentum through the $p$-wave $\pi$-$K$ interactions. In the ground state of the $\pi$-$K$ condensed phase in neutron-star matter without hyperon-mixing, the energy eigenstates are given by quasi-baryonic states with superposition of neutron and proton states under the $p$-wave $\pi^c$ condensates. The ground state is occupied solely by the lower energy eigenstates of the quasi-baryons, forming the one-Fermi sea, which may resolve the assumption of the universal strengths between different species of baryons for the UTBR. 

Throughout this paper, the $s$-wave KC has been considered for simplicity. Considering KC in hyperon-mixed matter, the $p$-wave $KNY$ interaction should be naturally taken into account as well as the $s$-wave $K$-$B$ interaction. It has been shown that collective modes composed of a pair of the proton-$\Lambda$-hole with $K^+$ quantum number and the $\Xi^-$-neutron hole with $K^-$ quantum number may be spontaneously created and lead to $p$-wave kaon condensation at densities where $\Lambda$ hyperons are more abundant than protons~\cite{muto2002}. These collective modes would also couple to $\pi^c$ and may reveal various aspects of $\pi$-$K$ condensation in hyperon-mixed matter. 

 As stated in Sec.~\ref{subsec:EOS}, the EOS in the present model faces with a causal limit at $\rho_{\rm B}\approx$1.03~fm$^{-3}$. In the recent work, one has improved the UTBR by assuming a Lorentz-scalar form so as to safely meet the causality condition over the relevant densities realized in compact stars~\cite{muto2024}. 
Another scenario for preventing the hadronic EOS from causality violation is a possible 
 connection of hadronic matter including the ($Y$+$K$) phase to quark matter at high densities.  
 In particular, there are extensive studies in which hadronic matter was connected to quark matter smoothly by crossover transition to obtain massive neutron stars compatible with observations~\cite{mht2013,baym2018,baym2019,kojo2021,fujimoto2022}. 
 In case of the model in this paper, the baryon density at causal limit  ($\approx$ 1.03~fm$^{-3}$) corresponds to the density 
 where repulsive core radius of baryons being set to 0.5~fm is attached together, implying the beginning of quark percolation. 
 The connection of hadronic phase including the ($Y$+$K$) phase with quark degrees of freedom at high densities is needed in future works. Role of meson condensation in both hadron phase and quark phase is also an important subject.

In order to clarify the existence of the ($Y$+$K$) phase, novel features associated with the ($Y$+$K$) phase and their implications for astrophysical phenomena should be explored. 
For example, the effect of strong magnetic field on the onset density of KC and the EOS in neutron-star matter has been studied with the quark-meson coupling model~\cite{ys2008}. It has been shown that strong magnetic field as large as $10^{18}$ G may alter the onset density of KC, matter composition, and the EOS significantly. For another example, effects of the symmetry energy on KC were studied for protoneutron stars with fixed entropy and trapped neutrinos with the modified quark-meson coupling model. 
It has been suggested that the melting of KC and a second formation of KC may be signalized by neutrinos during the cooling process~\cite{panda2014}.
Further, KC might reveal unique features as superconductivity through 
responses to rotation, strong magnetic field, and radial/nonradial oscillations of compact stars as a result of strong coupling to baryonic system. 

With regard to thermal evolution of neutron stars, several neutron stars have anomalously low temperature that suggests extraordinary rapid cooling processes~\cite{tsuruta1998,dohi2019,dohi2022,marino2024}. 
It has been shown that the relevant weak reactions and $K$-$B$ dynamics associated with KC can be described in a unified manner on the basis of chiral symmetry~\cite{t88}. The emissivities for these reactions should be obtained in conjunction with the quantities such as composition of matter for the ($Y$+$K$) phase. 
Cooling curves with the use of the emissivity in the ($Y$+$K$) phase and comparison with 
observation of surface temperatures should be elucidated in detail~\cite{bdmn2024}.

\vspace{-0.5cm}~
\section*{Acknowledgments}
The author expresses sincere thanks to T.~Tatsumi and Toshiki Maruyama for useful suggestions during the collaborative works. He is also grateful to H.~Sotani, N.~Yasutake, T.~Noda, A.~Dohi, and B.~Bhavnesh for discussion and interest in the present work. This work was financially supported by Chiba Institute of Technology.

\section{Appendix: Related quantities and consideration}
\label{sec:appendix}

\subsection{The baryon masses and the ``$K$-baryon sigma term''}
\label{subsec:appendixA}

Here the expressions of the ``$K$$b$ sigma terms'', which are defined as $\displaystyle\Sigma_{Kb}\equiv \frac{1}{2}(m_u+m_s)\langle b|(\bar u u +\bar s s)|b\rangle $, are given. The baryon rest masses $M_b$ ($b$=$p,n,\Lambda, \Sigma^-, \Xi^-$) are read from the last three terms in (\ref{eq:lagkb})  by 
\begin{eqnarray}
M_p &=& \bar M_B-2(a_1m_u+a_2m_s) \ , \cr
M_n &=& \bar M_B-2(a_1m_d+a_2m_s) \ , \cr
M_\Lambda &=&\bar M_B-1/3\cdot (a_1+a_2)(m_u+m_d+4m_s) \ , \cr
M_{\Sigma^-} &=& \bar M_B-2(a_1m_d+a_2m_u) \ , \cr
M_{\Xi^-} &=& \bar M_B-2(a_1m_s+a_2m_u)
\label{eq:fbmass}
\end{eqnarray}
with $ \bar M_B=M_B - 2a_3(m_u+m_d+m_s)+{\it \Delta} M(m_s)$, 
where the rest mass contribution of baryons, ${\it \Delta} M(m_s)$ in higher order with respect to $m_s$, has been taken into account. For the sake of brevity, ${\it \Delta} M(m_s)$ is assumed to be dependent upon only the strange quark mass $m_s$. 

The $\bar q q$ contents in the baryon $b$ are obtained by the use of the Feynman-Hellmann theorem, $ \langle b|\bar q q |b\rangle $=$\partial M_b/\partial m_q$, with the baryon masses given by (\ref{eq:fbmass}). The result is
\begin{eqnarray}
 \langle p|\bar u u |p\rangle &=&  \langle n|\bar d d |n\rangle = - 2(a_1+a_3) \ , \cr
  \langle p|\bar d d |p\rangle &=&  \langle n|\bar u u |n\rangle = - 2a_3 \ , \cr
    \langle p|\bar s s |p\rangle &=&  \langle n|\bar s s |n\rangle = - 2(a_2+a_3)+\Delta \ 
    \label{eq:qNcontent}
\end{eqnarray}
with $\Delta\equiv d{\it \Delta}M(m_s)/dm_s$, 
\begin{eqnarray}
 \langle \Lambda|\bar u u |\Lambda\rangle &=&  \langle \Lambda|\bar d d |\Lambda\rangle = - \frac{1}{3}(a_1+a_2)-2a_3 \ , \cr
  \langle \Lambda|\bar s s |\Lambda\rangle &=& - \frac{4}{3}(a_1+a_2)-2a_3 +\Delta \ , \cr
\langle \Sigma^-|\bar u u  |\Sigma^- \rangle &=& -2(a_2+a_3) \ , \langle \Sigma^-|\bar d d  |\Sigma^- \rangle = -2(a_1+a_3) \ , \cr
  \langle \Sigma^-|\bar s s |\Sigma^-\rangle &=& - 2a_3 +\Delta \ , \cr
\langle \Xi^-|\bar u u  |\Xi^- \rangle &=& -2(a_2+a_3) \ , \langle \Xi^-|\bar d d  |\Xi^- \rangle = -2a_3 \ , \cr
  \langle \Xi^-|\bar s s |\Xi^-\rangle &=& - 2(a_1+a_3)+\Delta \ .
    \label{eq:qYcontent}
\end{eqnarray}

The ``$K$-baryon sigma terms'' $\Sigma_{Kb}$ are
 given in terms of $a_1\sim a_3$ and $m_u$, $m_d$, $m_s$ by
\begin{subequations}\label{eq:Akbsigma}
\begin{eqnarray}
\Sigma_{Kn}&=&-(a_2+2\widetilde a_3)(m_u+m_s) = \Sigma_{K\Sigma^-} \ ,\label{eq:Akbsigma1} \\
\Sigma_{K\Lambda}&=& -\left(\frac{5}{6}a_1+\frac{5}{6}a_2+2\widetilde a_3\right)(m_u+m_s) \ , \label{eq:Akbsigma2} \\
\Sigma_{Kp}&=&-(a_1+a_2+2\widetilde a_3)(m_u+m_s) = \Sigma_{K\Xi^-}  \label{eq:Akbsigma3}
\end{eqnarray}
\end{subequations}
with $\widetilde a_3\equiv a_3-\Delta/4$ . 
\vspace{-0.1cm}

\subsection{$S$-wave on-shell $K$-$N$ scattering lengths}
\label{subsec:appendixB}

The $s$-wave on-shell $K$-$N$ scattering lengths $a(K N)$ ($K=K^-, K^+$, $N=p, n$) are read off from the self-energy of $K^-$ meson Eq.~(\ref{eq:selfk}) by setting $\omega_K\rightarrow m_K$. 
Taking into account the imaginary part of the $a(K N)$ only from the $\Lambda^\ast$-pole contribution, one obtains
\begin{widetext}
\begin{subequations}\label{eq:KNsc}
\begin{eqnarray}
a(K^-p)&=&\frac{1}{4\pi f^2(1+m_K/M_N)}\left(\Sigma_{Kp}+m_K+d_p m_K+\frac{g_{\Lambda^\ast}^2}{2}\frac{m_K^2}{M_{\Lambda^\ast}-M_N-m_K-i\gamma_{\Lambda^\ast}}\right) \ , \label{eq:KNsc1} \\
a(K^-n)&=&\frac{1}{4\pi f^2(1+m_K/M_N)}\left(\Sigma_{Kn}+\frac{1}{2}m_K+d_n m_K \right) \ ,  \label{eq:KNsc2} \\
 a(K^+ p)&=&\frac{1}{4\pi f^2(1+m_K/M_N)}\left(\Sigma_{Kp}-m_K+d_ p m_K+\frac{g_{\Lambda^\ast}^2}{2}\frac{m_K^2}{M_{\Lambda^\ast}-M_N
+m_K-i\gamma_{\Lambda^\ast}}\right) \label{eq:KNsc3} \ ,  \\
a(K^+ n)&=&\frac{1}{4\pi f^2(1+m_K/M_N)}\left(\Sigma_{Kn}-\frac{1}{2}m_K+d_n m_K \right) \ , \label{eq:KNsc4}
\end{eqnarray}
\end{subequations}
\end{widetext}
where $d_p\equiv (d_1+d_2)m_K/2$ = $d_{\Xi^-}$, $d_n \equiv d_1m_K/2$ = $d_{\Sigma^-}$, $d_\Lambda=(d_1+5d_2/6)m_K/2$.

The empirical values of the $K$-$N$ scattering lengths, $a^{\rm exp}(K^\pm N)$, are taken from Ref.~\cite{martin1981}: 
\begin{subequations}\label{eq:KNexp}
\begin{eqnarray}
a^{\rm exp}(K^- p)&=& -0.67+i 0.64~{\rm fm} \ , \label{eq:KNexp1} \\
 a^{\rm exp}(K^- n)&=&0.37+i 0.60~{\rm fm}  \ , \label{eq:KNexp2} \\
 a^{\rm exp}(K^+ p)&=& - 0.33~{\rm fm} \ , \label{eq:KNexp3} \\
 a^{\rm exp}(K^+ n)&=& - 0.16~{\rm fm} \ .\label{eq:KNexp4}
 \end{eqnarray}
 \end{subequations}
 From these values one obtains 
$g_{\Lambda^\ast}$ = 0.583, $\gamma_{\Lambda^\ast}$=12.4~MeV, $d_p$ = $0.351-\Sigma_{Kp}/m_K$, 
$d_n$ = $0.130-\Sigma_{Kn}/m_K$.
\vspace{-0.1cm}

\subsection{Expressions for quantities associated with saturation properties in SNM}
\label{subsec:appendixC}

The expressions for the incompressibility $K_0$, the symmetry energy $S_0$, and the slope $L$ in the (MRMF+UTBR+TNA) model are given as follows. 
\vspace{-0.5cm}~

\subsubsection{Incompressibility $K_0$}
\label{subsubsec:K}

\begin{eqnarray}
\hspace{-0.7cm}~K_0&=&9\rho_0\left(\partial^2{\cal E}/\partial\rho_{\rm B}^2\right)_{\rho_{\rm B} 
=\rho_0} \cr 
&=& K_0^{({\rm kin})}+K_0^{(\sigma)}+K_0^{(\omega)} + K_0^{(\rm UTBR)}+K_0^{(\rm TNA)} \ , 
\label{eq:K0total}
\end{eqnarray}
where
\begin{subequations}\label{eq:K0}
\begin{eqnarray}
K_0^{({\rm kin})}&=&9\rho_0 \frac{\pi^2}{2}\frac{1}{p_F \widetilde E(p_F)} \ , 
 \label{eq:K0kin} \\ 
K_0^{(\sigma)}&=& -\frac{(g_{\sigma N}M_N^\ast/\widetilde E(p_F))^2}{m_\sigma^2+g_{\sigma N}^2 I~(p_F)} \ , \label{eq:K0sigma} \\
K_0^{(\omega)} &=&\left(\frac{g_{\omega N}}{m_\omega}\right)^2 \ , \label{eq:K0omega} \\
K_0^{(\rm UTBR)} &=& (2688.12) \pi^{3/2}(\widetilde\lambda_r)^3\rho_0^2 \ , \label{eq:K0UTBR} \\
K_0^{(\rm TNA)} &=& 27\gamma_a \rho_0^2 e^{-\eta_a\rho_0}(6-6\eta_a\rho_0+\eta_a^2\rho_0^2) \label{eq:K0TNA} 
\end{eqnarray}
\end{subequations}
with $\widetilde E(p_F)\equiv \sqrt{p_F^2+M_N^{\ast 2}}$ , 
$M_N^\ast$ ( =$M_N-g_{\sigma N}\langle\sigma\rangle_0$) being the effective nucleon mass at $\rho_0$.  
The integral $I~(p_F)$ in Eq.~(\ref{eq:K0sigma}) is given by
\begin{equation}
I~(p_F)\equiv \frac{2}{\pi^2}\int_0^{p_F} dp\frac{p^4}{(p^2+M_N^{\ast 2})^{3/2}}=3\left(\frac{\rho_0^S}{M_N^\ast}-\frac{\rho_0}{\widetilde E(p_F)}\right) 
\label{eq:I}
\end{equation}
with $\rho_0^S$ being the nuclear scalar density at $\rho_0$ in SNM 
($\rho_0^S\equiv\rho_p^S+\rho_n^S$ with $\rho_p^S$ = $\rho_n^S$ for SNM). 

\subsubsection{Symmetry energy $S_0$}
\label{subsubsec:S}

\begin{eqnarray}
S_0&=&\frac{1}{8}\left(\frac{\partial^2 E}{\partial x_p^2}\right)_{\rho_B=\rho_0} \cr
&=&S_0^{(\rm kin)}+S_0^{(\rho)}+S_0^{(\rm TNA)} \ , 
\label{eq:s0}
\end{eqnarray}
where
\begin{subequations}\label{eq:S0}
\begin{eqnarray}
S_0^{(\rm kin)}&=&\frac{p_F^2}{6\widetilde E(p_F)} \ , \label{eq:Sokin} \\
S_0^{(\rho)}&=&\frac{1}{2}\left(\frac{g_{\rho N}}{m_\rho}\right)^2\rho_0 \ , \label{eq:S0rho} \\
S_0^{(\rm TNA)}&=& -2\gamma_a\rho_0^2e^{-\eta_a\rho_0} \ .\label{eq:S0TNA}
\end{eqnarray}
\end{subequations}

\subsubsection{Slope $L$ of the symmetry energy}
\label{subsubsec:L}

\vspace{-0.2cm}~
\begin{eqnarray}
L&\equiv& 3\rho_0\left(\frac{d S(\rho_{\rm B})}{d \rho_B}\right)_{\rho_B=\rho_0}=\frac{3}{\rho_0}P_{\rm PNM}(\rho_0) \cr
&=&L^{({\rm kin})}+L^{(\rho)}+L^{(\rm TNA)}~, 
\label{eq:L}
\end{eqnarray}
where
\begin{subequations}\label{eq:LL}
\begin{eqnarray}
 L^{({\rm kin})}&=&\frac{p_F^2}{6\widetilde E(p_F)}\Bigg\lbrack 2-\left(\frac{p_F}{\widetilde E(p_F)}\right)^2 \\
 &+&\frac{2}{\pi^2}\frac{(g_{\sigma N}M_N^{\ast})^2 (p_F/\widetilde E(p_F))^3}{m_\sigma^2+g_{\sigma N}^2 I(p_F)} \Bigg\rbrack \ , \label{eq:Lkin} \\
L^{(\rho)}&=&\frac{3}{2}\left(\frac{g_{\rho N}}{m_\rho}\right)^2\rho_0 \ , \label{eq:Lrho} \\
L^{(\rm TNA)}&=&6\gamma_a\rho_0^2 (\eta_a\rho_0 - 2)e^{-\eta_a\rho_0} \ . \label{eq:LTNA} 
\end{eqnarray}
\end{subequations}

\subsection{The relativistic mean-field model with the nonlinear self-interacting meson potentials}
\label{subsec:a4}

Here the RMF model with the NLSI terms [(MRMF+NLSI) model]  is overviewed in order to compare with the (MRMF+UTBR+TNA) model adopted throughout this paper.

The Lagrangian density for the RMF with the NLSI terms are given by
\begin{equation}
{\cal L}_{\rm MRMF+NLSI}={\cal L}_{\rm MRMF}+{\cal L}_{\rm NLSI}  \ , 
\label{eq:Lag-MRMF+NLSI}
\end{equation}
where ${\cal L} _{\rm MRMF}$ (= ${\cal L}_K+{\cal L}_{B, M}$) is the minimal RMF with ${\cal L}_K$ [Eq.~(\ref{eq:lagk})] and ${\cal L}_{B, M}$ [Eq.~(\ref{eq:lagbm})], and ${\cal L}_{\rm NLSI}$ consists of the nonlinear self-interacting meson field terms. For the latter, one takes the following form
\begin{equation}
{\cal L}_{\rm NLSI}=-U_\sigma(\sigma)+\frac{1}{4}c_\omega (\omega^\mu\omega_\mu)^2
+\lambda_{\omega\rho}(\omega^\mu\omega_\mu)(R^{\mu,a}R_{\mu, a}) \ , 
\label{eq:lagNLSI}
\end{equation}
where 
\begin{equation}
U_\sigma (\sigma)=\frac{b}{3}M_N(g_{\sigma N}\sigma)^3+\frac{c}{4}(g_{\sigma N}\sigma)^4 \ .
\label{eq:Usigma}
\end{equation}
The first term [Eq.~(\ref{eq:Usigma})] on the r.~h.~s. of Eq.~(\ref{eq:lagNLSI}) is introduced to give the empirical value of the incompressibility $K_0$ (=240~MeV), the second term to reproduce properties of stable and unstable nuclei systematically~\cite{st1994}, and the third term to adjust to the slope $L$. 

The energy density expression is given by simply replacement of the ${\cal E}~({\rm UTBR})+{\cal E}~({\rm TNA})$ terms in Eq.~(\ref{eq:total-edensity}) by ${\cal E}_{\rm NLSI}$ as
\begin{equation}
{\cal E}= {\cal E}_K+{\cal E}_{B,M}+{\cal E}_{\rm NLSI}+{\cal E}_e+{\cal E}_\mu 
\label{eq:e-total-NLSI}
\end{equation}
with
\begin{equation}
{\cal E}_{\rm NLSI}=U_\sigma (\sigma)+\frac{3}{4}c_\omega\omega_0^4+3\lambda_{\omega\rho}\omega_0^2 R_0^2 \ . 
\label{eq:e-density-NLSI}
\end{equation}

The classical kaon field equation for $\theta$ is the same as Eq.~(\ref{eq:keom2}). The equations of motion for the meson-mean fields are given as
\begin{subequations}\label{eq:NLSIeom}
\begin{eqnarray}
m_\sigma^2\sigma&=&-dU_\sigma/d\sigma + \sum_{b=p,n,\Lambda,\Sigma^-,\Xi^-}g_{\sigma b}\rho_b^s \ , \label{eq:NLSIeom1} \\
m_\sigma^{\ast 2}\sigma^\ast&=&\sum_{Y=\Lambda,\Sigma^-,\Xi^-}g_{\sigma^\ast Y}\rho_Y^s \ , \label{eq:NLSIeom2}\\
m_\omega^2\omega_0&=&-c_\omega \omega_0^3-2\lambda_{\omega\rho}\omega_0 R_0^2 \cr
&+&\sum_{b=p,n,\Lambda,\Sigma^-,\Xi^-} g_{\omega b}\rho_b \ ,  \label{eq:NLSIeom4}\\
m_\rho^2 R_0&=&-2\lambda_{\omega\rho}\omega_0^2 R_0 + \sum_{b=p,n,\Lambda,\Sigma^-,\Xi^-} g_{\rho b}{\hat I}_3^{(b)} \rho_b \ , \label{eq:NLSIeom5} \\
m_\phi^2\phi_0 &=&\sum_{Y=\Lambda,\Sigma^-,\Xi^-}g_{\phi Y}\rho_Y \ . \label{eq:NLSIeom6}
\end{eqnarray}
\end{subequations}

In the case of the SNM, the energy density, equations of motion for the meson mean-fields reduce to
\begin{equation}
{\cal E}= {\cal E}_{B,M}+U_\sigma (\sigma)+\frac{3}{4}c_\omega\omega_0^4 \ , 
\label{eq:e-total-NLSI-SNM}
\end{equation}
\begin{subequations}\label{eq:SNM-NLSIeom}
\begin{eqnarray}
m_\sigma^2\sigma&=&-dU_\sigma/d\sigma + \sum_{b=p,n} g_{\sigma b}\rho_b^s \ , \label{eq:SNM-NLSIeom1} \\
m_\omega^2\omega_0&=&-c_\omega \omega_0^3 + \sum_{b=p,n} g_{\omega b}\rho_b \ ,  \label{eq:SNM-NLSIeom4}  
\end{eqnarray}
\end{subequations}
with $R_0$ = 0.

The incompressibility $K_0$, the symmetry energy $S_0$, and the slope $L$ in the (MRMF+NLSI) model are given as follows:
\begin{widetext}
\begin{subequations}\label{eq:NLSI-phys}
\begin{eqnarray}
K_0 &=& 9\rho_0 \Bigg\lbrace \frac{\pi^2}{2}\frac{1}{p_F \widetilde E(p_F)}-\frac{\left(g_{\sigma N} M_N^\ast/\widetilde E(p_F)\right)^2}{m_\sigma^2+d^2 U_\sigma/d\sigma^2\vert_{\sigma=\langle\sigma\rangle_0}+g_{\sigma N}^2 I(p_F)} 
+\frac{g_{\omega N}^2}{m_\omega^2+3c_\omega \langle\omega_0\rangle_0^2} \Bigg\rbrace  \ , \label{eq:NLSI-phys-K} \\
S_0 &=& \frac{p_F^2}{6\widetilde E(p_F)} +\frac{1}{2}\left(\frac{g_{\rho N}}{m_\rho}\right)^2\rho_0\frac{1}{1+2\lambda_{\omega\rho} \langle\omega_0\rangle_0^2/m_\rho^2}  \ , \label{eq:NLSI-phys-S} \\
L &=& \frac{p_F^2}{6\widetilde E(p_F)} \Bigg\lbrace 2-\left(\frac{p_F}{\widetilde E(p_F)}\right)^2+2\left(\frac{g_{\sigma N}M_N^\ast}{\pi} \right)^2 \frac{\left(p_F/\widetilde E(p_F)\right)^3}{m_\sigma^2+d^2 U_\sigma/d\sigma^2\vert_{\sigma=\langle\sigma\rangle_0}+g_{\sigma N}^2 I(p_F)} \Bigg\rbrace \cr
&+&\frac{3}{2}g_{\rho N}^2\rho_0  \Bigg\lbrace\frac{1}{m_\rho^2+2\lambda_{\omega\rho} \langle\omega_0\rangle_0^2}-\frac{4\lambda_{\omega\rho}g_{\omega N}\rho_0\omega_0}{(m_\rho^2+2\lambda_{\omega\rho}\langle\omega_0\rangle_0^2)^2(m_\omega^2+3c_\omega\langle\omega_0\rangle_0^2)}\Bigg\rbrace  \ . \label{eq:NLSI-phys-L}
\end{eqnarray}
\end{subequations}
\end{widetext}

\begin{figure}[!]
\begin{center}
\includegraphics[height=0.38\textwidth]{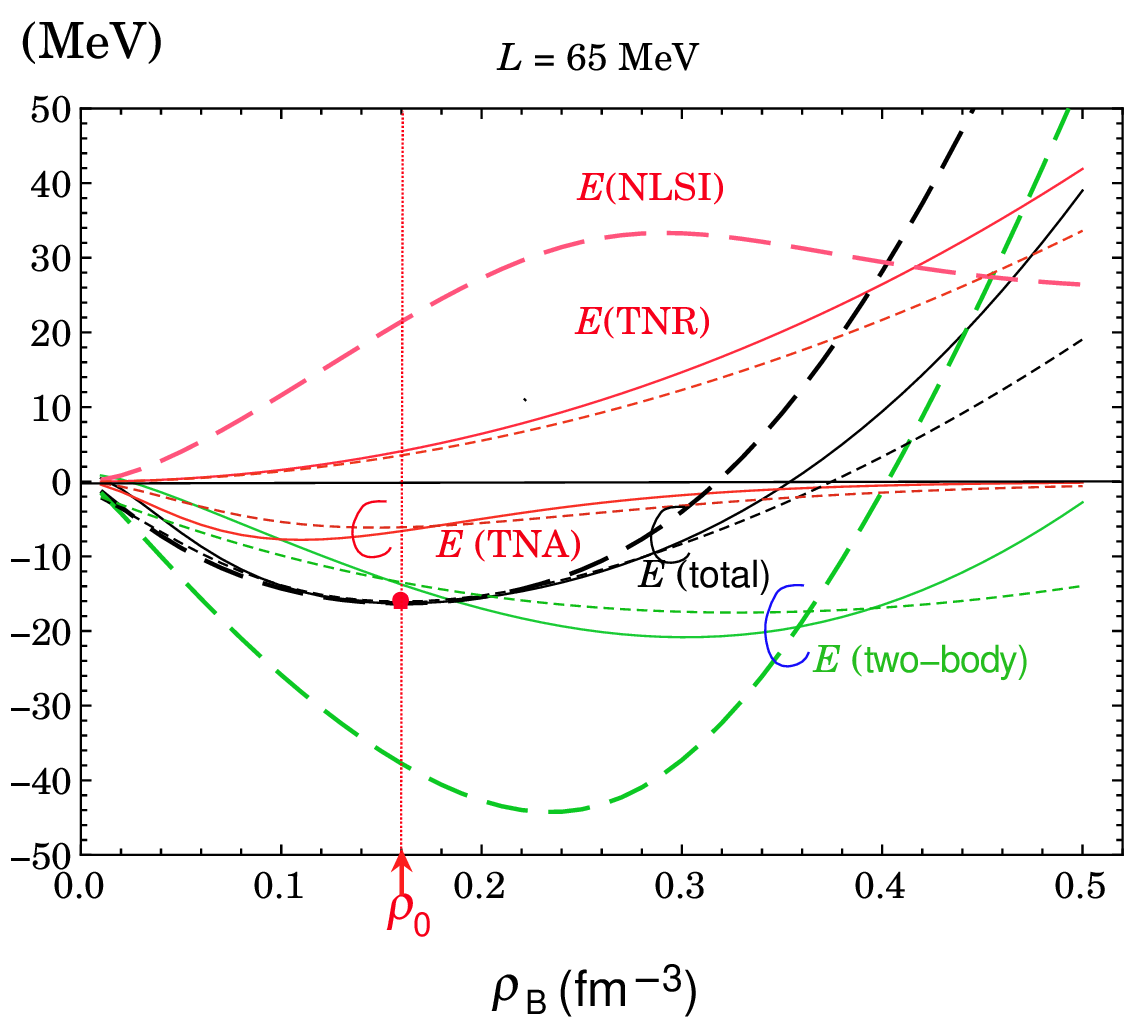}~
\end{center}~
\vspace{1.0cm}~
\caption{The result for the energy contributions with the use of the (MRMF+NLSI) model in the case of the slope $L$ = 65 MeV by the long dashed lines. For comparison, those obtained by the (MRMF+UTBR+TNA) model (this work) and those obtained from LP~(1981)\cite{lp1981} are shown by the solid lines and the dotted lines, respectively, which are the same as in Fig.~\ref{fig:esnm}. See the text for details.\\ }
\label{fig:eNLSI}
\end{figure}

In Fig.~\ref{fig:eNLSI}, the result for the energy contributions with the use of the (MRMF+NLSI) model as functions of $\rho_{\rm B}$ in SNM are shown by the long-dashed lines in the case of the slope $L$ = 65 MeV. $E$ (total) $(={\cal E}/\rho_{\rm B}$) is the total energy per nucleon, $E$ (NLSI) (=${\cal E}$~(NLSI)/$\rho_{\rm B}$) is the contribution from the NLSI terms, and $E$(two-body) (=${\cal E}_{B,M}/\rho_{\rm B}$) the sum of kinetic and two-body interaction energies, the expression of which is equal to Eq.~(\ref{eq:edSNM}) for the (MRMF+UTBR+TNA) case. The parameters are chosen as $b$ = 2.77$\times 10^{-3}$, $c$ = $-$3.62$\times 10^{-3}$, $c_\omega$ = 1.025, and $\lambda_{\omega\rho}$ = 3.82$\times 10^2$ for $L$ = 65 MeV.~ 
The coupling constants $g_{\sigma N}$, $g_{\omega N}$, $c_\omega$, and the meson mean fields, $\langle\sigma\rangle_0$,  $\langle\omega_0\rangle_0$, are determined from the saturation conditions, 
\begin{eqnarray*}
&&E~({\rm total})\vert_{\rho_{\rm B} = \rho_0}-M_N=-B_0, \cr
&&\partial E({\rm total})/\partial\rho_{\rm B}\vert_{\rho_{\rm B}=\rho_0} = 0
\end{eqnarray*}
 with  
\begin{equation*}
E~({\rm total}) = {\cal E}/\rho_{\rm B} 
=\left({\cal E}_{B,M}+U_\sigma (\sigma)+\frac{3}{4}c_\omega\omega_0^4\right)/\rho_{\rm B} \ ,
\end{equation*}
together with Eqs.~(\ref{eq:SNM-NLSIeom1}), (\ref{eq:SNM-NLSIeom4}), and (\ref{eq:NLSI-phys-K}). 
It is to be noted that the total energy per nucleon, $E$ (total), does not have isospin-dependent terms including the coupling constants $g_{\rho N}$ and $\lambda_{\omega\rho}$ in the SNM, since the rho-meson mean field $R_0$ vanishes due to isospin symmetry. 
Therefore, these equations are decoupled with the remaining equations, (\ref{eq:NLSI-phys-S}) and (\ref{eq:NLSI-phys-L}) with respect to the  $S_0$ and $L$, from which the coupling constant $g_{\rho N}$ and $\lambda_{\omega\rho}$ are determined. Thus, $E$(two-body) does not depend on the $g_{\rho N}$ and $\lambda_{\omega\rho}$ or the choice of the slope $L$. 

One can see that the saturation is fulfilled after large cancellation between the repulsive NLSI terms and two-body attraction. Further, the energy contribution from the NLSI terms has a peak at $\rho_{\rm B} \approx$ 0.29~fm$^{-3}$, and it monotonically decreases with increase in density. In contrast, the repulsive energy due to the two-body $B$-$B$ interaction becomes dominant at high densities ($\rho_{\rm B}\gtrsim$ 0.5~fm$^{-3}$) and controls the stiffness of the EOS. 

\bibliography{mybibfile-muto}

\end{document}